\documentclass[twoside,twocolumn,9pt]{article}
\usepackage{extsizes}
\usepackage[super,sort&compress,comma]{natbib}
\usepackage[version=3]{mhchem}
\usepackage[left=1.5cm, right=1.5cm, top=1.785cm, bottom=2.0cm]{geometry}
\usepackage{balance}
\usepackage{mathptmx}
\usepackage{sectsty}
\usepackage{graphicx}
\usepackage{lastpage}
\usepackage[format=plain,justification=justified,singlelinecheck=false,font={stretch=1.125,small,sf},labelfont=bf,labelsep=space]{caption}
\usepackage{float}
\usepackage{fancyhdr}
\usepackage{fnpos}
\usepackage[english]{babel}
\addto{\captionsenglish}{%
  
}
\usepackage{array}
\usepackage{droidsans}
\usepackage{charter}
\usepackage[T1]{fontenc}
\usepackage[usenames,dvipsnames]{xcolor}
\usepackage{setspace}
\usepackage[compact]{titlesec}
\usepackage{hyperref}
\usepackage{epstopdf}
\usepackage{amsmath,amssymb}
\usepackage{ulem}

\definecolor{cream}{RGB}{222,217,201}
\definecolor{darkgreen}{RGB}{0,125,65}
\definecolor{revmagenta}{RGB}{150,0,150}

\usepackage{booktabs}
\usepackage{makecell}

\newcommand{\medbullet}{\mathbin{\vcenter{\hbox{\scalebox{0.8}{$\bullet$}}}}}
\begin{document}

\pagestyle{fancy}
\thispagestyle{plain}
\fancypagestyle{plain}{\renewcommand{\headrulewidth}{0pt}}

\makeFNbottom
\makeatletter
\renewcommand\LARGE{\@setfontsize\LARGE{15pt}{17}}
\renewcommand\Large{\@setfontsize\Large{12pt}{14}}
\renewcommand\large{\@setfontsize\large{10pt}{12}}
\renewcommand\footnotesize{\@setfontsize\footnotesize{7pt}{10}}
\makeatother

\renewcommand{\thefootnote}{\fnsymbol{footnote}}
\renewcommand\footnoterule{\vspace*{1pt}%
\color{cream}\hrule width 3.5in height 0.4pt \color{black}\vspace*{5pt}}
\setcounter{secnumdepth}{5}

\makeatletter
\renewcommand\@biblabel[1]{#1}
\renewcommand\@makefntext[1]%
{\noindent\makebox[0pt][r]{\@thefnmark\,}#1}
\makeatother
\renewcommand{\figurename}{\small{Fig.}~}
\sectionfont{\sffamily\Large}
\subsectionfont{\normalsize}
\subsubsectionfont{\bf}
\setstretch{1.125}
\setlength{\skip\footins}{0.8cm}
\setlength{\footnotesep}{0.25cm}
\setlength{\jot}{10pt}
\titlespacing*{\section}{0pt}{4pt}{4pt}
\titlespacing*{\subsection}{0pt}{15pt}{1pt}

\fancyfoot{}
\fancyfoot[LO,RE]{\vspace{-7.1pt}\includegraphics[height=9pt]{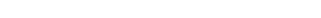}}
\fancyfoot[CO]{\vspace{-7.1pt}\hspace{13.2cm}\includegraphics{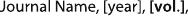}}
\fancyfoot[CE]{\vspace{-7.2pt}\hspace{-14.2cm}\includegraphics{head_foot/RF}}
\fancyfoot[RO]{\footnotesize{\sffamily{1--\pageref{LastPage} ~\textbar \hspace{2pt}\thepage}}}
\fancyfoot[LE]{\footnotesize{\sffamily{\thepage~\textbar\hspace{3.45cm} 1--\pageref{LastPage}}}}
\fancyhead{}
\renewcommand{\headrulewidth}{0pt}
\renewcommand{\footrulewidth}{0pt}
\setlength{\arrayrulewidth}{1pt}
\setlength{\columnsep}{6.5mm}
\setlength\bibsep{1pt}

\makeatletter
\newlength{\figrulesep}
\setlength{\figrulesep}{0.5\textfloatsep}
\newcommand{\topfigrule}{\vspace*{-1pt}\noindent{\color{cream}\rule[-\figrulesep]{\columnwidth}{1.5pt}} }
\newcommand{\botfigrule}{\vspace*{-2pt}\noindent{\color{cream}\rule[\figrulesep]{\columnwidth}{1.5pt}} }
\newcommand{\dblfigrule}{\vspace*{-1pt}\noindent{\color{cream}\rule[-\figrulesep]{\textwidth}{1.5pt}} }
\makeatother

%%%TITLE, AUTHORS AND ABSTRACT%%%
\twocolumn[
  \begin{@twocolumnfalse}
{\includegraphics[height=30pt]{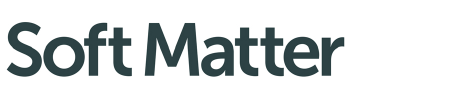}\hfill\raisebox{0pt}[0pt][0pt]{\includegraphics[height=55pt]{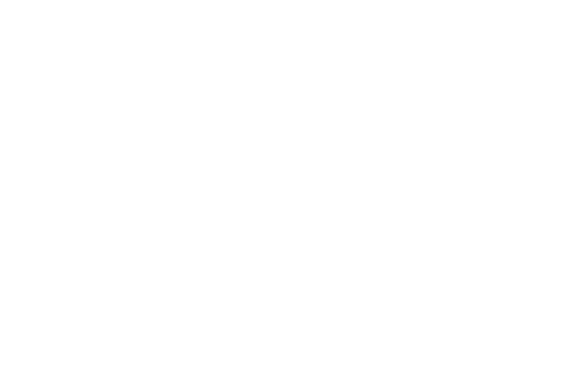}}\\[1ex]
\includegraphics[width=18.5cm]{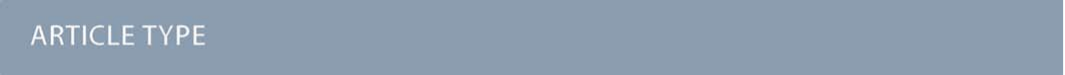}}\par
\vspace{1em}
\sffamily
\begin{tabular}{m{4.5cm} p{13.5cm} }

\includegraphics{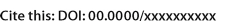} & \noindent\LARGE{\textbf{Multiscale dynamics in yield stress fluids during homogeneous and shear-banded yielding$^\dag$}} \\
\vspace{0.3cm} & \vspace{0.3cm} \\

 & \noindent\large{Paolo Edera,\textit{$^{a,b,\ddag}$} Matteo Brizioli,\textit{$^{a,\ddag}$} Mahnoosh Madani,\textit{$^{c,d}$} Elie N'gouamba,\textit{$^{e}$} Philippe Coussot,\textit{$^{e}$} V{\'e}ronique Trappe,\textit{$^{f}$} George Petekidis,\textit{$^{c,d}$} Fabio Giavazzi,\textit{$^{a,\S}$} and Roberto Cerbino\textit{$^{a,g,\S}$}} \\

\includegraphics{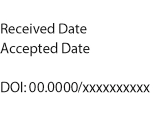} & \noindent\normalsize{Whether homogeneous and shear-banded yielding involve distinct microscopic mechanisms remains unresolved. We combine oscillatory rheology with gradient-resolved particle tracking to compare a ductile Carbopol microgel that yields homogeneously with a more brittle dense emulsion that develops stationary shear bands. In both systems, we separate long-range correlated ballistic flow from local Fickian diffusion and find that the two act on the same tracers, rather than on distinct persistent populations. This finding challenges the distinct-population interpretation of earlier experiments, which could access only dynamics averaged over particles and along the velocity-gradient direction (S. Aime \textit{et al.}, \textit{Nat. Phys.}, 2023, \textbf{19}, 1673). We distinguish these contributions in Carbopol by varying the tracer radius $a$ and the observation-window size $L$ independently. The ballistic velocity is independent of $a$ but increases with $L$, indicating correlations beyond the field of view; the diffusion coefficient $D$ scales as $1/a$ and is independent of $L$. At fixed driving frequency and tracer size, emulsion measurements collapse onto a single $D(\gamma_L)$ curve, whether taken across the banded profile or at fixed height under varying applied amplitude. Local strain amplitude, rather than stress, therefore governs the mean diffusive mobility within our resolution. The increase of $D$ with $\gamma_L$ is steeper than in Carbopol, consistent with the emulsion's greater \textit{brittility}. Diffusive displacements are nearly Gaussian in Carbopol, with no measurable cooperativity, while in the emulsion they are strongly non-Gaussian and cooperative. There, the four-point susceptibility is largest at the strain characterising the interband region. Homogeneous and shear-banded yielding therefore share the same dynamical ingredients, while differing in the statistics and spatial organisation of the residual diffusive motion.} \\

\end{tabular}

 \end{@twocolumnfalse} \vspace{0.6cm}
 ]
%%%END OF TITLE, AUTHORS AND ABSTRACT%%%
\renewcommand*\rmdefault{bch}\normalfont\upshape
\rmfamily
\section*{}
\vspace{-1cm}

\footnotetext{\textit{$^{a}$~Dipartimento di Biotecnologie Mediche e Medicina Traslazionale, Universit\`a degli Studi di Milano, I-20054 Segrate, Italy.}}
\footnotetext{\textit{$^{b}$~Molecular, Macromolecular Chemistry and Materials, ESPCI Paris, CNRS, PSL Research University, 75005 Paris, France.}}
\footnotetext{\textit{$^{c}$~Department of Material Science and Technology, University of Crete, Heraklion 71300, Crete, Greece.}}
\footnotetext{\textit{$^{d}$~Foundation for Research and Technology Hellas, Institute of Electronic Structure and Laser, Heraklion 71110, Crete, Greece.}}
\footnotetext{\textit{$^{e}$~Laboratoire Navier, Ecole des Ponts, Univ.~Gustave Eiffel, CNRS, 77420 Champs-sur-Marne, France.}}
\footnotetext{\textit{$^{f}$~Department of Physics, University of Fribourg, Ch.~du Mus\'ee 3, 1700 Fribourg, Switzerland.}}
\footnotetext{\textit{$^{g}$~Faculty of Physics, University of Vienna, Boltzmanngasse 5, 1090 Vienna, Austria.}}
\footnotetext{\dag~Electronic supplementary information (ESI) available. See DOI: 10.1039/xxxxxxxxxx}
\footnotetext{\ddag~These authors contributed equally to this work.}
\footnotetext{\S~Corresponding authors: fabio.giavazzi@unimi.it, roberto.cerbino@univie.ac.at.}

%%%%%%%%%%%%%%%%%%%%%%%%%%%%%%%%%%%%%%%%%%%%%%%%%%%
\section{Introduction}
\label{sec:intro}

Yield stress fluids (YSFs), such as dense microgel pastes, concentrated emulsions, colloidal glasses, foams and many biological tissues, behave as soft solids at rest and flow as viscoplastic liquids above a characteristic yield stress~\cite{bonn2017yield,nicolas2018deformation,Joshi2018YieldSF,coussot2002viscosity,dahbi2010rheology,cerbino2023introduction,berthier2025yielding}. Their transformation from solid-like to liquid-like under mechanical rather than thermal driving controls the processing and shelf-life of products ranging from paints and foods to cements and drilling muds~\cite{nelson2017design}. Moreover, it provides an experimentally accessible realisation of the disordered solids of statistical physics: the arrested state is a glass, and its shear-driven fluidisation is a nonequilibrium transition whose brittle or ductile character is predicted to depend on the degree of annealing~\cite{berthier2025yielding,divoux2024ductile,ozawa2018random}.

Yielding nevertheless remains only partly understood, because the underlying physics couples at least three length scales: the microscopic scale of the elementary constituents, where plasticity emerges as irreversible rearrangements; the mesoscopic scale, where the shear rate may become inhomogeneous and shear bands form; and the macroscopic scale, where the bulk stress--strain response is measured. These scales are usually probed by different experiments and different analyses, and their interplay has remained difficult to access. In the following we summarise what is known at each scale and identify the open questions that motivate the present work.

\subsection{Macroscopic scale: universal and non-universal features of the oscillatory rheology}
At the macroscopic scale, a common probe of yielding is oscillatory shear rheology, in which an oscillating stress $\sigma(t)=\sigma_0\sin(\omega t)$ (or, equivalently, an oscillating strain $\gamma(t)=\gamma_0\sin(\omega t)$) is applied and the first-harmonic in-phase and out-of-phase responses, the storage and loss moduli $G'$ and $G''$, are recorded as functions of amplitude. As the amplitude is increased, YSFs almost universally exhibit ``type III'' sweeps~\cite{hyun2011review}: at low strain, a linear response with $G'\gg G''$; at intermediate strain, an overshoot in which $G''$ rises to a maximum $G''_{\mathrm{peak}}$ before decreasing; and, at large strain, a terminal regime where $G''>G'$ and both moduli decay as power laws of the strain amplitude, $G''\propto\gamma_0^{-\nu}$ and $G'\propto\gamma_0^{-2\nu}$, with a terminal exponent $\nu$ typically between 0 and 1~\cite{miyazaki2006nonlinear,koumakis2012direct}. Three numbers extracted from such a sweep recur throughout this work: the loss tangent measured at low strain, $\tan\delta_0=G''_{\mathrm{lin}}/G'_{\mathrm{lin}}$, which quantifies how solid-like the material is before yielding; the normalised overshoot $G''_{\mathrm{peak}}/G''_{\mathrm{lin}}$; and the terminal exponent $\nu$.

This phenomenology is common to a wide range of metastable complex fluids, and the $G''$ overshoot has long been understood as the signature of a transition from a predominantly solid-like response at low amplitude to a liquid-like one above yielding---the oscillatory counterpart of the stress overshoot observed in steady start-up shear---both in theoretical treatments based on mode-coupling and soft-glassy-rheology approaches~\cite{miyazaki2006nonlinear} and in experiments on model hard- and soft-particle glasses~\cite{koumakis2012direct}. Despite this common phenomenology, type-III YSFs differ substantially in how abruptly they yield, and several scalar indicators have been proposed to quantify that abruptness (Table~S1). Donley \textit{et al.}~\cite{donley2020elucidating} made this connection quantitative by resolving the overshoot into recoverable and unrecoverable strain within a single oscillation cycle, and proposed a viscoplastic fragility index $N_F$ that measures how sharp this crossover is. This decomposition is formulated within the sequence-of-physical-processes analysis of oscillatory data~\cite{rogers2011sequence}, later formalised into a set of model-independent transient viscoelastic functions~\cite{rogers2017search}. Building on this, Kamani and Rogers~\cite{kamani2024brittle} introduced a dimensionless \textit{brittility} factor $B_t$, ranging from $B_t=1$ (fully ductile) to $B_t\to\infty$ (fully brittle), which relates the $G''$ overshoot of oscillatory sweeps with the stress overshoot of steady-shear startup~\cite{ozawa2018random,barlow2020ductile,vasisht2020emergence,ozawa2022rare,benzi2023continuum} and with delayed yielding in creep~\cite{leocmach2014creep,cochran2024slow}; $B_t$ has been interpreted as the ratio of a mesoscopic local modulus to the bulk modulus, implying that brittle YSFs develop a pronounced separation between local and bulk mechanical response. Most recently, Keane \textit{et al.}~\cite{keane2025universal} connected this picture to the linear response itself, reporting a universal relationship between $\tan\delta_0$ and the normalised overshoot $G''_{\mathrm{peak}}/G''_{\mathrm{lin}}$ across materials spanning several decades in modulus.

Unlike conventional hard materials, where brittleness corresponds to limited plastic deformation before fracture and ductility to extensive plasticity, the operational distinction between ``brittle'' and ``ductile'' yielding in YSFs is far less clear-cut~\cite{divoux2024ductile,cerbino2023introduction,berthier2025yielding}. Throughout this work we adopt the Kamani--Rogers definition~\cite{kamani2024brittle}, which assigns each sample a definite position on a ductile-to-brittle axis by means of a single macroscopic observable, $B_t$. All of these markers, however, are defined and measured at the scale of the whole sample, and none of them specifies what the constituent particles do as the material yields. 

The \textbf{first question} we address is therefore which microscopic processes are encoded in the macroscopic markers of yielding---the $G''$ overshoot, the terminal exponent $\nu$, and the brittility factor $B_t$; answering it requires measuring the particle-scale dynamics in the same samples, and under the same oscillatory protocol, from which those markers are extracted.

\subsection{Microscopic scale: yielding as a dynamic transition}
\label{sec:intro_micro}

At the microscopic scale, the amorphous architecture of YSFs poses a specific difficulty. Unlike crystalline solids, whose plasticity is mediated by topological defects such as dislocations, amorphous materials possess no long-range order, and despite sustained effort no universally agreed structural precursor of plastic events has emerged~\cite{denisov2015sharp,nicolas2018deformation,parmar2019strain}. Yielding is therefore usually described as a \textit{dynamical} transition, with an order parameter built from the mobility of the elementary constituents and their ability to participate in irreversible rearrangements~\cite{ozawa2018random,kawasaki2016macroscopic,leishangthem2017yielding,parley2022mean,bhaumik2021role,priezjev2018yielding,yeh2020glass,barlow2020ductile,pollard2022yielding,paulsen2019memory,keim2024mechanical}.

Experimentally, a direct way to quantify this irreversible mobility in an oscillatory protocol is the \textit{echo} scheme originally introduced in the light-scattering community~\cite{hebraud1997yielding,hohler1997periodic,petekidis2002rearrangements,laos_dws_colloidal_gels_2014,smith2007yielding,Derakhshandeh2013Ageing}: particle configurations are compared at times that are integer multiples of the mechanical period, $t=n\,T$, so that any affine contribution returns to zero and only non-affine, irreversible rearrangements remain. Echo protocols in various implementations---diffusing-wave spectroscopy (DWS)~\cite{hebraud1997yielding,hohler1997periodic,petekidis2002rearrangements}, confocal microscopy~\cite{knowlton2014microscopic}, X-ray photon correlation spectroscopy (XPCS)~\cite{rogers2014echoes,rogers2018microscopic}, dynamic light scattering (DLS)~\cite{aime2018microscopic,aime2019probingI,larobina2021enhanced,aime2023unified}, and differential dynamic microscopy (DDM)~\cite{aime2019probingII,edera2021deformation,richards2021characterising}---consistently show that once the oscillation amplitude exceeds a threshold, shear-induced mobility sets in and grows steeply with strain, over the same range of amplitudes in which $G'$ and $G''$ signal macroscopic yielding.

What that mobility actually consists of, however, remains contested. Early echo-DWS studies of emulsions~\cite{hebraud1997yielding}, foams~\cite{hohler1997periodic}, and hard-sphere glasses~\cite{petekidis2002rearrangements} suggested Fickian (diffusive) dynamics. Colloidal polycrystals sheared in the plastic regime instead showed ballistic dynamics~\cite{tamborini2014plasticity}; XPCS on colloidal gels and nanoemulsions revealed superdiffusive or compressed-exponential behaviour~\cite{rogers2014echoes,rogers2018microscopic,aime2018microscopic,larobina2021enhanced}; and echo-DDM on Carbopol microgels recovered Fickian diffusion~\cite{edera2021deformation}. Non-diffusive dynamics have recently been observed by Chen \textit{et al.}~\cite{chen2020microscopic} in stress-relaxation experiments: in rheo-XPCS on a nanocolloidal soft glass held at fixed macroscopic strain after a yielding step, the correlation functions measured along the flow direction decay at a rate proportional to $q$ rather than to $q^2$, the signature of convective rather than diffusive motion, with a velocity that tracks the rate of macroscopic stress decay. Complementary evidence comes from creep and recovery protocols combining rheology with XPCS. Donley \textit{et al.}~\cite{donley2023investigation} find that the loss of correlation of the microstructure is controlled by the accumulated unrecoverable strain, but that it proceeds more slowly than the $\gamma_\mathrm{unrec}^{-2}$ decay expected for a homogeneous affine deformation; they interpret the excess correlation as evidence that the irreversible rearrangements are confined to part of the sample, the remainder retaining its configuration across yielding.

A recent unifying interpretation by Aime \textit{et al.}~\cite{aime2023unified} posits that these observations reflect the \textit{coexistence} of two populations of particles within the sample: one slow and ballistic, attributed to solid-like regions, and one fast and diffusive, attributed to fluidised regions. In their lattice model, the two states are spatially organised, with slow- and fast-relaxing sites forming a coarse-grained mosaic. Because each site contains many particles, a partition into slow and fast regions is also a partition of the particles: a given particle belongs to whichever population its region does. The relative weight of the two populations was argued to vary with strain amplitude and to map, via a Van-der-Waals-like state diagram, onto a rounded first-order yielding transition, thereby reconciling first-order~\cite{kawasaki2016macroscopic,jaiswal2016mechanical,aime2023unified} and continuous~\cite{hima2014experimental,leishangthem2017yielding,parley2022mean} transition scenarios previously reported in simulations and experiments.

The coexistence scenario makes a definite and consequential prediction: across the yielding transition the particles should partition into two families with distinct relaxation modes, so that yielding proceeds through a genuine two-phase coexistence and the Van-der-Waals-like state diagram acquires a literal meaning. This prediction has not been tested directly, largely because techniques such as DLS and DDM average over all particles by construction and cannot reveal how the two relaxation processes are distributed among them. 

The \textbf{second question} we address is therefore whether the predicted partition into two families is actually realised, which requires following both processes on the same individual tracers within a single sample.

\subsection{Mesoscopic scale: role of shear bands}
At the mesoscopic scale, yielding in soft solids may be accompanied by the formation of shear bands, spatial regions that flow at rates different from the bulk, which redistribute the imposed deformation along the velocity-gradient direction~\cite{fielding2014shear,divoux2016shear,ovarlez2009phenomenology,bonn2017yield}. Simulations of oscillatory-sheared amorphous solids consistently find shear bands near yielding~\cite{kawasaki2016macroscopic,radhakrishnan2016shear,leishangthem2017yielding,priezjev2018yielding,parmar2019strain,bhaumik2021role,pollard2022yielding,cochran2024slow,benzi2023continuum}, and the appearance of bands is routinely associated with brittle yielding, in line with the random-critical-point picture of Ozawa \textit{et al.}~\cite{ozawa2018random} and its disorder-controlled variant~\cite{ozawa2022rare}, with the brittle-failure theory of Pollard and Fielding~\cite{pollard2022yielding} and Barlow \textit{et al.}~\cite{barlow2020ductile}, and with the delayed-yielding scenario of Cochran \textit{et al.}~\cite{cochran2024slow}.

Because $G'(\gamma_0)$ and $G''(\gamma_0)$ are by construction gap-averaged, shear bands leave only indirect signatures on the first-harmonic response, and local observables are required to detect them unambiguously. In steady shear, wide-gap rheometry, NMR imaging, ultrasound imaging and confocal microscopy have shown that bands are characterised by a \textit{cooperativity} of plastic events at the microscopic scale: a plastic rearrangement at one location increases the probability of rearrangements at distant locations, with a cooperativity length that scales with the deviation from the yield point~\cite{goyon2008spatial,bocquet2009kinetic,parisi2017shear}. Under oscillatory shear, particle image velocimetry has resolved shear-banded profiles in wormlike micellar solutions, with the onset of banding coinciding with the appearance of higher harmonics in the stress response~\cite{dimitriou2012rheopiv}; general criteria for the onset of banding in time-dependent protocols (step stress, strain ramp and shear startup) have been derived theoretically by Moorcroft and Fielding~\cite{moorcroft2013criteria}; and rheo-microscopy of Pluronic F127 micellar cubic gels has revealed a transient banding instability coinciding with the onset of bulk nonlinearity~\cite{agrawal2025connecting}.

For amorphous solids, a mean-field theory by Parisi \textit{et al.}~\cite{parisi2017shear} predicts that shear banding coincides with the emergence of a diverging dynamical correlation length, measurable via the four-point susceptibility $\chi_4$. A closely related prediction follows from the lattice model of Aime and Truzzolillo~\cite{aime2025emergent}, in which a dynamic correlation length emerges from the coupling between disorder and bifurcating dynamics, peaks at yielding, and grows as the mechanical disorder is reduced, so that abrupt yielding is associated with the most extended correlated dynamics. Neither prediction has been tested directly. In the earlier experiments of Ref.~\cite{aime2023unified}, fast- and slow-relaxing regions were reported in several soft-repulsive colloids but no shear bands were detected, the DLS measurement averaging along the velocity-gradient direction, which is precisely the direction along which bands develop. 

The \textbf{third question} we address is therefore whether shear banding is accompanied by enhanced dynamical cooperativity and, if so, at which local strain within the banded profile it emerges. Answering it requires access to the plastic dynamics that is resolved both at the particle scale and across the gap, which is the principal methodological contribution of the present work.

\subsection{Scope}

Motivated by these three questions, we use a rheo-microscope~\cite{edera2021deformation,villa2022quantitative} to carry out a gradient-resolved study of the shear-induced microscopic dynamics of two contrasting model YSFs: a dense Carbopol microgel that yields homogeneously, and a dense oil-in-water emulsion that yields with stationary shear bands. Stroboscopic particle tracking in the echo protocol gives access to the non-affine displacements of individual tracers, from which we extract the shear-induced mean-squared displacement, the coarse-grained velocity field, the displacement statistics and the four-point susceptibility $\chi_4$; differential dynamic microscopy provides an independent, reciprocal-space check on the same image sequences.

To address the first question, of which microscopic processes are encoded in the macroscopic markers, the two samples are placed on the ductile-to-brittle axis by bulk oscillatory rheology, so that any difference in their particle-scale response can be referred to a difference in $B_t$ and their microscopic dynamics compared at matched local strain. To address the second, of whether the two relaxation processes are carried by distinct particle families, we vary independently the tracer radius and the size of the imaged region in the Carbopol, thereby testing how the two relaxation processes are distributed among the particles. To address the third, of whether shear banding is accompanied by enhanced cooperativity, we repeat the echo measurement plane by plane across the gap, so that the dynamics measured at each height can be linked to the local strain amplitude and compared with the gradient of the local strain amplitude, obtained independently at the same positions.

%How this work builds on, and departs from, our earlier rheo-microscopy is discussed in Section~\ref{sec:disc_priorwork}.

%%%%%%%%%%%%%%%%%%%%%%%%%%%%%%%%%%%%%%%%%%%%%%%%%%%
\section{Experimental section}
\label{sec:experimental}

\subsection{Samples}
\label{sec:samples}

%We investigated two dense, soft-glassy YSFs. The first is a 
Carbopol microgel at a mass concentration $c=5$~wt\% was prepared by dispersing Carbopol 971P~NF powder (Lubrizol) in Milli-Q water, neutralising with aqueous NaOH to pH $7.0\pm0.2$, and stirring gently for a few days until fully homogeneous. The protocol follows established preparations of Carbopol microgels~\cite{piau2007carbopol,geraud2013confined,oelschlaeger2022imaging} and is described in detail in our previous work~\cite{edera2021deformation,n2019elastoplastic}. At this concentration Carbopol forms a dense, space-filling packing of swollen polyelectrolyte microgels, with hydrodynamic radii of a few micrometres and a broad size distribution~\cite{piau2007carbopol,oelschlaeger2022imaging}. The sample is optically transparent and mechanically reproducible from day to day when loaded with the pre-shear protocol of Section~\ref{sec:shearprotocol}.

The second system is a direct (oil-in-water) emulsion at an oil volume fraction of 87\%, prepared by dispersing a silicone oil (viscosity $\simeq 10^{-2}$~Pa\,s) in a water/glycerol mixture (48/52~wt/wt) containing 3~wt\% tetradecyltrimethylammonium bromide (TTAB) as surfactant, following the procedure of Refs.~\cite{n2019elastoplastic,ngoua2020}. The water/glycerol composition is chosen to match the refractive index of the silicone oil, which renders the concentrated sample optically transparent. The mean droplet radius, $R_d = 0.45\pm0.15$~\textmu m, was obtained by image analysis of a dilute sample on a microscope slide.

Both samples were seeded with polystyrene colloidal tracers (Microparticles GmbH) at a volume fraction $\phi_t=0.05\%$. In the Carbopol, three tracer radii were used---$a=0.5$, $1.0$ and $2.5$~\textmu m---which allows characterising the dependence of the microscopic dynamics on tracer size over a factor of five. In the emulsion, all measurements were performed with the intermediate size, $a=1.0$~\textmu m. The tracer-size analysis is therefore restricted to the Carbopol, and the comparison between the two samples is made at fixed tracer size, $a=1.0$~\textmu m. Control amplitude sweeps confirm that tracers at this concentration do not perceptibly modify the bulk $G'(\gamma_0)$ or $G''(\gamma_0)$. 

\subsection{Rheo-microscopy}
\label{sec:rheomicroscopy}
Most shear experiments were performed with a custom-built stress-controlled shear cell integrated on an inverted optical microscope~\cite{villa2022quantitative}. The plates were separated by a gap $h$, and the top plate was oscillated with displacement amplitude $A_0$, so that the nominal strain amplitude imposed on the sample is $\gamma_0=A_0/h$. All measurements were performed at a temperature of $T = 25 \pm 0.5$~$^\circ$C. The setup imposes an oscillatory shear while the sample is imaged, giving simultaneous access to the bulk mechanical response and to the mesoscopic and microscopic dynamics. Its optical configuration allows imaging at different positions along the velocity-gradient direction, which is essential to disentangle the contribution of each plane to the overall sample dynamics, especially in the presence of shear banding. Such access to non-idealities of the shear profile, such as shear bands and wall slip, is precluded in standard commercial rheometry, which returns gap-averaged quantities only; rheo-imaging platforms of various kinds are for this reason increasingly adopted to overcome this limitation~\cite{besseling2009quantitative,divoux2016shear,villa2022quantitative}. Rheo-scattering setups, on the other hand, retain a bulk mechanical measurement but average the sample information over the entire thickness of the gap~\cite{aime2019probingI,aime2023unified}.

A comprehensive description of the apparatus and of the data-acquisition protocols is given in Ref.~\cite{villa2022quantitative}, while the image-processing and analysis procedures are detailed in Refs.~\cite{edera2021deformation,villa2022quantitative}. A limited number of experiments were additionally conducted with a commercial strain-controlled shear cell, described in a previous publication~\cite{edera2021deformation}. The bulk rheological characterisation was instead performed on a separate commercial rheometer, as described in Section~\ref{sec:rheology}.

All the experiments presented in this work were performed at a constant oscillatory frequency $\omega/(2\pi)=1$~Hz. We use $x$ to denote the flow direction, $y$ the vorticity direction, and $z$ the velocity-gradient direction (normal to the plates).

\subsubsection{Imaging}
Imaging is performed through the bottom glass plate with a $20\times$ objective (NA~0.45), with the condenser diaphragm fully opened so that the combined illumination--detection numerical aperture is $\simeq0.52$, restricting the depth of field to $\simeq 20$~\textmu m. While this configuration cannot match the optical sectioning of a confocal microscope, particularly in background rejection, it is sufficient to access different planes within the sample, and it avoids the complications of fluorescence imaging, such as photobleaching over long acquisitions. Images are captured by a Ximea MQ042MG-CM USB3.0 camera, giving an effective pixel size at the sample $d_\mathrm{eff}=0.55$~\textmu m. Image acquisition is triggered by the shear-cell electronics and is therefore phase-locked to the applied stress at every $z$.

\subsubsection{Pre-conditioning}
\label{sec:shearprotocol}
Before every measurement, the sample is rejuvenated by applying an oscillatory deformation at fixed frequency $\omega/(2\pi)=1$~Hz, with an amplitude decreasing progressively from a large value---corresponding to a strain amplitude $\gamma_0\simeq200\%$--$250\%$, far into the nonlinear regime---down to the target value at which data acquisition begins. The sweep lasts approximately 400 cycles. This pre-shearing protocol brings the sample to a well-defined, reproducible state at the beginning of the measurement, as verified for the emulsion in Fig.~S5, and avoids the well-known transients that follow a sudden decrease of the amplitude to its target value~\cite{hima2014experimental}. The relaxation time governing these transients grows with the strain amplitude and diverges in the vicinity of the yield strain, so that stepping directly to the target amplitude would require a prohibitively long wait before stationary conditions are recovered.

\subsubsection{Deformation profile.}
\label{sec:defprofile}
For each target amplitude, we perform a fast $z$-scan with vertical resolution $\Delta z=20$~\textmu m from the top plate to the bottom plate. At each plane we acquire a short movie spanning 4--5 oscillation periods at 45 frames per cycle (ROI linear size 512 pixels, corresponding to $282\times282$~\textmu m$^2$), from which the oscillating in-plane displacement field $\Delta x(z,t)$ of the tracers is obtained by image cross-correlation~\cite{edera2021deformation,villa2022quantitative}. The displacement is harmonic, $\Delta x(z,t) = A(z)\sin(\omega t+\phi(z))$; we report the amplitude $A(z)$ and phase $\phi(z)$ as well as the local strain amplitude $\gamma_{L}(z)$, which is estimated as the amplitude of the relative displacement $(\Delta x(z_{i+1},t)-\Delta x(z_{i},t))$ between two consecutive imaged planes, divided by the distance $\Delta z=z_{i+1}-z_{i}$ between the two planes. We stress that $\gamma_L$ is obtained as the amplitude of the \textit{difference} between the two displacement signals, and not as the difference between their amplitudes: because the phase $\phi(z)$ itself varies across the gap, the two definitions do not coincide, and only the former returns the true local strain amplitude. The displacement fields at different heights are not acquired simultaneously, but because the acquisition is phase-locked, $\gamma_L(z)$ can still be reconstructed provided the response is stationary.

\subsubsection{Echo protocol and microscopic dynamics.}
\label{sec:echo}
After the $z$-scan, we acquire a long ``echo'' sequence at the mid-plane of the gap unless otherwise stated. In the echo protocol, $n_\mathrm{echo}=10$ images are captured per oscillation period over $\simeq 400$ cycles, so that ten stroboscopic sub-sequences--- each obtained by considering one image per cycle, at a fixed phase ---can be extracted. Within each stroboscopic sub-sequence, the affine contribution to the displacement is zero by construction, and only non-affine, irreversible rearrangements survive.

Residual mechanical drift of the apparatus is corrected in two stages. First, for each sub-sequence, a global background image is subtracted and a rigid intensity-based registration (ImageJ StackReg plugin~\cite{thevenaz1998pyramid}) removes any global translation. Second---and more importantly---we systematically work \textit{in the centre-of-mass frame of the imaged tracers}: the spatial average of the single-particle velocity at each stroboscopic step is removed before any ensemble averaging. Additional controls on the impact of this subtraction are reported in Section~\ref{sec:results_superposition} and in the ESI (Fig.~S3).

As a control on the spatial extent of the non-affine dynamics, we repeated the echo measurement in the Carbopol while varying the ROI linear size $L$ from 256 to 1024~pixels (corresponding to $140\times140$ to $560\times560$~\textmu m$^2$) and the position of the ROI centre.

\subsubsection{Particle tracking, velocity maps and differential dynamic microscopy.}
\label{sec:analysis}
Tracer positions are identified using a customised version of the MATLAB particle-tracking code from the Kilfoil group~\cite{pelletier2009microrheology}, which routinely achieves $\sim 50$~nm localisation. Typically, $\sim 500$ trajectories are reconstructed within each stroboscopic sub-sequence over an ROI of 1024 pixels, leading to a total of $\sim 5000$ single-particle trajectories $[x^{(i)}(n),y^{(i)}(n)]$ for each imposed strain amplitude. From the trajectories we compute (i) the stroboscopic mean-squared displacement in the vorticity ($y$) direction: %insensitive to affine shear along $x$
\begin{equation}
\mathrm{MSD}(\Delta n|\gamma_L) = \langle | y^{(i)}(n+\Delta n) - y^{(i)}(n) |^2 \rangle_{i,n},
\label{eq:MSD}
\end{equation}
and (ii) a coarse-grained velocity map $\mathbf{v}(\mathbf{x}_m,n)$ obtained by considering a fixed delay $\Delta n=20$ between frames and averaging single-particle displacements over a square interrogation window centred at $\mathbf{x}_m$ (side 200~pixels, overlap $80\%$, $\sim20$ particles per window). Because the analysis is performed in the centre-of-mass frame, each velocity map has zero spatial average, $\frac{1}{M}\sum_{m=1}^M \mathbf{v}(\mathbf{x}_m,n)\simeq0$. From the time-averaged (stationary) velocity field $\mathbf{v}_s(\mathbf{x}_m)=\langle \mathbf{v}(\mathbf{x}_m,n)\rangle_{n}$ we estimate the root-mean-square velocity along the vorticity direction $\hat{\mathbf{y}}$,
\begin{equation}
v_\mathrm{rms}(\gamma_L) = \left[\frac{1}{M}\sum_{m=1}^M |\mathbf{v}_s(\mathbf{x}_m)\cdot{\hat{\mathbf{y}}}|^2 \right]^{1/2},
\label{eq:vrms}
\end{equation}
which reports on the non-affine, spatially varying component of the velocity field only. (iii) Using the coarse-grained velocity maps together with the stroboscopic displacements, we further estimate the short-wavelength dynamics through the residual MSD,
\begin{equation}
    \mathrm{MSD}_r(\Delta n|\gamma_L)=\langle |\Delta y_r^{(i)}(n,\Delta n)|^2\rangle_{i,n},
\label{eqMSDVelDef_res}
\end{equation}
where $\Delta y_r^{(i)}(n,\Delta n)= y^{(i)}(n+\Delta n)-y^{(i)}(n)-\mathbf{v}_{s}(\mathbf{x}^{(i)})\cdot{\hat{\mathbf{y}}}\,\Delta n$ is the residual displacement along the vorticity direction and $\mathbf{v}_{s}(\mathbf{x}^{(i)})$ is the value of the time-averaged velocity field at $\mathbf{x}^{(i)}$, obtained by linear interpolation of the coarse-grained map. (iv) Finally, we compute the non-Gaussian parameter $\alpha_2$ associated with the short-wavelength dynamics along the vorticity direction,
\begin{equation}
    \alpha_2(\Delta n|\gamma_L)=\frac{1}{3}\frac{\langle |\Delta y_r^{(i)}(n,\Delta n)|^4\rangle_{i,n}}{\langle |\Delta y_r^{(i)}(n,\Delta n)|^2\rangle_{i,n}^2}-1.
\label{eqNonGaus}
\end{equation}
Unless otherwise stated, all the dynamical observables reported below---$\mathrm{MSD}$, $\mathrm{MSD}_r$, $D$, $v$ and $v_\mathrm{rms}$---refer to the vorticity direction.
As an independent, reciprocal-space check on our real-space results, we also apply differential dynamic microscopy (DDM)~\cite{cerbino2008differential,giavazzi2009scattering,aime2019probingII,edera2021deformation} to the same echo sequences, extracting the intermediate scattering function $f(q,\Delta n)$ along the vorticity direction.

\subsubsection{Four-point dynamic correlations.}
\label{sec:chi4}
Following Ref.~\cite{berthier2011dynamic}, to quantify cooperativity in the short-wavelength non-affine dynamics, we build the mobility field
\begin{equation}
c(\mathbf{x}|n,\Delta n) = \sum_i\big[ o^{(i)}(n,\Delta n)-\langle o^{(i)}(n,\Delta n)\rangle_{n}\big]\delta(\mathbf{x}-\mathbf{x}^{(i)}(n)),
\label{eq:mobility}
\end{equation}
where $o^{(i)}(n,\Delta n)$ is an overlap function, which quantifies single-particle mobility along the vorticity direction between cycle $n$ and $n+\Delta n$,
\begin{equation}
o^{(i)}(n,\Delta n) = \exp\!\left[-\frac{|\Delta y_r^{(i)}(n,\Delta n)|^2}{d^2}\right].
\label{eq:overlap}
\end{equation}
Here $d = 0.15$~\textmu m is the characteristic displacement below which the particle is considered immobile. This value is chosen to lie safely above the particle-tracking localisation noise---whose magnitude is set by the short-$\Delta n$ plateau $\mathrm{MSD}_0$ of the measured mean-squared displacements (Fig.~S3), and is essentially independent of the strain amplitude---while remaining well below the displacements accumulated over the range of $\Delta n$ at which $\chi_4$ is evaluated. The overlap function therefore discriminates genuine rearrangements from tracking noise. From the mobility maps, we compute the four-point dynamic correlation function~\cite{berthier2011dynamic,keys2007measurement}
\begin{equation}
G_4(\Delta\mathbf{r},\Delta n) = \langle c(\mathbf{x}+\Delta\mathbf{r}|n,\Delta n)\,c(\mathbf{x}|n,\Delta n)\rangle_{n,\mathbf{x}},
\label{eq:G4}
\end{equation}
which measures the spatial and temporal correlation of particle motion at a certain separation distance $\Delta r$ and for a certain number of elapsed cycles $\Delta n$. 
The four-point susceptibility
\begin{equation}
\chi_4(\Delta n|\gamma_L) = \int d^2\Delta\mathbf{r}\,\bigl[G_4(\Delta\mathbf{r},\Delta n)-G_4(\Delta r_\infty,\Delta n)\bigr],
\label{eq:chi4}
\end{equation}
is then reported after subtracting a baseline measured at $|\Delta\mathbf{r}|=\Delta r_\infty \ge 200$~\textmu m that accounts for any residual long-wavelength correlation. As a peak observable we report $\chi_4^\ast(\gamma_L) \equiv \max_{\Delta n}\chi_4(\Delta n|\gamma_L)$, which is proportional to the number of particles involved in cooperative rearrangements.

\subsection{Rheology}
\label{sec:rheology}
For reference and cross-validation, we also performed a macroscopic rheological characterisation with a commercial rheometer (Anton Paar MCR~501, roughened cone-plate geometry, 50~mm diameter, $1^\circ$ angle) operating in strain-controlled mode. This geometry provides a nominally homogeneous shear field and therefore an independent check on our rheo-microscopy outputs. In addition to oscillatory amplitude sweeps, steady-shear flow curves were also measured.

From the amplitude sweeps we estimated the brittility factor $B_t$ by fitting the Kamani--Rogers model of Ref.~\cite{kamani2024brittle} to the experimental storage and loss moduli, using $B_t$ as the sole free parameter. The remaining model parameters for each material are extracted independently from the flow curves and from the complex moduli in the linear regime (see Fig.~S1).

%%%%%%%%%%%%%%%%%%%%%%%%%%%%%%%%%%%%%%%%%%%%%%%%%%%
\section{Results}
\label{sec:results}

The two samples we study exhibit macroscopic amplitude sweeps with the characteristic features of type-III YSFs (Fig.~S1): at low $\gamma_0$, a plateau with $G'_\mathrm{lin}\gg G''_\mathrm{lin}$; at intermediate $\gamma_0$, a peak in $G''$; at high $\gamma_0$, a crossover to the terminal regime $G''>G'$ with both moduli decaying as power laws. 

Despite this common phenomenology, the two samples have very different brittility factors (see Section~\ref{sec:rheology}): the Carbopol has $B_t\simeq 1.0$ (weakly pronounced $G''$ overshoot, small terminal exponent $\nu\simeq 0.44$%~\cite{miyazaki2006nonlinear}
), while the emulsion has $B_t\simeq 6.5$ ($\nu\simeq 0.80$). The other indicators introduced in Section~\ref{sec:intro} follow the same ordering: the inverse loss tangent $(\tan\delta_0)^{-1}$, the normalised overshoot $G''_{\mathrm{peak}}/G''_{\mathrm{lin}}$ (related to one another through Ref.~\cite{keane2025universal}) and the viscoplastic fragility index $N_F$~\cite{donley2020elucidating} all attain higher values for the emulsion than for the Carbopol (ESI, Table~S1). As we will show, this macroscopic distinction is mirrored both in the mesoscale deformation profile (homogeneous \textit{vs} banded) and in the microscopic statistics (Gaussian and non-cooperative \textit{vs} non-Gaussian and cooperative).

\subsection{Homogeneous yielding: Carbopol microgel}
\label{sec:results_carbopol}

\subsubsection{The deformation profile remains homogeneous.~~}
A vertical scan of the oscillating displacement field at the target amplitude is shown in Fig.~S2. The amplitude $A(z)$ is linear in $z$, $A(z)\simeq A_0\,z/h$, with apparent slip at both plates consistent with zero within $5$\% of the nominal oscillation amplitude $A_0$ of the moving plate. The local strain amplitude $\gamma_L(z)$ is therefore spatially uniform and coincides with the imposed nominal strain amplitude $\gamma_0 = A_0/h$ over the observable gap, independent of the applied $\gamma_0$ (from $\gamma_0\simeq 0.02$ to $\gamma_0\simeq 2.5$). %. 
Beyond characterising the sample, this result is a stringent test of the shear cell itself: even at the largest amplitudes explored, the cell does not impose any spurious localisation of the deformation.

\begin{figure*}[t]
\centering
\includegraphics[width=0.75\linewidth]{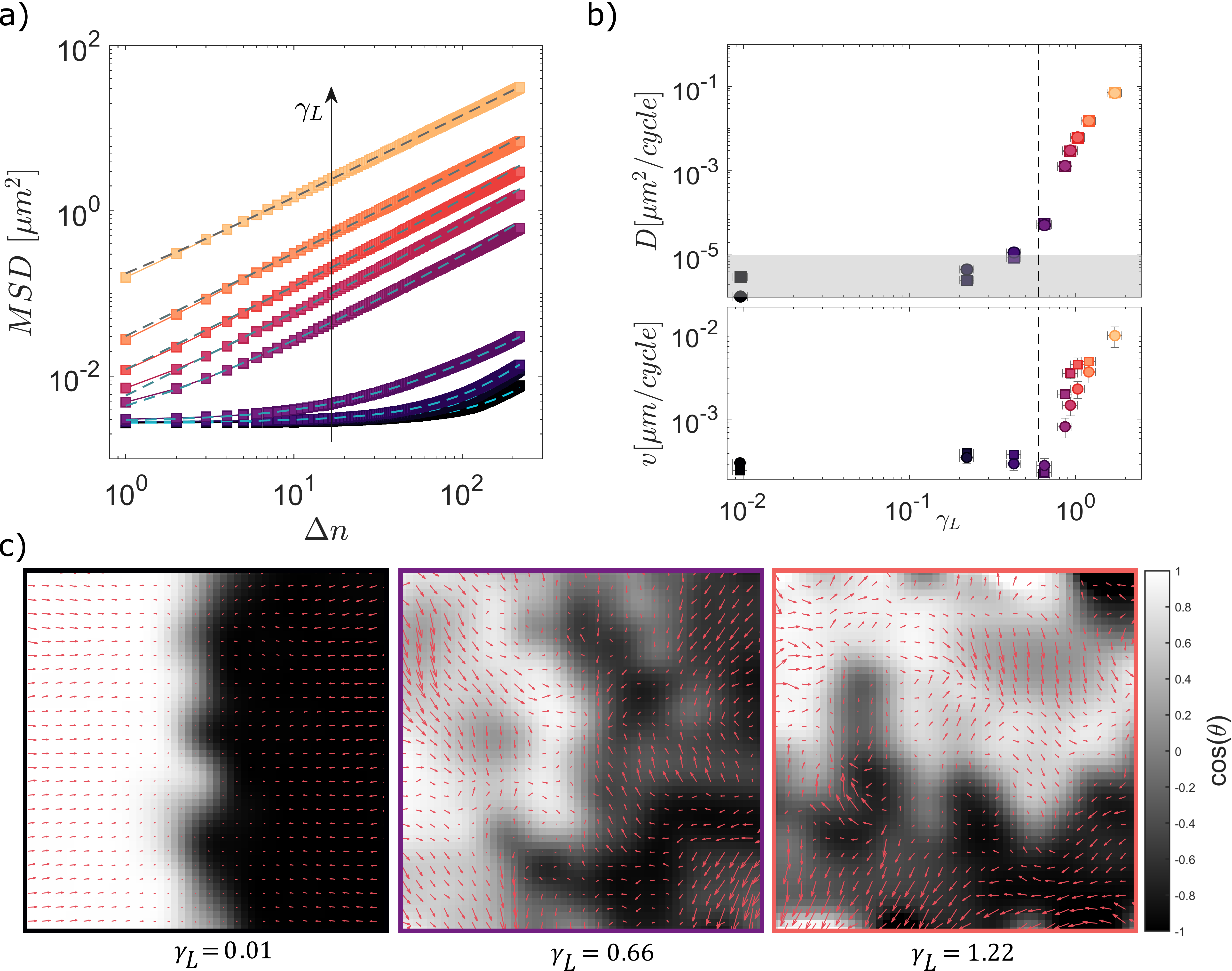}
\caption{\textbf{Homogeneous yielding in the dense Carbopol microgel ($c = 5$~wt\%).}
(a) Stroboscopic mean-squared displacement (MSD) of the tracers (radius $a=1.0$~\textmu m) in the vorticity direction as a function of the number of elapsed cycles $\Delta n$, for local strain amplitudes $\gamma_L$ ranging from $0.01$ to $1.77$ (from dark to light). Dashed lines are the best-fitting curves to the data using Eq.~(\ref{eq:twoprocess}), $\mathrm{MSD}(\Delta n) = \mathrm{MSD}_0 +2 D\,\Delta n + v^2\,\Delta n^2$.
(b) Shear-induced diffusion coefficient $D$ (top) and characteristic velocity (bottom), both as functions of the local strain amplitude $\gamma_L$. The vertical lines indicate the strain for which $G'=G''$ in bulk rheology. In the top panel, squares are obtained by fitting Eq.~(\ref{eq:twoprocess}) to the data of panel~(a); the circles are obtained independently from the residual MSD, $\mathrm{MSD}_r$, according to Eq.~(\ref{eqMSDVelDef_res}); the shaded area marks the limit of reliability of the estimate. In the bottom panel, squares are the fit values $v$ from panel~(a), circles the estimated $v_\mathrm{rms}$ obtained directly from the coarse-grained velocity maps of panel~(c) using Eq.~(\ref{eq:vrms}). The two estimates agree within error at all strain amplitudes, confirming that the ballistic component of the MSD originates from the spatially correlated, persistent flow shown in (c) rather than from independent, tracer-scale motion.
(c) Time-averaged velocity maps $\mathbf{v}_s(\mathbf{x}_m) = \langle \mathbf{v}(\mathbf{x}_m,n)\rangle_{n}$ computed on a $560 \times 560$~\textmu m$^2$ field of view at mid-plane, for $\gamma_L = 0.01$ (linear regime, left), $\gamma_L = 0.66$ (yielding onset, centre), and $\gamma_L = 1.22$ (flow regime, right). Red arrows, scaled for readability, represent the local velocity and are overlaid on maps of $\cos\theta$ (greyscale), where $\theta$ is the local flow orientation with respect to the horizontal axis. Below yielding the velocity field remains correlated over distances extending beyond the field of view; at higher strain the correlation length decreases and the flows become finer-grained.}
\label{fgr:carbo}
\end{figure*}

\subsubsection{The mean-squared displacement exhibits a diffusive and a ballistic component.~~}
Fig.~\ref{fgr:carbo}a shows the stroboscopic vorticity-direction MSD for the Carbopol at strain amplitudes ranging from $\gamma_L=0.01$ to $1.77$. As the strain amplitude increases, the MSD emerges earlier from its plateau and displays two successive scalings: linear in $\Delta n$ for small $\Delta n$, and super-linear for large $\Delta n$. This is well captured by the two-process model
\begin{equation}
\mathrm{MSD}(\Delta n|\gamma_L) = \mathrm{MSD}_0(\gamma_L) + 2D(\gamma_L)\,\Delta n + v^2(\gamma_L)\,\Delta n^2
\label{eq:twoprocess}
\end{equation}
(dashed lines in Fig.~\ref{fgr:carbo}a). Here, $\mathrm{MSD}_0$ is the plateau value observable at short $\Delta n$. For the Carbopol, this plateau is compatible with the particle-tracking localisation noise, which is essentially independent of strain amplitude (see Fig.~S3), whereas in the emulsion it acquires a different meaning, discussed in Section~\ref{sec:results_emulsion}. $D$ is an effective shear-induced diffusion coefficient, and $v$ is a velocity associated with a ballistic process that becomes more apparent at large $\Delta n$. As shown in Fig.~\ref{fgr:carbo}b (top panel), $D(\gamma_L)$ becomes measurable well below the crossover point of $G'$ and $G''$, increasing by four orders of magnitude as the system is driven into the non-linear regime. Conversely, at small strain amplitudes the dynamics is predominantly ballistic, as is evident from the MSD curves of Fig.~\ref{fgr:carbo}a, with a velocity that stays approximately constant before rising steeply where the shear-induced diffusivity becomes dominant (Fig.~\ref{fgr:carbo}b, bottom panel).

    The fact that the MSD has a ballistic contribution at large $\Delta n$ is a key observation of the present work: it explicitly identifies \textit{two} processes whose signatures are resolved in real space, as reported in detail in the following sections.
\begin{figure*}[t]
\centering
\includegraphics[width=0.9\linewidth]{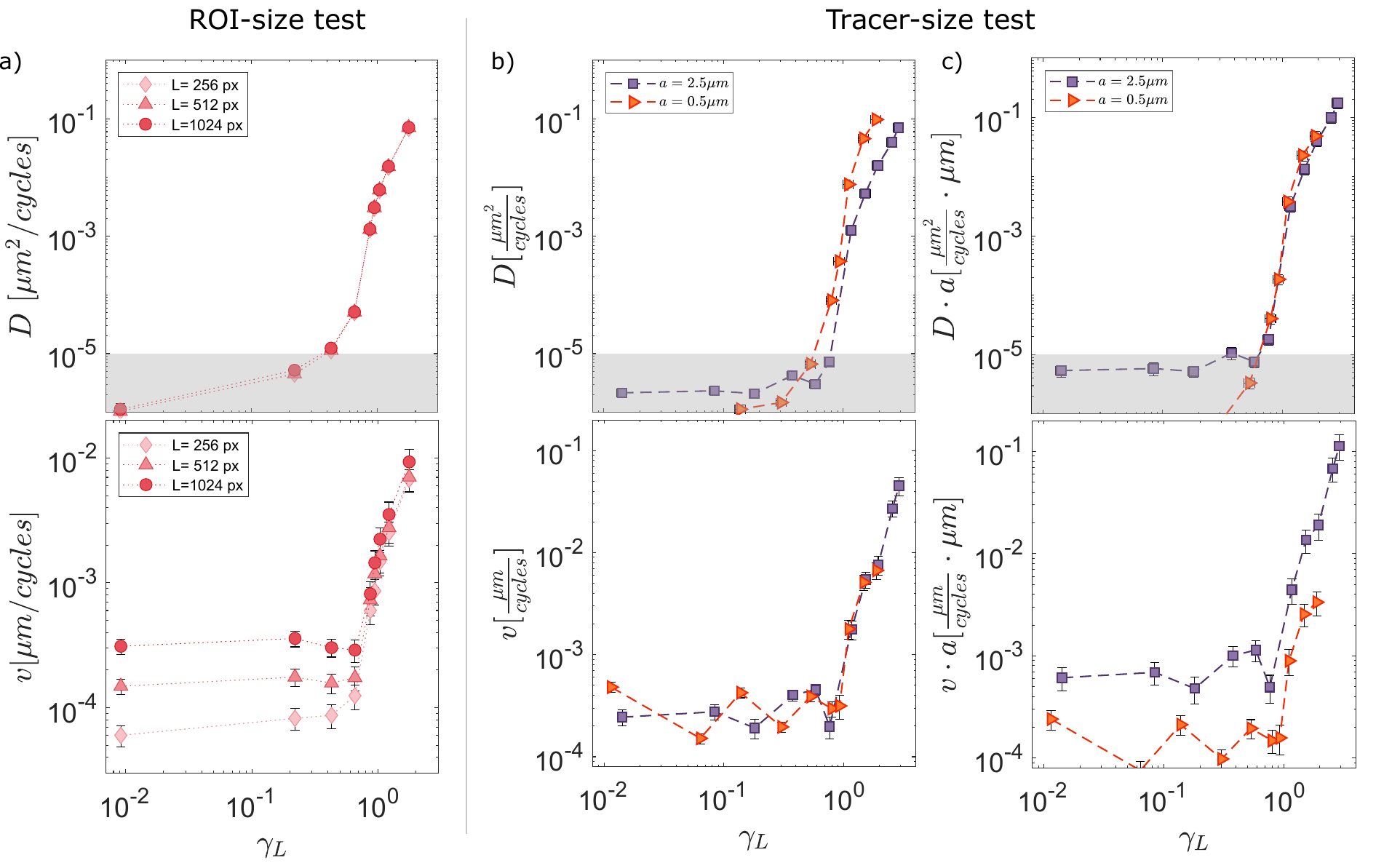}
\caption{\textbf{Effect of tracer size and region of interest on the two relaxation processes, in the Carbopol.} (a) Diffusion coefficient $D$ (top) and ballistic velocity $v$ (bottom) as functions of $\gamma_L$, computed on three different region-of-interest (ROI) sizes, $L = 256$, $512$ and $1024$ pixels (corresponding to $140$, $280$ and $560$~\textmu m, respectively). $D$ is independent of $L$ within experimental uncertainty, showing that it is not set by the size of the observation window. $v$ scales linearly with $L$ at small and intermediate $\gamma_L$ (a correlated process whose correlation length $\xi$ exceeds $L$), and the dependence becomes milder at the largest amplitudes, consistent with $\xi$ becoming comparable to or smaller than $L$. At the lowest $\gamma_L$ the three curves are offset by roughly half a decade, $v$ growing by a factor $\simeq5$ for a factor $4$ increase in $L$; the offset closes progressively as $\gamma_L$ increases. (b) Diffusion coefficient $D$ (top) and ballistic velocity $v$ (bottom), as functions of the local strain amplitude $\gamma_L$, for the Carbopol seeded with two tracer radii, $a = 0.5$~\textmu m (triangles) and $a = 2.5$~\textmu m (squares). (c) The same data with both observables multiplied by the tracer radius. $D\cdot a$ collapses onto a single master curve, indicating Stokes--Einstein-like scaling $D\propto 1/a$, whereas $v\cdot a$ splits the two datasets apart by the ratio of the radii---the signature of a velocity that is already independent of $a$ and therefore requires no rescaling. Shaded areas indicate the measurement-reliability limit for $D$. Together, panels~(a)--(c) demonstrate that the diffusive process is local and microscopic while the ballistic process is non-local, with a characteristic length scale exceeding the tracers and the ROI.}
\label{fgr:trc}
\end{figure*}

\subsubsection{Velocity maps: the ballistic process is long-ranged and spatially correlated.~~}
\label{sec:results_superposition}
A first key question is whether the two processes in Eq.~\ref{eq:twoprocess} are spatially homogeneous or exhibit distinct spatial patterns. To answer this, we construct coarse-grained velocity maps $\mathbf{v}(\mathbf{x}_m,n)$, which allow us to identify the presence of non-affine ballistic components. Remarkably, for a given strain amplitude, the maps computed within a single echo sub-sequence considering different starting points $n$ or different delays $\Delta n$ are nearly identical, indicating a stationary flow field. We therefore work with the time-averaged field $\mathbf{v}_s(\mathbf{x}_m)=\langle \mathbf{v}(\mathbf{x}_m,n)\rangle_{n}$: since its spatial average vanishes by construction, any non-zero $\mathbf{v}_s(\mathbf{x}_m)$ is the signature of a persistent, spatially organised flow.

Representative maps $\mathbf{v}_s(\mathbf{x}_m)$ are shown in Fig.~\ref{fgr:carbo}c, where the length of the arrows represents the magnitude of the local in-plane velocity and their orientation $\theta$ defines the direction of the flow, visually highlighted through a colour map based on $\cos\theta$. At $\gamma_L=0.01$ (linear regime), the field reveals correlated and anti-correlated domains across the field of view, indicating persistent and organised flows whose spatial correlation length exceeds the ROI size ($\sim 560$~\textmu m). At $\gamma_L=0.66$ the two domains observed at low strain become partially smeared out, but long-range correlations remain clearly visible, showing that the flow retains its persistent, spatially organised character well into the nonlinear regime. As $\gamma_L$ increases further to $1.22$ the correlation length decreases and the flow pattern becomes finer-grained (Fig.~\ref{fgr:carbo}c, right), though still well structured on length scales substantially larger than the tracer size. 

Crucially, the root-mean-square velocity $v_\mathrm{rms}$ extracted from these maps quantitatively matches the coefficient $v$ obtained by fitting Eq.~(\ref{eq:twoprocess}) to the MSD (Fig.~\ref{fgr:carbo}b, bottom), confirming that the ballistic contribution to the MSD originates from these persistent, long-wavelength flows and not from independent, tracer-scale motion. The arrows in Fig.~\ref{fgr:carbo}c report the full in-plane velocity and lengthen appreciably with $\gamma_L$, whereas $v_\mathrm{rms}$ and $v$ refer to the vorticity component alone, which varies far more weakly over the same range. The agreement between the two estimates is therefore not affected. Any residual deviation between the two estimates at the highest strains is explained by the fact that, when diffusion dominates over the ballistic contribution, the velocity extracted from the MSD becomes intrinsically less accurate and is affected by the diffusive component of the dynamics.

Given this agreement, in what follows we drop the distinction between the two estimators and denote simply by $v$ the characteristic velocity obtained from the velocity maps. Likewise, $D$ denotes the diffusion coefficient estimated from the residual mean-squared displacement, Eq.~(\ref{eqMSDVelDef_res}), whose dependence on $\Delta n$ is well described by $\mathrm{MSD}_r(\Delta n)=\mathrm{MSD}_0+2D\Delta n$. As shown in Fig.~\ref{fgr:carbo}b (top panel), the two estimates of the diffusion coefficient are in excellent agreement (see also ESI Fig.~S3).

\subsubsection{ROI-size test: $D$ is local, $v$ is non-local.~~}
The two processes have very different spatial features. We tested this quantitatively by extracting $D$ and $v$ separately on ROIs of three different linear sizes $L$ (Fig.~\ref{fgr:trc}a). While the diffusion coefficient $D$ is independent of $L$ within error, as expected for a process whose characteristic scale lies well below the observation window, the ballistic velocity $v$ is proportional to $L$ for small and intermediate $\gamma_L$, as expected for a correlated flow whose correlation length $\xi$ exceeds $L$. At the smallest amplitudes the three curves are offset by roughly half a decade: $v$ increases by a factor $\simeq5$ when $L$ is increased by a factor $4$, from $256$ to $1024$~px, so that $v\propto L^{\beta}$ with $\beta\simeq1.1$, i.e.\ an rms velocity that grows in proportion to the box size as long as $L<\xi$. The offset shrinks monotonically as $\gamma_L$ grows, with $\beta$ falling to $\simeq0.25$ at the largest amplitudes explored, consistent with $\xi$ becoming comparable to or smaller than the ROI so that $v$ approaches its $L$-independent limit. In that regime the smallest ROI necessarily displays the steepest apparent rise with $\gamma_L$, since it starts furthest below that common limit. This is consistent with the shrinking correlation structures visible in Fig.~\ref{fgr:carbo}c. The observation of this long-wavelength motion, even after subtracting the centre-of-mass displacement, indicates that it corresponds to a genuine physical process, not an artefact, occurring at length scales greater than the image field of view. Moreover, this demonstrates that the residual MSD (Eq.~\ref{eqMSDVelDef_res}) on which we compute $D$ is a robust, ROI-independent quantity.

\subsubsection{Tracer-size test: $D$ is Stokes--Einstein-like, $v$ is not.~~}
Fig.~\ref{fgr:trc}b shows $D$ and $v$ for the Carbopol seeded with two tracer sizes, $a=0.5$~\textmu m and $a=2.5$~\textmu m, at matched strain amplitudes. These are the two extremes of the three radii introduced in Section~\ref{sec:samples}, chosen to maximise the lever arm in $a$; the intermediate size, $a=1.0$~\textmu m, is the one used for all other measurements reported here, in both samples. The diffusion coefficient $D$ depends systematically on $a$. Assuming a power-law dependence $D\propto a^{-b}$, minimisation of the rms logarithmic difference between the scaled curves $D\cdot a^b$ yields $b=0.95\pm0.15$, statistically indistinguishable from $b=1$. Accordingly, plotting $D\cdot a$ as a function of $\gamma_L$ leads to a good collapse of the data for different tracer sizes onto a single curve (Fig.~\ref{fgr:trc}c, top). This scaling $D\sim1/a$ is strikingly reminiscent of thermal diffusion described by the Stokes--Einstein relation. The resemblance is structural rather than literal. The tracers are elastically trapped in the matrix, which limits their thermal motion to a plateau of $k_BT/(\pi a G'_\mathrm{lin})$. For Carbopol, this is of order $10^{-5}$~\textmu m$^2$, three orders of magnitude below $\mathrm{MSD}_0$. A bounded plateau does not grow with $\Delta n$, so thermal motion cannot contribute to the $2D\Delta n$ term of Eq.~(\ref{eq:twoprocess}), which increases without limit. The measured Fickian growth is therefore entirely shear-induced. The $1/a$ scaling points to a shear-activated energy scale associated with processes below the tracer size, rather than to $k_BT$. In line with earlier passive-microrheology evidence~\cite{brizioli2022reciprocal,edera2021deformation}, this is consistent with the hypothesis that the energy driving the diffusive process is released at scales smaller than the smallest tracer, i.e. at the scale of the microgel mesh; we return to this point in Section~\ref{sec:disc_diffusive}. In contrast, the ballistic velocity $v$ is independent of $a$ (Fig.~\ref{fgr:trc}b, bottom): the ballistic process carries tracers of different sizes with the same velocity, as expected for a flow whose correlation length exceeds both the tracer and the ROI.

\subsubsection{DDM confirms the superposition.~~}
\label{sec:results_ddm}
To independently assess this two-process picture, we compute the intermediate scattering function $f(q,\Delta n)$ using DDM\cite{aime2019probingII,edera2021deformation} on the same echo sequences. We found that $f(q,\Delta n)$ is well-described by a \textit{product} of two Fourier transforms of the underlying diffusive and ballistic propagators, consistent with the two processes acting simultaneously on the same particles. Specifically, we modelled the intermediate scattering function as $f(q,\Delta n)=P(\Gamma_2(q)\Delta n)\exp(-\Gamma_1(q)\Delta n)$, where the exponential term accounts for the diffusion, while $P(\Gamma_2(q)\Delta n)$ is a closed-form analytical expression describing the ballistic contribution, obtained assuming a Flory--Schulz distribution of particle velocities \cite{wilson2011differential}. As shown in Fig.~S3, the two estimated relaxation rates exhibit the expected scaling: $\Gamma_1(q)$ follows the quadratic scaling characteristic of diffusion, $\Gamma_1(q)=Dq^2$, whereas $\Gamma_2(q)$ scales linearly with $q$, allowing a characteristic velocity to be extracted.

The shear-induced diffusion coefficient obtained from DDM is in remarkably good agreement with the particle-tracking value. Moreover, DDM velocity matches the one obtained from particle tracking with the intensity-based registration alone, without centre-of-mass subtraction. This is expected, since DDM operates directly on the images and cannot remove the mean displacement of the tracers; the two routes agree once referred to the same frame (ESI, Fig.~S3).

A direct comparison of representative DDM results obtained using our superposition model and a \textit{coexistence} model~\cite{aime2023unified}, based on the \textit{sum} of ballistic and diffusive contributions, $f(q,\Delta n)=w\,P(\Gamma_2(q)\Delta n)+(1-w)\exp(-\Gamma_1(q)\Delta n)$, is reported in Fig.~S9. Partly owing to the additional fitting parameter $w$, the coexistence model still provides a reasonably good fit to the experimental ISFs. However, while it yields a fairly consistent estimate of the diffusive component, the determination of the subdominant ballistic contribution becomes highly unreliable (ESI, Fig.~S9), likely because of cross-talk among the parameters $v$, $\sigma_v$ and $w$. Reciprocal-space data alone would therefore make it difficult to choose between the two models; it is the agreement with the particle-tracking data that singles out the superposition.

\subsection{Shear-banded yielding: dense emulsion}
\label{sec:results_emulsion}
\begin{figure}[t!]
\centering
\includegraphics[width=0.9\columnwidth]{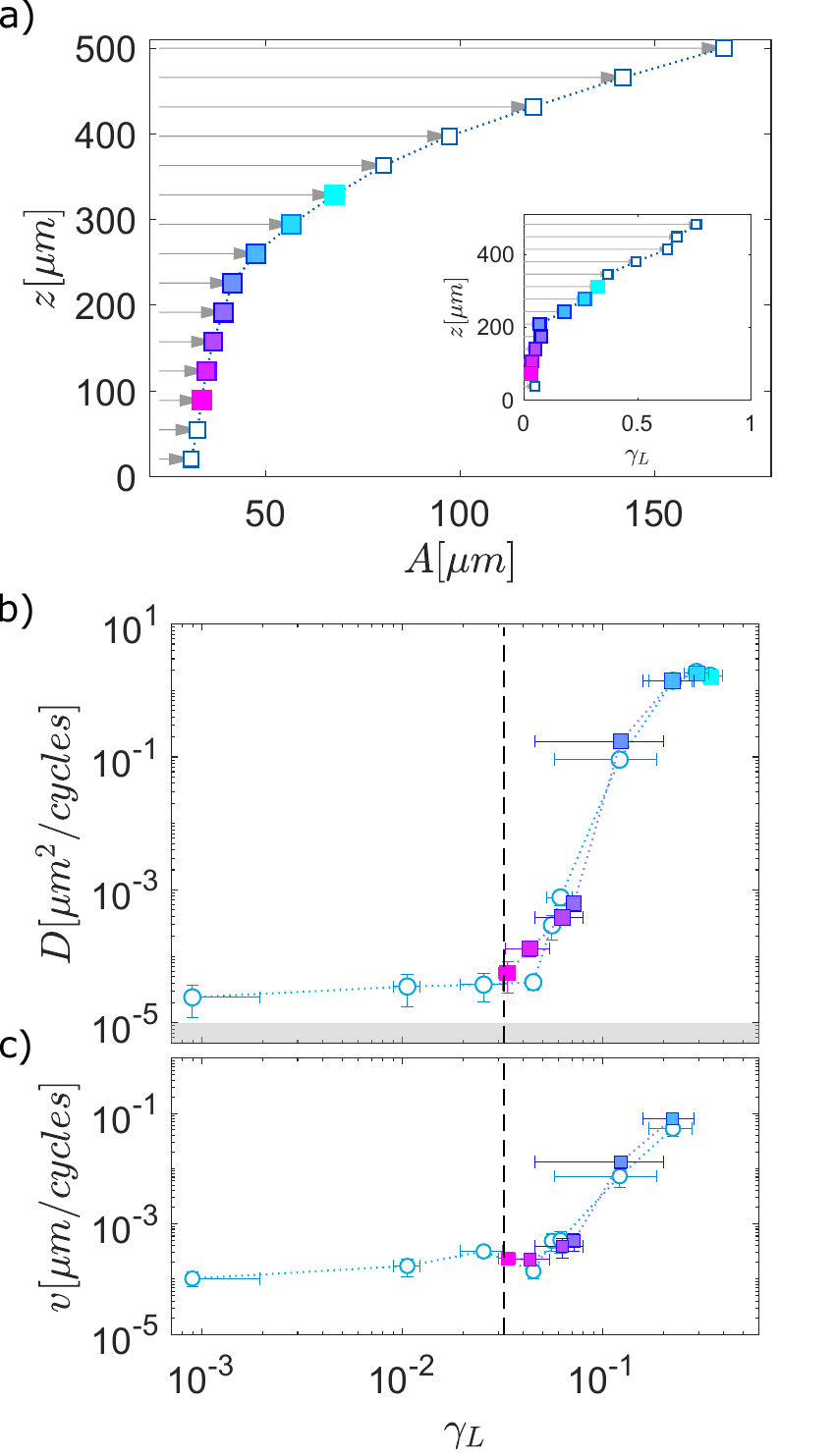}
\caption{\textbf{Shear-banded yielding in the dense oil-in-water emulsion.}
(a) Displacement amplitude $A(z)$ (main panel) and local strain amplitude $\gamma_L(z)$ (inset), defined as in Section~\ref{sec:defprofile}, as functions of the vertical position $z$ across the gap, measured for an imposed top-plate displacement $A_0 = 250$~\textmu m (corresponding to nominal gap-averaged strain $\gamma_0 \simeq 0.5$). The apparent wall-slip velocities at the top and bottom plate amount to $\simeq 35$\% of the local fluid velocity. Colour coding (cyan to magenta) indicates the height $z$ used in panels~(b)--(c). (b) Shear-induced diffusion coefficient $D$, and (c) rms ballistic velocity $v$, both plotted against the \textit{local} strain amplitude $\gamma_L$. The vertical lines indicate the strain amplitude for which $G'=G''$ in bulk rheology. Blue circles: data acquired at a fixed height $z = h/2$ with varying nominal amplitude $A_0$. Squares (coloured as in panel~a): data acquired at a fixed nominal amplitude $A_0 = 250$~\textmu m but varying height $z$, corresponding to different local strain amplitudes inside the same banded profile. The two datasets collapse onto a single master curve: the microscopic dynamics is controlled primarily by $\gamma_L$. Shaded areas indicate measurement-reliability limits for each observable. Particle radius is $a=1.0$ \textmu m.}
\label{fgr:emul}
\end{figure}

We now turn to the dense emulsion, whose macroscopic amplitude sweep is qualitatively similar to that of the Carbopol (Fig.~S1) but whose brittility factor $B_t$ is substantially larger (see above). As in the Carbopol, the tracer dynamics can be described as the superposition of two processes, with the shear-induced diffusion coefficient $D$ and the ballistic velocity $v$ rising sharply at a characteristic $\gamma_L$ of $0.1$--$0.15$ (Fig.~\ref{fgr:emul}b--c and Fig.~S4). The velocity maps of the emulsion (ESI, Fig.~S4) also reveal persistent, long-wavelength flows, but their root mean square amplitude is a factor $\sim 5$ smaller than in the Carbopol at matched $D$.

\subsubsection{Emergence of stationary shear bands.~~}
The mesoscopic behaviour of the emulsion is qualitatively different from that of the Carbopol. Fig.~\ref{fgr:emul}a shows the displacement amplitude profile $A(z)$ for an imposed top-plate displacement $A_0=250$~\textmu m. The profile is strongly non-linear, with the local strain amplitude $\gamma_L(z)$ (inset) varying across the gap, from $\gamma_L\simeq 0.6$ near the top plate to $\gamma_L\simeq 0.02$ near the bottom plate, suggesting the occurrence of shear banding. The data of Figs.~S2 and~S5 strongly support this picture: at sufficiently large stresses, two distinct bands can be identified across the gap, namely a solid-like band close to the bottom and a fluidised band in the upper part of the cell, separated by a relatively smooth crossover region. Throughout, we locate the interband boundary at the maximum of $\partial_z\gamma_L(z)$, i.e.\ at the position of greatest curvature of $A(z)$. %We note that this is not the inflection point of $A(z)$: even for a constant phase, where $\gamma_L\simeq A'(z)$, the condition $A''=0$ makes $\partial_z\gamma_L$ vanish rather than peak, and a varying phase makes $\gamma_L$ no longer a simple derivative of $A(z)$.
 The width of the crossover is not resolved by our $z$-scan and we do not attempt to quantify it here. As the stress increases, the high-strain-amplitude band gradually expands and occupies an increasing fraction of the gap at the expense of the low-strain-amplitude band~\cite{divoux2016shear}. Notably, for each imposed strain amplitude, the profile remains stationary within the experimental time scale, and its formation is fully repeatable upon rejuvenation, both in the position of the bands within the gap and in the corresponding strain amplitudes (Fig.~S5). %The same figure also shows that the profile does not depend on the direction in which the amplitude is varied: ascending and descending scans superpose, both within a single series and between two series separated by a rejuvenation step. We therefore detect no measurable thixotropic or history dependence of the banded profile over the range explored.

Alongside the non-homogeneous profiles, the deformation profiles present marked wall slip, in line with the well-known difficulty of fully suppressing slip in microgel and emulsion pastes~\cite{meeker2004slip,seth2011micromechanical,zhang2017wall,vleminckx2020effect,woodbridge2026subyield}, despite the use of ground-glass slides. However, we stress that our definition of \textit{local} strain $\gamma_L(z)$ is insensitive to the presence of slip.

Edge disturbances are known to induce apparent, quasi-bulk shear banding in complex fluids~\cite{hemingway2020interplay,chan2023perspective}. We exclude this origin for the bands reported here: the imaged region lies at least $1$~cm from the free edge, and the stationarity and quantitative repeatability of the profile documented above would not be expected for an edge-driven disturbance.

\subsubsection{$D$ and $v$ depend only on the local strain.~~}
\label{sec:results_localstrain}
One of the main methodological advantages of the gradient-resolved echo is that we can measure the shear-induced $D$ and $v$ at different planes $z$ within the same banded profile. Fig.~\ref{fgr:emul}b--c shows $D(\gamma_L)$ and $v(\gamma_L)$ for two experimental strategies: (blue open circles) fixed height $z=h/2$ and varying nominal strain amplitude $A_0$; (squares, coloured as in Fig.~\ref{fgr:emul}a) fixed nominal amplitude $A_0 = 250$~\textmu m and varying height $z$, which corresponds to different local strain amplitudes $\gamma_L(z)$. Remarkably, the two datasets collapse onto a single master curve for both $D(\gamma_L)$ and $v(\gamma_L)$. \textit{The shear-induced microscopic dynamics is therefore solely governed by the local strain, within our experimental resolution}; the information that a given layer is part of a macroscopic banded structure does not enter explicitly. The second dataset makes this point directly: the shear stress is uniform across the gap in this geometry, so all the squares correspond to one and the same applied stress, yet they span local strain amplitudes differing by more than an order of magnitude and follow the same curve as the fixed-height data. The macroscopic stress is therefore not the relevant control parameter at the microscopic level~\cite{coussot2009macroscopic,goyon2008spatial}. Finally, the dependence of $D$ on $\gamma_L$ is substantially steeper in the emulsion than in the Carbopol (Fig.~\ref{fgr:emul}b \textit{vs} Fig.~\ref{fgr:carbo}b), consistent with its larger brittility.

\begin{figure*}[t]
\centering
\includegraphics[width=0.9\linewidth]{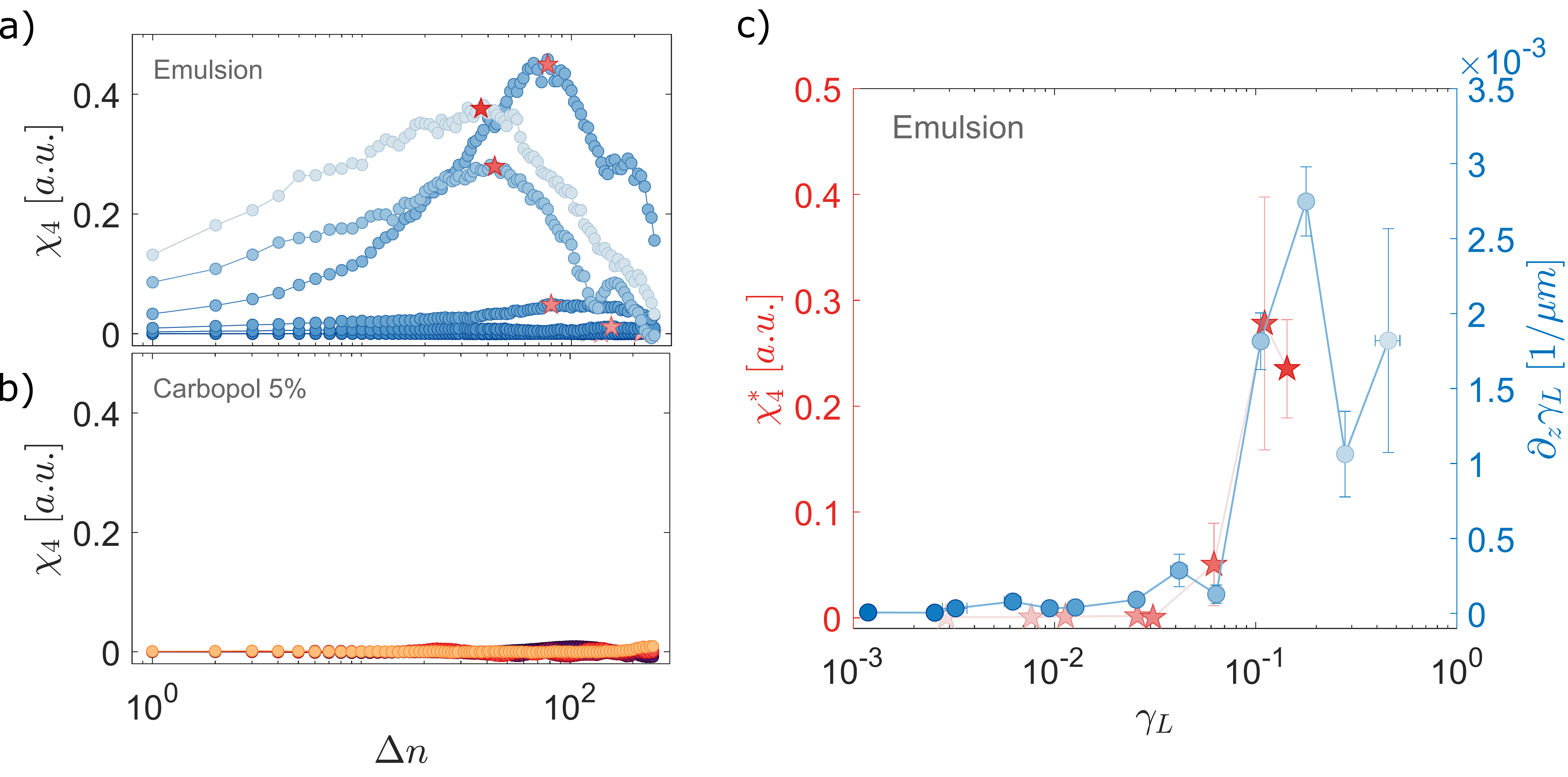}
\caption{\textbf{Cooperativity of the diffusive process and its emergence at the interband strain.}
(a) Four-point susceptibility $\chi_4(\Delta n)$ of the emulsion, computed from the four-point dynamic correlation function $G_4$ according to Eqs.~(\ref{eq:G4}--\ref{eq:chi4}), as a function of the number of elapsed cycles $\Delta n$, in a single representative experiment performed at a fixed observation plane while varying the imposed stress amplitude, corresponding to local strain amplitudes $\gamma_L$ within the range $[0.003,\,0.15]$. For every $\gamma_L$ above yielding, $\chi_4(\Delta n)$ develops a pronounced peak $\chi_4^\ast$ (stars), whose magnitude measures the size of cooperatively rearranging regions. Because this panel reports a single run, the individual $\chi_4^\ast$ values shown here can exceed the run-averaged values plotted in panel (c); the complete set of single-run values is given in Fig. S8c.
(b) Same observable for the Carbopol, for local strain amplitudes $\gamma_L$ ranging from $0.01$ to $1.77$ (from dark to light). $\chi_4(\Delta n)$ remains flat and small at all strain amplitudes explored: the diffusive dynamics of the Carbopol is non-cooperative.
(c) Test of the prediction of Parisi \textit{et al.}~\cite{parisi2017shear}. Red stars: peak four-point susceptibility $\chi_4^\ast$ for the emulsion, averaged over the two independent echo experiments reported in Fig. S8c, plotted as a function of the local strain amplitude $\gamma_L$. Blue circles: gradient of the local strain amplitude, $\partial_z \gamma_L(z)$, measured in a series of five independent $z$-scans (Fig.~S8b), similar to the one reported in Fig.~\ref{fgr:emul}. Over the range where both are measured ($\gamma_L\lesssim 0.16$), the two observables track each other, both largest where the gradient of the shear-banded profile is steepest ($\gamma_L\simeq 0.10$--$0.15$), indicating that the diffusive component of the microscopic dynamics becomes cooperative at the strain that marks the interband region. Particle radius is $a=1.0$ \textmu m for all samples. }
\label{fgr:chi4}
\end{figure*}

\subsubsection{Heterogeneous, non-Gaussian, cooperative dynamics.~~}
While the observables $D$ and $v$ display qualitatively similar behaviour in the two samples once plotted against $\gamma_L$, a marked difference emerges in the statistics of particle displacement. Beyond the long-time linear dependence of the $\mathrm{MSD}_r$ on $\Delta n$, we find that the probability distribution functions (PDFs) of the particle displacements for the emulsion display highly non-Gaussian, exponential-like tails at all strain amplitudes. This contrasts with the Carbopol, whose dynamics is Fickian and close to Gaussian. The difference can be quantified through the non-Gaussian parameter $\alpha_2$, Eq.~(\ref{eqNonGaus}): for the emulsion, $\alpha_2$ increases significantly at intermediate strain amplitudes and exhibits peaks one to two orders of magnitude larger than in the Carbopol, where $\alpha_2<1.5$ at all strain amplitudes explored (see Fig.~S6). The values we find for the Carbopol are comparable to those reported for model hard-sphere glasses under oscillatory shear, where the non-Gaussian parameter peaks at a similar magnitude close to the yield strain and decreases at larger amplitudes~\cite{koumakis2016amorphous}.

The origin of this behaviour can be investigated by inspecting single-particle trajectories. Comparing trajectories in the emulsion and in the Carbopol at strain values leading to the same effective diffusion coefficient $D$ (see Fig.~S7), we observe that Carbopol particles undergo small positional fluctuations with little interparticle variability, whereas trajectories in the emulsion display pronounced heterogeneity, with particles exhibiting localised motion as well as large positional fluctuations and intermittent rearrangements, which leads to the highly non-Gaussian statistics. This heterogeneity is nevertheless not the signature of two distinct populations: the same particle may alternate intermittently between localised and highly mobile dynamics along a single trajectory.

Furthermore, some tracers display rare, recurrent back-and-forth jumps between metastable positions. These events account for the fitted short-time plateau value $\mathrm{MSD}_0$, which, unlike in the Carbopol, cannot be attributed to localisation noise and instead increases with the applied strain amplitude (Fig.~S4). Because the echo protocol cancels the affine contribution, these excursions are genuine non-affine rearrangements rather than elastic deformation. They are also, in a specific sense, reversible: the tracer returns repeatedly to positions it has already occupied, instead of drifting away from them.

It is worth being precise about what such a plateau can and cannot mean. A tracer returning to exactly the same position at the end of every cycle would contribute nothing whatsoever to a stroboscopic mean-squared displacement, at any $\Delta n$. A finite $\mathrm{MSD}_0$ therefore does not measure reversible motion \textit{within} a cycle (to which the echo protocol is blind by construction) but departures from exact cycle-to-cycle repetition. Individual trajectories show directly what these are. In Fig.~S7 the squared displacement of representative emulsion tracers at $\gamma_L=0.12$ alternates repeatedly between two well-separated levels, while other tracers in the same field of view show no such switching, so that only a minority of tracers is involved. The excursions correspond to separations of order $2$~\textmu m between the two metastable positions, comparable to the droplet diameter $2R_d=0.9$~\textmu m, so that each jump amounts to a local rearrangement of the surrounding droplets.

This picture is supported by simulations of cyclically sheared amorphous solids. The elementary unit of plasticity in such systems is a bistable local rearrangement switching hysteretically between two configurations~\cite{mungan2019networks}; the number and activity of these units grow as the yield amplitude is approached, and they are found to be spatially clustered~\cite{szulc2022cooperative}. We stress that this clustering is not a partition of the sample into two populations in the sense of Section~\ref{sec:disc_superposition}: the bistable units are a minority of localised rearrangements embedded in a matrix that is itself advected by the ballistic flow, and the same tracer may switch between participating and not participating as the drive proceeds. What is spatially heterogeneous is the \textit{activity} of a single dynamics, not the identity of two coexisting phases. Deterministic period-locking is moreover fragile: weak noise makes a system locked in a limit cycle hop between neighbouring ones~\cite{regev2013onset}, which is the relevant situation for a thermal sample driven at finite rate rather than quasistatically. Consistently, a stroboscopic mean-squared displacement that saturates below yield at a value increasing with driving amplitude is also found in these simulations~\cite{regev2019critical}. We therefore interpret $\mathrm{MSD}_0$ as a signature of \textit{reversible plasticity}~\cite{keim2014mechanical,reichhardt2023reversible}, in the sense of rearrangements that are revisited rather than accumulated.

While this intermittent behaviour indicates the presence of spatiotemporal fluctuations in single-particle trajectories, an important question is whether these fluctuations are spatially correlated, thereby reflecting cooperative behaviour. To quantify the degree of cooperativity in the dynamics, we compute the four-point dynamic susceptibility $\chi_4$, Eq.~(\ref{eq:chi4}). Small values of $\chi_4$ indicate independent displacements, while a pronounced peak signals heterogeneous dynamics characterised by spatial clustering of particles with similar mobility. Fig.~\ref{fgr:chi4}a shows $\chi_4$ as a function of the number of elapsed cycles $\Delta n$ for the emulsion at a series of $\gamma_L$: above yielding, $\chi_4(\Delta n)$ develops a pronounced peak $\chi_4^\ast$, whose magnitude and position depend on $\gamma_L$, whereas no measurable peak is found at any $\gamma_L$ for the corresponding Carbopol data (Fig.~\ref{fgr:chi4}b).

The non-affine shear-induced diffusive dynamics is therefore non-cooperative and Gaussian in the Carbopol and strongly cooperative and non-Gaussian in the emulsion.

\subsubsection{$\chi_4^\ast$ is largest at the interband strain.~~}
No dynamical heterogeneity is observed in the Carbopol, which also shows no banding, whereas the emulsion displays both. This suggests that the cooperativity of the diffusive process is related to the presence of a banded deformation profile, a hypothesis we now test quantitatively. Plotting the gradient of the local strain amplitude, $\partial_z\gamma_L(z)$, against $\gamma_L$ itself, we find that deviations from a linear deformation profile ($\gamma_L(z)=\mathrm{const}$, $\partial_z\gamma_L(z)=0$) are largest between $\gamma_L\simeq0.10$ and $0.15$ in every experimental run (Fig.~S8). The onset of shear banding is therefore itself controlled by the local strain amplitude: once $\gamma_L$ exceeds a threshold value, the deformation ceases to be sustained uniformly and the profile separates into two bands.

In Fig.~\ref{fgr:chi4}c we report both this gradient (blue circles) and the peak four-point susceptibility $\chi_4^\ast$ (red stars) as functions of the local strain, averaged over five and two independent experimental runs, respectively. The two observables track each other, both largest between $\gamma_L\simeq0.10$ and $0.15$, the strain at which the gradient of the deformation profile is largest. Note that the $\chi_4(\Delta n)$ curves of Fig.~\ref{fgr:chi4}a correspond to a single representative run, so the individual peak values visible there are not directly comparable with the run-averaged $\chi_4^\ast$ of Fig.~\ref{fgr:chi4}c.

The two measurements are independent: $\chi_4$ comes from amplitude sweeps at a fixed observation plane, the gradient from $z$-scans of the deformation profile. We chose the fixed-plane protocol because $\gamma_L(z)$ is reconstructed offline, so that during acquisition one cannot know which local strain a given plane is experiencing, while a converged $\chi_4$ demands a long echo sequence at every point. Sweeping the amplitude at a fixed plane instead lets the local strain traverse the interband value continuously, every point being assigned its measured $\gamma_L$ \textit{a posteriori}. A direct axial mapping of $\chi_4$ is left to future work, possibly making use of the stress- and strain-controlled setup of Ref.~\cite{kalafatakis2025}.

It is essential here that the analysis is carried out in terms of $\gamma_L$ rather than of the nominal strain: the apparent wall slip in the emulsion is large ($\simeq 35\%$ of the local fluid velocity, Fig.~\ref{fgr:emul}a), so $\gamma_0=A_0/h$ is a poor proxy for the deformation actually experienced at a given plane. Because $\gamma_L$ is obtained from the relative displacement of neighbouring planes, it is insensitive to slip (Section~\ref{sec:results_emulsion}), and the collapse of Fig.~\ref{fgr:emul}b--c establishes that it remains the controlling variable even in the presence of pronounced slip. The strain at which $\chi_4^\ast$ is largest therefore coincides with the strain that characterises the interband region. We stress that this is an identification of a \textit{strain}, not a direct spatial localisation. The collapse of $D$ and $v$ against $\gamma_L$ does not by itself guarantee that a correlation measure such as $\chi_4$ is also a function of $\gamma_L$ alone: identical mean mobility can coexist with different spatial correlations. The data of Fig.~\ref{fgr:chi4}c are in fact equally compatible with a cooperativity controlled by the gradient $\partial_z\gamma_L$ itself. Both readings place the enhanced cooperativity at the interband region, but only a gradient-resolved measurement of $\chi_4$ could separate them.

This result indicates enhanced cooperativity as the shear bands develop inside the gap, consistent with the prediction of Parisi \textit{et al.}~\cite{parisi2017shear} that shear banding in amorphous solids is accompanied by the growth of a dynamical correlation length. Our data further locate that signature: the ballistic component is spatially correlated in both samples, including the Carbopol, where no bands form, so its correlation is not by itself a signature of banding. It is the \textit{diffusive} component that becomes cooperative, and it does so at the strain marking the interband region.

%%%%%%%%%%%%%%%%%%%%%%%%%%%%%%%%%%%%%%%%%%%%%%%%%%
\section{Discussion}
\label{sec:discussion}

\subsection{Superposition \textit{vs} coexistence}
\label{sec:disc_superposition}
As recalled in Section~\ref{sec:intro_micro}, the echo experiments of Aime \textit{et al.}~\cite{aime2023unified} identify two relaxation processes---slow/ballistic and fast/diffusive---and attribute them to two \textit{coexisting} populations of particles. The data we report here are equally compatible with the presence of two relaxation processes, but our real-space analysis indicates that they are superposed on the \textit{same} particles rather than carried by distinct populations: the ballistic process has a correlation length that exceeds the ROI and a magnitude independent of tracer size (Fig.~\ref{fgr:trc}), while the diffusive process is Stokes--Einstein-like and ROI-independent. Neither process is confined to a subset of tracers: the coarse-grained velocity field is smooth across the entire field of view, so every tracer is carried by the ballistic flow, while the diffusive contribution is present at every tracer size examined. What discriminates between the two scenarios is this continuity of the velocity field: a ballistic subpopulation interspersed with a diffusive one would not produce a smooth, spatially continuous field advecting every tracer within it. The size scalings of Fig.~\ref{fgr:trc} are complementary, but weaker, evidence: $D\propto1/a$ and a tracer-size-independent $v$ characterise the two \textit{processes}, and would hold under either scenario. In this case, DDM analysis (ESI, Figs.~S3 and~S9) cannot clearly discriminate between the two scenarios, as the intermediate scattering function can be equally well described by either a \textit{product} or a \textit{sum} of two decays. The latter corresponds to the decomposition expected for two distinct fractions of the sample that relax at different rates and do not interconvert over the measurement timescale. A similar ambiguity affects the rheo-XPCS measurements of Donley \textit{et al.}~\cite{donley2023investigation}: the excess correlation they measure relative to the homogeneous affine prediction establishes that the distribution of irreversible single-particle displacements is not the affine one, but, being an ensemble average over the illuminated volume that carries no label identifying which particle moved, it does not by itself separate a broad distribution to which every particle contributes from a bimodal partition into rearranging and non-rearranging regions. Resolving that distinction is precisely what tracer-level measurements afford. We note that our observables are averaged over the duration of the echo sequence, so that a two-state dynamics fluctuating faster than this window would also produce both contributions for every tracer. What our measurements exclude is therefore a partition into persistent populations, one exclusively ballistic and one exclusively diffusive; they do not exclude a rapid exchange of individual tracers between dynamical states.

This distinction has consequences for the state-diagram description of yielding. A superposition picture does not, by itself, require a first-order coexistence region: both processes are ``on'' at the same time once the amplitude exceeds a threshold, and the ratio between them sets the effective relaxation. It is therefore possible that the Van-der-Waals mapping of Aime \textit{et al.} reflects the evolution of \textit{relative weights} of two simultaneously active processes rather than the evolution of populations across a coexistence region, and that the apparent first-order character of the transition in brittle samples is a reflection of the non-linear coupling between the two processes near the threshold. A careful reanalysis of published DLS data along the lines we propose here---fitting a product of two relaxations and testing the $q$-dependence of the two rates---would settle this question directly, as would an equivalent study of a sample that \textit{does} exhibit shear bands, such as the present emulsion. This is not in conflict with Section~\ref{sec:results_ddm}: what cannot discriminate between the two scenarios is the fit of an intermediate scattering function at a given $q$, whereas the test proposed here is the $q$-dependence of the two extracted rates, which distinguishes a product from a sum.

\subsection{Implications for constitutive modelling}
\label{sec:disc_constitutive}
If the shear-induced relaxation is a superposition of two processes rather than a sum of two populations, then constitutive models built on a single relaxation time are not excluded on structural grounds: the two processes do not correspond to two coexisting states requiring separate state variables. Such a model must nonetheless reproduce the phenomenology reported here. The Miyazaki relation $\Gamma=\Gamma_0+K|\dot\gamma|^n$~\cite{miyazaki2006nonlinear} could be re-written in terms of the unrecoverable strain rate~\cite{donley2020elucidating,kamani2024brittle,kamani2023understanding}---the quantity that, in the recovery-rheology picture, genuinely drives plastic rearrangements---rather than the total strain rate. This would preserve the single-time structure of the Miyazaki model while reconciling it with the $G''$ overshoot and the brittility picture. The proposal is supported by rheo-XPCS measurements on concentrated colloidal glasses, in which the microscopic relaxation was found to correlate with the unrecoverable rather than the total strain~\cite{donley2023investigation}, albeit in a creep and recovery rather than an oscillatory protocol. Our observation of simultaneous diffusion and ballistic drift, with two different spatial ranges and two different amplitude dependences, provides a stringent constraint that any such model must reproduce.

\subsection{Origin of the ballistic component}
\label{sec:disc_ballistic}
The long-wavelength, spatially correlated velocity field that drives the ballistic component of the MSD is a striking feature of our data. Its correlation length exceeds the ROI at small $\gamma_L$ and decreases with increasing $\gamma_L$ (Fig.~\ref{fgr:carbo}c), a behaviour qualitatively reminiscent of the Eshelby-cascade organisation seen in two-dimensional simulations of amorphous plasticity under athermal quasistatic driving~\cite{nicolas2018deformation,lemaitre2009rate}. Lemaître and Caroli~\cite{lemaitre2009rate} find that the size of avalanches decreases with increasing rate in the athermal limit; since our measurements are performed at fixed frequency, increasing $\gamma_L$ increases the strain rate, so the trend we observe maps onto theirs, in the sense that an avalanche size and a velocity correlation length both decrease as the rate increases. The correspondence concerns the spatial organisation and its rate dependence, not the temporal statistics of individual events, and suggests that the long-wavelength flows resolved here may carry a signature of rate-dependent avalanching. We note that our velocity field is stationary over the echo sequence (Section~\ref{sec:results_superposition}), whereas avalanches are intermittent by nature. A time-averaged field retains structure only if the plastic activity is spatially pinned rather than randomly distributed, so the stationarity we observe constrains the avalanche interpretation rather than supporting it directly. Intermittent, avalanche-like dynamics below and near the yield stress have likewise been reported in hard-sphere colloidal glasses, where they are progressively suppressed once the sample flows~\cite{jacob2019rheological}.

Independent support for a long-wavelength, non-diffusive component comes from a protocol in which macroscopic flow is absent altogether. Chen \textit{et al.}~\cite{chen2020microscopic} combined XPCS with \textit{in situ} rheometry on a charged-silica nanocolloidal glass sheared through yielding and then held at fixed macroscopic strain while the stress relaxed. Along the flow direction the correlation functions decay at a rate proportional to $q$ rather than $q^2$, the signature of convective motion, with a peak velocity that scales with the rate of stress decay and is close to the elastic-recoil estimate $(h/2G')|\mathrm{d}\sigma/\mathrm{d}t|$. Two aspects bear on our interpretation of $v$. First, the plates are held fixed throughout, so the internal flow is sustained by the relaxation of stored elastic stress rather than by an imposed wall displacement, which supports an internal, plasticity-driven origin for long-wavelength flows in yielded soft solids. Second, their measurement is averaged over the gap, and the authors note that a symmetric triangular flow and a narrow slip plane at one wall are indistinguishable in their data---precisely the limitation that the gradient-resolved echo removes. The two protocols differ, so the comparison bears on the existence and internal origin of the convective component rather than on a quantitative correspondence.

An alternative, also consistent with our data, is that the ballistic component reflects residual inhomogeneities of the boundary conditions (e.g. small wall-slip gradients, minor plate misalignment) amplified by the long correlation length of soft-glassy flow. Parallel-superposition-rheometry experiments on Carbopol and a body lotion under a subyield stress bias have shown that apparent unrecoverable drift can be quantitatively accounted for by a Navier slip contribution $\dot\gamma_\mathrm{slip}=\beta\,\sigma_{xy}$, with a slip mobility scaling as $\beta\propto 1/h$ with the plate gap $h$~\cite{woodbridge2026subyield,zhang2017wall}. In the Carbopol the measured profile is linear across the gap, with apparent wall-slip velocities at $z=0$ and $z=h$ consistent with zero (Section~\ref{sec:results_carbopol} and ESI, Fig.~S2); in the emulsion an apparent slip is measured, but the local-strain master curve $D(\gamma_L)$ is insensitive to the absolute plate displacement by construction.

We therefore ask whether an \textit{undetected} residual slip, lying below our measurement uncertainty on $A(z=0)$ and $A(z=h)$, could still account for the ballistic contribution to the MSD. A spatially uniform slip would produce a rigid plug-flow oscillation of the entire field of view and would be removed exactly by our centre-of-mass analysis, leaving no signature in $v$; only a spatially heterogeneous residual slip could contribute. Two observations bound such a scenario. First, a $v$ dominated by residual Navier slip would scale linearly with $\sigma_0$, i.e.\ as $\gamma_L^{1-\nu}$ in the terminal regime, predicting a growth of less than a factor $1.5$ per doubling of $\gamma_L$ for the Carbopol~5\% ($\nu\simeq0.44$); the measured $v$ follows no such law, being nearly independent of $\gamma_L$ at small and intermediate amplitudes and then rising far more steeply (Fig.~\ref{fgr:carbo}b, bottom). Second, the spatial organisation of the residual velocity field (Fig.~\ref{fgr:carbo}c) evolves systematically with $\gamma_L$, whereas the plate roughness that would set any heterogeneous-slip pattern is fixed.

Taken together, these observations disfavour residual slip as the dominant origin of $v$, but they do not establish the avalanche scenario positively; we present the two as alternatives that our data constrain without separating. A gap-thickness scan testing the $\beta\propto1/h$ prediction, together with superposition rheology, would place a direct upper bound on any residual slip. Bulk harmonic analysis cannot settle the question: even harmonics have long been attributed to wall slip~\cite{graham1995wall} but also arise in strictly no-slip flows from elasticity, shear thinning and inertia~\cite{atalik2004occurrence}, while a slip symmetric between the two half-cycles contributes only odd harmonics. Discriminating residual slip from bulk nonlinearity therefore requires the kind of gradient-resolved, real-space measurement we perform here.

Finally, all measurements reported here were performed at a single driving frequency, $\omega/(2\pi)=1$~Hz. The rate dependence implied by both the constitutive and the avalanche scenarios discussed above is therefore not tested by our data, and is a natural direction for future work.

\subsection{Origin of the diffusive component}
\label{sec:disc_diffusive}
The diffusive process obeys $D\propto1/a$ across a factor of five in tracer radius (Fig.~\ref{fgr:trc}c). Generally, a probe acquires a $1/a$ mobility when the forces driving it are uncorrelated over its own size, so that it couples to the surrounding material through Stokes friction alone; were the agitation organised on a scale comparable to $a$, the probe would sample it coherently and the $1/a$ dependence would be lost. Observing the scaling over the whole range $0.5$--$2.5$~\textmu m therefore suggests an upper bound of order $0.5$~\textmu m on the length scale at which the energy driving the diffusive process is released. The bound is not model-free: it assumes that the probe couples to the surrounding material through Stokes friction and that the driving forces are uncorrelated over the probe size.

That bound is not trivially satisfied, because the Carbopol microgels are several micrometres in hydrodynamic radius, i.e. larger than two of the three tracers we used. Had the shear-induced agitation been organised on the scale of the microgels themselves, $D$ would have crossed over as $a$ passed through a few micrometres. No such crossover is seen. This suggests that the elementary plastic events occur well inside the individual microgels, plausibly on the scale of their internal polymer mesh, the microgels behaving as swollen, mutually interpenetrating objects rather than as discrete solid grains.

Seen from a micrometric probe, cyclic shear thus acts on the diffusive contribution as a bath: it sets the intensity of the agitation, which grows by four orders of magnitude with $\gamma_L$, but not the scale of the elementary events, which remains sub-micrometric throughout, compatibly with the softness of the microgels. This is what separates the two processes so cleanly in the Carbopol: the ballistic component is governed by a correlation length far larger than any tracer, and is accordingly blind to $a$ but sensitive to $L$; the diffusive component is driven by events far smaller than any tracer, and is accordingly sensitive to $a$ but independent of $L$.

It is worth being explicit about which length the Stokes--Einstein scaling constrains, since the emulsion is strongly cooperative while the Carbopol is not. The $1/a$ dependence fixes the scale of the \textit{elementary} plastic events, whereas $\chi_4$ measures the spatial correlation of the \textit{mobility}, that is, of the local rate at which those events occur. The two are independent: a tracer may be driven by many small, mutually uncorrelated rearrangements, and therefore diffuse with a Stokesian $1/a$ mobility, while the rate of those rearrangements varies coherently over regions much larger than the tracer. The same combination is familiar in supercooled liquids, where cage-scale elementary events coexist with dynamical heterogeneity extending over many particle diameters~\cite{berthier2011dynamic}. What changes between our two samples is therefore not the microscopic mechanism of the diffusive process but its spatial organisation. It is to that organisation, and to its relation with shear banding, that we now turn.

\subsection{Shear banding as a microscopic-cooperativity transition}
\label{sec:disc_shearbanding}
The two samples differ in the cooperativity $\chi_4^\ast$ of the diffusive process, which in the banded sample is largest at the interband strain and in the homogeneous one is absent at all strains. This is consistent with the mean-field prediction of Parisi \textit{et al.}~\cite{parisi2017shear} that shear banding in amorphous solids is a critical phenomenon accompanied by a growing dynamical correlation length, although our accessible range of local strain does not extend far enough beyond the maximum to test whether that length actually diverges. The same conclusion follows from the lattice model of Aime and Truzzolillo~\cite{aime2025emergent}, in which the correlation length of shear-induced dynamic heterogeneities peaks at yielding and grows as the coupling disorder is reduced, so that samples yielding abruptly are those developing the most extended correlated dynamics. Our two samples realise the two limits of this scenario: no measurable cooperativity in the ductile Carbopol, and a pronounced $\chi_4^\ast$ in the brittle emulsion.

Our data therefore support the emergence of a growing dynamical length scale with increasing brittility, while clarifying its microscopic origin. The spatially correlated mobility we detect in the emulsion does not correspond to a partition of the sample into two distinct dynamical states: both relaxation processes remain active on every tracer, and the same tracer visits both extremes of mobility along a single trajectory. What grows with brittility is thus the spatial correlation of a single, strongly heterogeneous dynamics, rather than the size of coexisting solid-like and fluidised domains. We stress that the correlation length probed by the ROI dependence of $v$ is that of the ballistic velocity field, which \textit{decreases} with increasing $\gamma_L$, and is distinct from the correlation length of mobility heterogeneities probed by $\chi_4$, which is the quantity predicted to peak at yielding in Ref.~\cite{aime2025emergent}. We further note that the model does not presently produce steady shear bands, which its authors identify as requiring gradient terms in the local relaxation rate.

Our approach provides a concrete picture of the ``local non-universality'' of yielding in shear-banded systems noted by Coussot and Ovarlez~\cite{coussot2009macroscopic} and by the Bocquet--Colin--Ajdari fluidity model~\cite{bocquet2009kinetic,goyon2008spatial}: a single macroscopic sample lives, at different $z$, at different points of a phase diagram parameterised by $\gamma_L$, and the strains that characterise the ``interface'' between the two bands are those at which cooperativity is greatest. We emphasise that this two-state description applies to the \textit{flow profile}, at the scale of the bands, and not to the particle populations discussed in Section~\ref{sec:disc_superposition}, where our controls argue against any partition of the sample into two dynamical families for a given local shear rate.

\subsection{Brittility in real space}
\label{sec:disc_brittle}
Our real-space picture assigns a clear physical meaning to the ductile-to-brittle axis. In the ductile limit (low $B_t$), the deformation profile remains homogeneous, the shear-induced $D$ rises moderately steeply, and the diffusive process is Gaussian and non-cooperative: each sample layer behaves independently. In the brittle limit (high $B_t$), the deformation profile breaks into bands, the shear-induced $D$ rises more steeply, and the diffusive process becomes cooperative and non-Gaussian at the interband strain.

This picture is consistent with the random-critical-point scenario of Ozawa \textit{et al.}~\cite{ozawa2018random,ozawa2022rare}, where brittle yielding occurs when quenched disorder is small and the Eshelby interactions develop a long-range organised character, and with the slow-fatigue and delayed-yielding picture of Cochran \textit{et al.}~\cite{cochran2024slow}. In the same spirit, and adopting the same ductile-to-brittle framing independently, rheo-microscopy of Pluronic F127 micellar cubic gels---a high-yield-stress system with an abrupt solid-to-liquid transition---reveals a transient banding instability coinciding with the onset of bulk nonlinearity, together with macroscopic fracture at the largest amplitudes~\cite{agrawal2025connecting}. That fracture is observed there but not in our emulsion suggests that banding and fracture are distinct, and possibly sequential, manifestations of brittle yielding; a systematic connection to the fracture mechanics of soft solids~\cite{long2021fracture} remains to be established.

This picture also ties the $G''$ overshoot, one macroscopic signature of $B_t$, to the spatial organisation of plastic rearrangements, by linking $B_t$ to the terminal exponent $\nu$ and both in turn to $\chi_4^\ast$. The empirical universality observed by Keane \textit{et al.}~\cite{keane2025universal}, that the linear loss tangent $\tan\delta_0$ captures the nonlinear yielding behaviour, is consistent with our findings if one regards $B_t$ as a macroscopic proxy for the disorder landscape that, in turn, controls the cooperativity signature of the microscopic dynamics---an identification supported by the lattice model of Aime and Truzzolillo~\cite{aime2025emergent}, in which mechanical disorder is precisely the parameter separating gradual from abrupt yielding.

We note that the constitutive interpretation underlying that universality has recently been contested~\cite{woodbridge2026subyield}; we use $B_t$ purely as a macroscopic ordering parameter, which does not require the subyield-flow assumption, so the ordering of our two samples on the $B_t$ axis and its correlation with shear banding and cooperativity are robust to the outcome of that debate. We stress, however, that our two samples differ in composition as well as in brittility, so the correlation we report between $B_t$, shear banding and cooperativity is suggestive rather than causal; establishing $B_t$ as the controlling parameter would require a series of samples spanning intermediate $B_t$ at fixed chemistry, which we leave to future work.

%%%%%%%%%%%%%%%%%%%%%%%%%%%%%%%%%%%%%%%%%%%%%%%%%%%
\section*{Conclusions}
\label{sec:conclusions}
We have used rheo-microscopy with gradient-resolved particle tracking to compare two model yield stress fluids that lie at opposite ends of a macroscopic ductile-to-brittle (brittility) axis. In both samples the shear-induced microscopic dynamics is a superposition of two processes: a long-wavelength ballistic flow and a Fickian diffusion. In the Carbopol, where the tracer-size dependence was measured, the ballistic flow is size-independent and the diffusion Stokes--Einstein-like; the tracer-size control was not repeated in the emulsion. Independent controls on the region-of-interest size (isolating the non-local ballistic process) and on the tracer size (isolating the Stokes--Einstein-like diffusion) disfavour the coexistence of two particle populations---a description that reciprocal-space data alone cannot exclude---and favour a superposition on the same particles. In each sample the shear-induced diffusion coefficient $D$ is controlled primarily by the local strain $\gamma_L$, even in the emulsion, where $\gamma_L$ varies by more than an order of magnitude across the gap because of stationary shear bands; the two samples nonetheless follow distinct $D(\gamma_L)$ curves, that of the brittle emulsion being the steeper.

What distinguishes them at the microscopic level is the statistics of the diffusive process: the ductile Carbopol has near-Gaussian, non-cooperative dynamics, while the brittle emulsion develops non-Gaussian tails and a four-point susceptibility $\chi_4^\ast$ that is largest at the local strain characterising the interband region, providing direct experimental evidence of the enhanced cooperativity predicted to accompany shear banding in amorphous solids. Two samples cannot on their own establish $B_t$ as the controlling parameter, but they show what the ductile and brittle limits look like in real space, and thereby give concrete physical content to what it means for a yield-stress soft solid to be ``brittle''. They also indicate that constitutive descriptions need not invoke two coexisting populations, but must reproduce two simultaneous relaxation processes, acting on the same particles, with different spatial ranges and different amplitude dependences.

\section*{Author contributions}
Conceptualisation: P.E., M.B., V.T., G.P., P.C., F.G., R.C.; Methodology: P.E., M.B., V.T., G.P., F.G., R.C.; Investigation: P.E., M.B., M.M.; Sample preparation and formulation: M.M., E.N., P.C.; Data curation: P.E., M.B.; Formal analysis: P.E., M.B.; Software: P.E., M.B.; Validation: P.E., M.B., M.M.; Visualisation: P.E., M.B.; Supervision: P.C., V.T., G.P., F.G., R.C.; Funding acquisition: V.T., G.P., F.G., R.C.; Writing---original draft: P.E., M.B., V.T., F.G., R.C.; Writing---review and editing: all authors.

\section*{Conflicts of interest}
There are no conflicts to declare.

\section*{Data availability}
The analysis codes and representative datasets supporting the findings of this paper are publicly available on the Zenodo repository, \href{https://doi.org/10.5281/zenodo.10559405}{DOI: 10.5281/zenodo.10559405}. Raw image sequences (several TB) can be made available upon reasonable request to the corresponding authors.

\section*{Acknowledgements}
We are grateful to Stefano Aime, Ezequiel Ferrero, Stefano Villa and Giuliano Zanchetta for fruitful discussions. The authors thank Lubrizol for kindly supplying the Carbopol
(microgel) powder. This research was funded in whole or in part by the Austrian Science Fund (FWF) [Grant DOI 10.55776/PIN9311124] (Transforming Gels Through Training—TRAINGEL). For open access purposes, the authors have applied a CC BY public copyright license to any author accepted manuscript version arising from this submission. This research was funded by the European Union through the Horizon Europe Twinning project FORGREENSOFT (Grant Agreement No. 101078989). Views and opinions expressed are however those of the author(s) only and do not necessarily reflect those of the European Union. Neither the European Union nor the granting authority can be held responsible for them. P.E. acknowledges financial support from CNRS (France) and V.T. from the Swiss National Science Foundation [Grant No. 197340].

\balance

% Main-text bibliography, embedded from the supplied RSC BibTeX style.
\providecommand{\noopsort}[1]{}\providecommand{\singleletter}[1]{#1}%
\providecommand*{\mcitethebibliography}{\thebibliography}
\csname @ifundefined\endcsname{endmcitethebibliography}
{\let\endmcitethebibliography\endthebibliography}{}

\clearpage
\renewcommand{\thefigure}{S\arabic{figure}}
\renewcommand{\thetable}{S\arabic{table}}
\renewcommand{\theequation}{S\arabic{equation}}
\setcounter{figure}{0}
\setcounter{table}{0}
\setcounter{equation}{0}

\pagestyle{fancy}
\thispagestyle{plain}
\fancypagestyle{plain}{\renewcommand{\headrulewidth}{0pt}}

\makeFNbottom
\makeatletter
\renewcommand\LARGE{\@setfontsize\LARGE{15pt}{17}}
\renewcommand\Large{\@setfontsize\Large{12pt}{14}}
\renewcommand\large{\@setfontsize\large{10pt}{12}}
\renewcommand\footnotesize{\@setfontsize\footnotesize{7pt}{10}}
\makeatother

\renewcommand{\thefootnote}{\fnsymbol{footnote}}
\renewcommand\footnoterule{\vspace*{1pt}%
\color{cream}\hrule width 3.5in height 0.4pt \color{black}\vspace*{5pt}}
\setcounter{secnumdepth}{5}

\makeatletter
\renewcommand\@biblabel[1]{#1}
\renewcommand\@makefntext[1]%
{\noindent\makebox[0pt][r]{\@thefnmark\,}#1}
\makeatother
\renewcommand{\figurename}{\small{Fig.}~}
\sectionfont{\sffamily\Large}
\subsectionfont{\normalsize}
\subsubsectionfont{\bf}
\setstretch{1.125}
\setlength{\skip\footins}{0.8cm}
\setlength{\footnotesep}{0.25cm}
\setlength{\jot}{10pt}
\titlespacing*{\section}{0pt}{4pt}{4pt}
\titlespacing*{\subsection}{0pt}{15pt}{1pt}

\fancyfoot{}
\fancyfoot[LO,RE]{\vspace{-7.1pt}\includegraphics[height=9pt]{head_foot/LF}}
\fancyfoot[CO]{\vspace{-7.1pt}\hspace{13.2cm}\includegraphics{head_foot/RF}}
\fancyfoot[CE]{\vspace{-7.2pt}\hspace{-14.2cm}\includegraphics{head_foot/RF}}
\fancyfoot[RO]{\footnotesize{\sffamily{1--\pageref{LastPage} ~\textbar  \hspace{2pt}\thepage}}}
\fancyfoot[LE]{\footnotesize{\sffamily{\thepage~\textbar\hspace{3.45cm} 1--\pageref{LastPage}}}}
\fancyhead{}
\renewcommand{\headrulewidth}{0pt}
\renewcommand{\footrulewidth}{0pt}
\setlength{\arrayrulewidth}{1pt}
\setlength{\columnsep}{6.5mm}
\setlength\bibsep{1pt}

%%%TITLE, AUTHORS%%%
\twocolumn[
  \begin{@twocolumnfalse}
{\includegraphics[height=30pt]{head_foot/SM}\hfill\raisebox{0pt}[0pt][0pt]{\includegraphics[height=55pt]{head_foot/RSC_LOGO_CMYK}}\\[1ex]
\includegraphics[width=18.5cm]{head_foot/header_bar}}\par
\vspace{1em}
\sffamily
\begin{tabular}{m{4.5cm} p{13.5cm} }

\includegraphics{head_foot/DOI} & \noindent\LARGE{\textbf{Electronic Supplementary Information for ``Multiscale dynamics in yield stress fluids during homogeneous and shear-banded yielding''}} \\
\vspace{0.3cm} & \vspace{0.3cm} \\

 & \noindent\large{Paolo Edera,\textit{$^{a,b,\ddag}$} Matteo Brizioli,\textit{$^{a,\ddag}$} Mahnoosh Madani,\textit{$^{c,d}$} Elie N'gouamba,\textit{$^{e}$} Philippe Coussot,\textit{$^{e}$} V{\'e}ronique Trappe,\textit{$^{f}$} George Petekidis,\textit{$^{c,d}$} Fabio Giavazzi,\textit{$^{a,\S}$} and Roberto Cerbino\textit{$^{a,g,\S}$}} \\

\end{tabular}

 \end{@twocolumnfalse} \vspace{0.6cm}
  ]
%%%END OF TITLE %%%

\renewcommand*\rmdefault{bch}\normalfont\upshape
\rmfamily
\section*{}
\vspace{-1cm}

\footnotetext{\textit{$^{a}$~Dipartimento di Biotecnologie Mediche e Medicina Traslazionale, Universit\`a degli Studi di Milano, I-20054 Segrate, Italy.}}
\footnotetext{\textit{$^{b}$~Molecular, Macromolecular Chemistry and Materials, ESPCI Paris, CNRS, PSL Research University, 75005 Paris, France.}}
\footnotetext{\textit{$^{c}$~Department of Material Science and Technology, University of Crete, Heraklion 71300, Crete, Greece.}}
\footnotetext{\textit{$^{d}$~Foundation for Research and Technology Hellas, Institute of Electronic Structure and Laser, Heraklion 71110, Crete, Greece.}}
\footnotetext{\textit{$^{e}$~Laboratoire Navier, Ecole des Ponts, Univ.~Gustave Eiffel, CNRS, 77420 Champs-sur-Marne, France.}}
\footnotetext{\textit{$^{f}$~Department of Physics, University of Fribourg, Ch.~du Mus\'ee 3, 1700 Fribourg, Switzerland.}}
\footnotetext{\textit{$^{g}$~Faculty of Physics, University of Vienna, Boltzmanngasse 5, 1090 Vienna, Austria.}}
\footnotetext{\ddag~These authors contributed equally to this work.}
\footnotetext{\S~Corresponding authors: fabio.giavazzi@unimi.it, roberto.cerbino@univie.ac.at.}

%%%%%%%%%%%%%%%%%%%%%%%%%%%%%%%%%%%%%%%%%%%%%%%%%%
\setcounter{figure}{0}
%%%%%%%%%%%%%%%%%%%%%%%%%%%%%%%%%%%%%%%%%%%%%%%%%%

\begin{figure*}[htbp]
\centering
\includegraphics[width=0.9\linewidth]{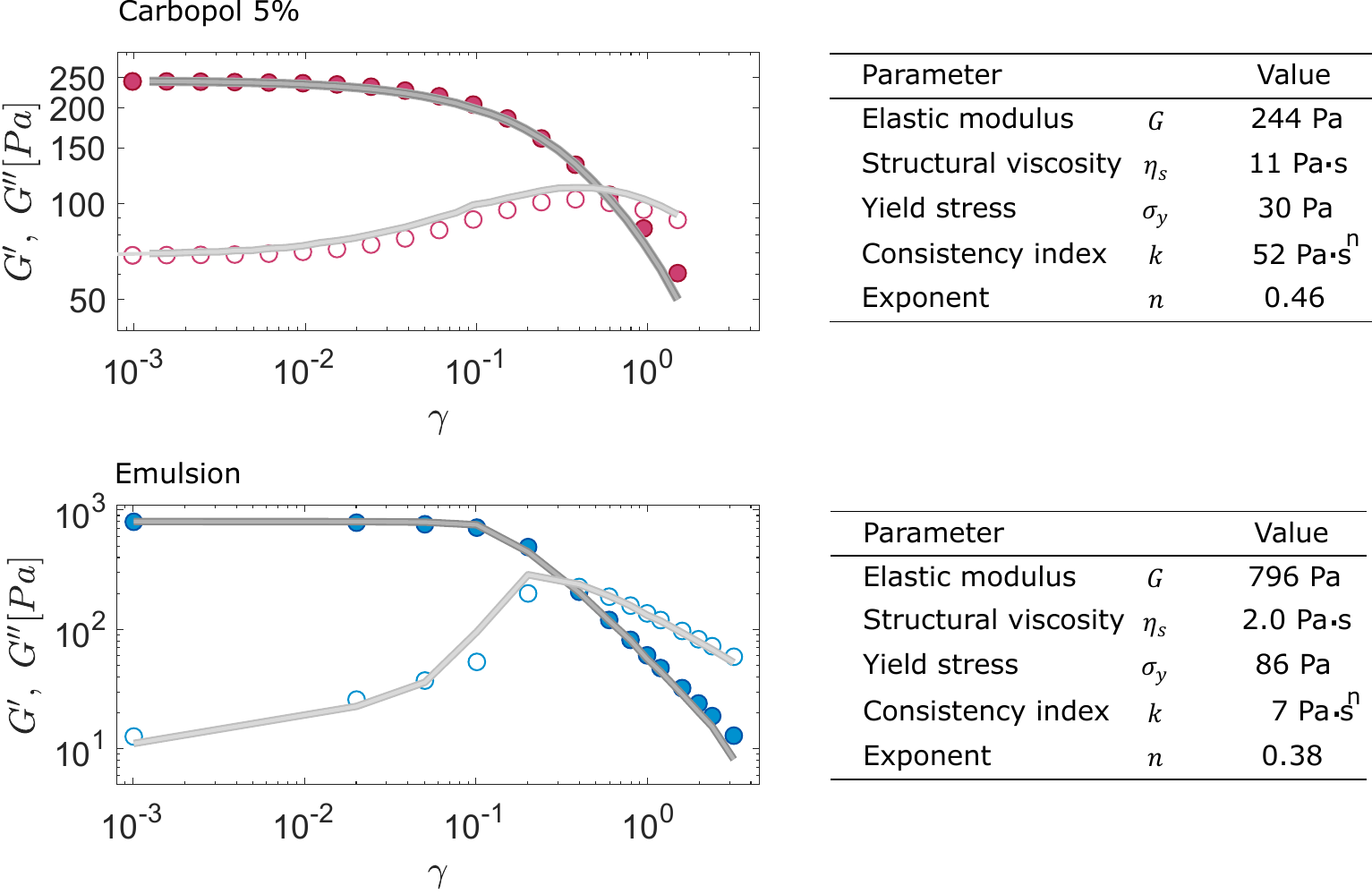}
\caption{LAOS amplitude sweeps for Carbopol 5\% (top panel) and the oil-in-water emulsion (bottom panel) at frequency $f=1Hz$ ($\omega=2\pi \ \mathrm{rad}/s)$. The measurements were performed on a commercial Anton Paar MCR~501 rheometer exploiting a cone-plate (50~mm, $1^\circ$) geometry. Solid and open circles correspond to the First-harmonic storage ($G'$), and loss ($G''$) moduli, respectively. Solid lines represent the best-fitting curve to the data using the brittility model presented in Ref.\cite{kamani2024brittle}. From the fits, we estimate a brittility factor $B_t=1$ for the Carbopol  and $B_t=6.5$ for the emulsion. The Brittility factor is the sole free parameter of the model; the other parameters listed in the table are fixed to the values obtained from independent rheological measurements. Specifically, the elastic modulus $G$ and the structural viscosity $\eta_s$ are estimated from the linear regime amplitude sweep, while the yield stress $\sigma_y$, the consistency index $k$ and the exponent $n$ are estimated from the steady shear flow curve (Herschel-Bulkley model).}
\label{fig:S1}
\end{figure*}

\begin{table*}%[h!]
    \centering
     
    \renewcommand{\arraystretch}{1.3}
 \setlength{\arrayrulewidth}{0.3pt}
     \begin{tabular}{c|ccccc}
        \Xhline{1.2pt}
         & $B_t$\cite{kamani2024brittle} & $\nu$\cite{miyazaki2006nonlinear} & $N_F$\cite{donley2020elucidating} & $(\tan\delta_0)^{-1}$\cite{keane2025universal} &
         $G^{\prime\prime}_{\mathrm{peak}}/G^{\prime\prime}_{\mathrm{lin}}$\cite{keane2025universal} \\
        \Xhline{1pt}
        Carbopol 5\% & 1 & 0.44  & 0.38 & 3.3 & 1.5 \\
        %\Xhline{0.1pt}
        Emulsion & 6.5 & 0.80& 25 & 63 & 20 \\
        \Xhline{1.2pt}
    \end{tabular}
    \caption{\textbf{Estimated brittility indicators}: (i) brittility factor $B_t$ estimated following Ref.\cite{kamani2024brittle}, which can be interpreted as the ratio of some mesoscopic local modulus to the bulk modulus $B_t=G_{local}/G$, indicating a separation of the mechanical response between different length scales. (ii) The flow exponent $\nu$   for the terminal regime \cite{miyazaki2006nonlinear} is derived from a global fit of the terminal regime of the moduli as a function of the strain, $G^{\prime \prime}=|\gamma|^{-\nu}$ and $G^{\prime}=|\gamma|^{-2\nu}$. This experiment describes the shear-thinning behaviour of the sample at high strain, mirroring the sample sensitivity to the strain amplitude. (iii) The viscoplastic fragility $N_F$ is a phenomenological quantitative indicator of the abruptness of the transition \cite{donley2020elucidating}. For simplicity we adopted the following definition: $N_F=\text{max} \left[ \frac{1}{G''_{lin}}\cdot \frac{\partial G''}{\partial{\log\gamma_0}} \right]$. Here $G''_{lin}$ is the loss modulus in the linear regime. $N_F$ quantifies the rapid increase in macroscopic unrecoverable deformation before the $G''$ peak. (iv)  Inverse of $\tan \delta_0=G_{lin}^{\prime \prime}/G_{lin}^{\prime}$ computed in the linear viscoelastic regime, and (v) overshoot $G^{\prime\prime}_{\mathrm{peak}}$ normalised by the loss modulus in the linear regime $G^{\prime\prime}_{\mathrm{lin}}$. These two quantities correlate as reported in Ref.\cite{keane2025universal}.}
\end{table*}

\begin{figure*}[htbp]
\centering
\includegraphics[width=1\linewidth]{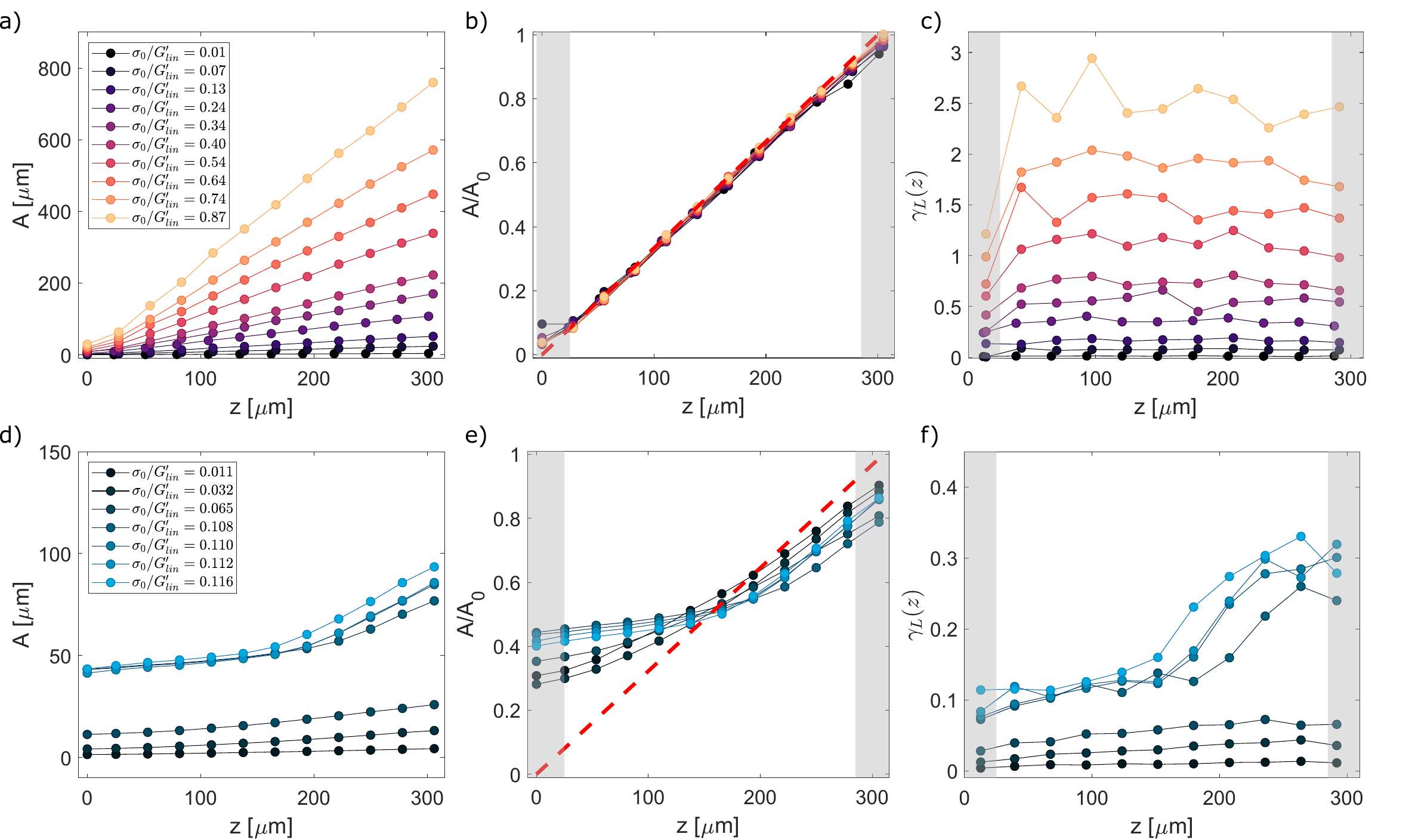}
\caption{\textbf{Deformation profiles and local strain.} Panels (a-c) refers to Carbopol $5\%$ and panels (d-f) to emulsion.
(a) Deformation profiles $A(z)$ for different imposed stress amplitudes increasing from black to yellow. (b) Rescaled amplitudes $A(z)$ with the top plate displacement (bulk) $A_0$ as measured from the shear-cell\cite{villa2022quantitative}. The red dashed line stands for the ideal case of a homogeneous profile without shear bands and wall-slip. (c) Effective strain amplitude $\gamma_L(z)$ in the gap estimated exploiting Eq.~(1) of the main text. (d) Representative deformation profile for emulsion. Colours from black to light blue correspond to increasing stress amplitude. (e) normalised deformation profile over the bulk deformation $A_0$. (f) Estimated strain amplitude $\gamma_L(z)$ across the gap.The width of gray regions in panels (b-c) and (e-f) roughly corresponds to the depth of field $\simeq 20$~\textmu m of the microscope.}
\label{fig:S2}
\end{figure*}

\begin{figure*}[htbp]
\centering
\includegraphics[width=1\linewidth]{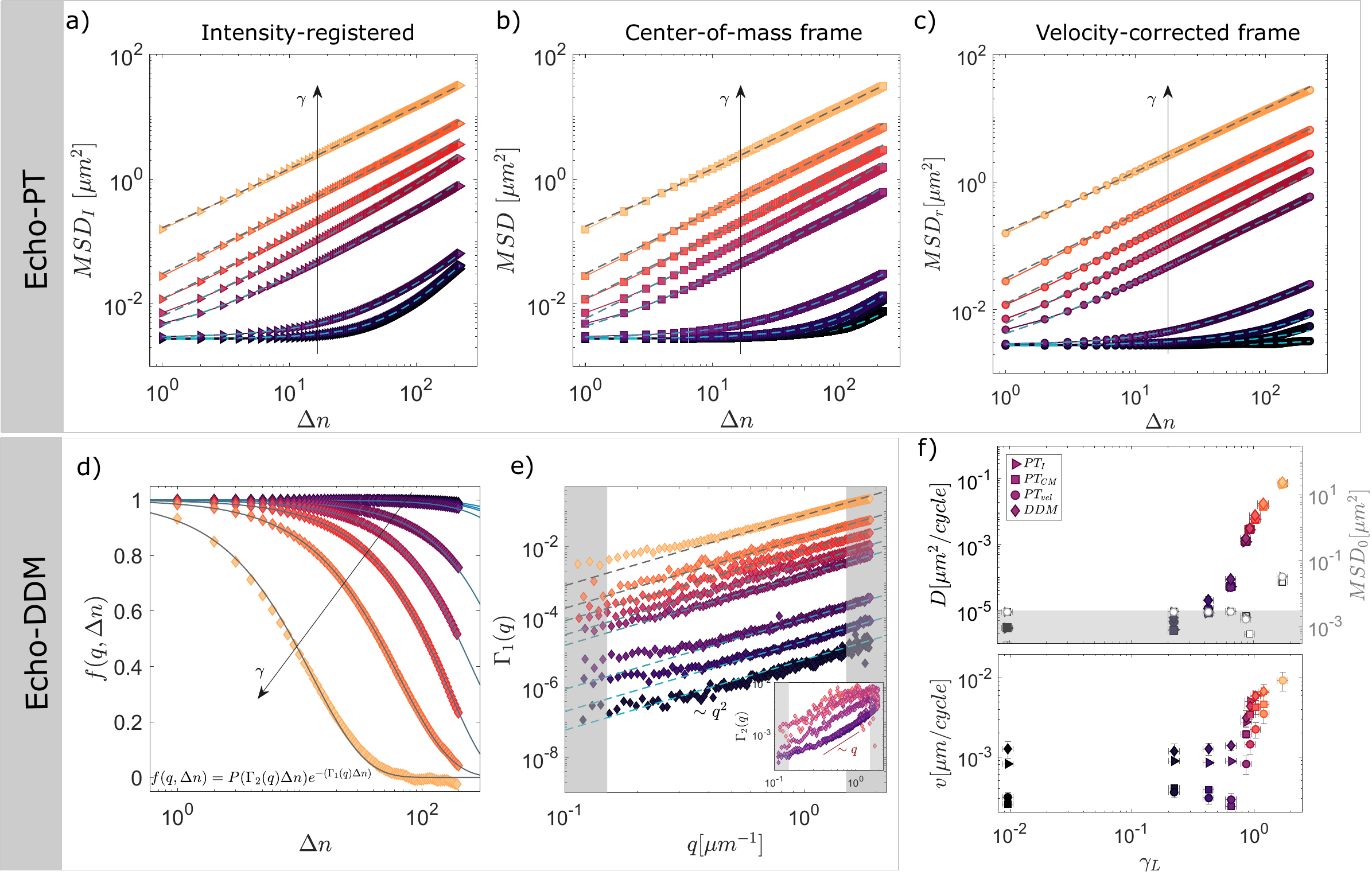}
\caption{\textbf{Echo analysis for Carbopol 5\%}: (a-c) Mean squared displacement of the tracer particles (radius $a_1=1.0$ \textmu m) as a function of the elapsed number of cycles $\Delta n$ at different strain amplitudes (increasing from dark blue to yellow, corresponding to $\gamma_L\in[0.01,1.77]$). MSDs are computed by (a) applying only the initial intensity-based image registration, (b) working in the particle centre-of-mass frame, and (c) subtracting the measured velocity field from the particle trajectories, leading to the residual MSD, $\mathrm{MSD}_r$. Dashed lines represent the best fitting curve using a superposition model $\mathrm{MSD}=\mathrm{MSD}_0
+2\,D\Delta n+ v^2 \Delta n^2 $ for panels (a and b), while $\mathrm{MSD}_r$ are fitted with $\mathrm{MSD}_r=\mathrm{MSD}_0
+2\,D\Delta n$. (d) Intermediate scattering functions $f(q,\Delta n)$ at $q=1.01$ \textmu m$^{-1}$  for different strain amplitudes (same colour code as in (a--c)). Continuous lines are the best-fitting curves using as model $f(q,\Delta n)=P(\Gamma_2(q)\Delta n)\exp(-\Gamma_1(q)\Delta n)$, where the exponential term accounts for the diffusion, while the first term $P(\Gamma_2(q)\Delta n)$ accounts for the velocity contribution. Specifically, we assumed a Flory--Schulz distribution of the velocity leading to $P(\Gamma_2(q)\Delta n)=\frac{(Z+1)}{Z\Gamma_2(q)\Delta n}\frac{\sin(Z\tan^{-1}\theta)}{(1+\theta^2)^{Z/2}}$, where $Z$ is connected to the variance trough the relation $\sigma=v(Z+1)^{-1/2}$, while $\theta=(\Gamma_2\Delta n)/(Z+1)$\cite{wilson2011differential}. The two rates $\Gamma_1(q)$ and $\Gamma_2(q)$, both expressed in inverse cycles, estimated from the fits for the different strain amplitudes are reported in panel (e) and inset, respectively. The dashed lines in (e) correspond to the best fitting curve assuming a quadratic model $\Gamma_1(q)=Dq^2$. The grey shaded region marks the $q$-range where the estimates are not reliable. (f) Comparison of the effective shear-induced diffusion coefficient (coloured symbols) and $\mathrm{MSD}_0$ (grey symbols) (top), and velocity (bottom) extracted from the different analysis procedures. No appreciable differences are observed for the shear-induced diffusion coefficient. In contrast, the velocities obtained from particle tracking without centre-of-mass subtraction and from DDM are systematically larger than those obtained from the rms of the velocity maps and from particle tracking in the centre-of-mass frame. This discrepancy originates from residual affine motion contributions that are not completely removed by the global image registration procedure.}
\label{fig:S3}
\end{figure*}

\begin{figure*}[htbp]
\centering
\includegraphics[width=0.85\linewidth]{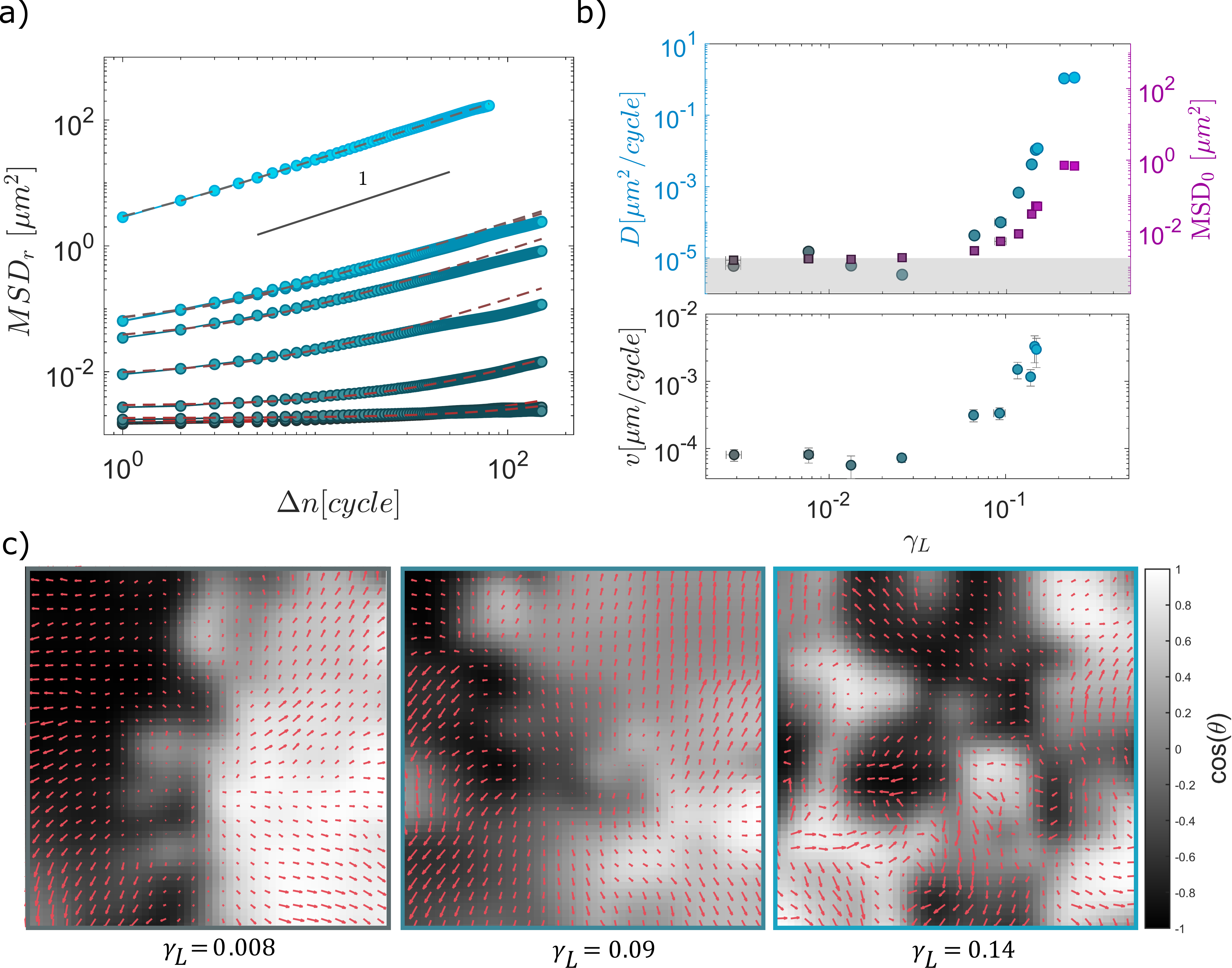}
\caption{ \textbf{Echo particle tracking for the emulsion}: (a) $\mathrm{MSD}_r$ of the tracer particles (radius $a_1=1.0$ \textmu m) in the vorticity direction, as a function of the number of elapsed cycles $\Delta n$, for local strain amplitudes $\gamma_L$ ranging from 0.003 to 0.24 (from black to light blue). Dashed lines are the best-fitting curves to the data using $\mathrm{MSD}_r(\Delta n) = \mathrm{MSD}_0 +2 D\,\Delta n$. (b) Shear-induced diffusion coefficient $D$ (top) and root-mean-square ballistic velocity $v$ (bottom) as functions of the local strain amplitude $\gamma_L$. (c) Time-averaged velocity maps of the emulsion. Velocity fields  $v_s(\mathbf{x}_m)$ (red arrows) overlaid on a colour map built on the value assumed by $\cos\theta$, where $\theta$ is the field orientation with respect the $\mathrm{x}$-axis, for three local strain amplitudes $\gamma_L = 0.008$ (left), $\gamma_L = 0.09$ (centre), $\gamma_L = 0.14$ (right). The fields reveal the same qualitative organisation as in the Carbopol (Fig.~1c of the main text). Quantitatively, the rms amplitude is $\sim 5\times$ smaller than in the Carbopol at comparable $D$.}
\label{fig:S4}
\end{figure*}

\begin{figure*}[htbp]
\centering
\includegraphics[width=0.9\linewidth]{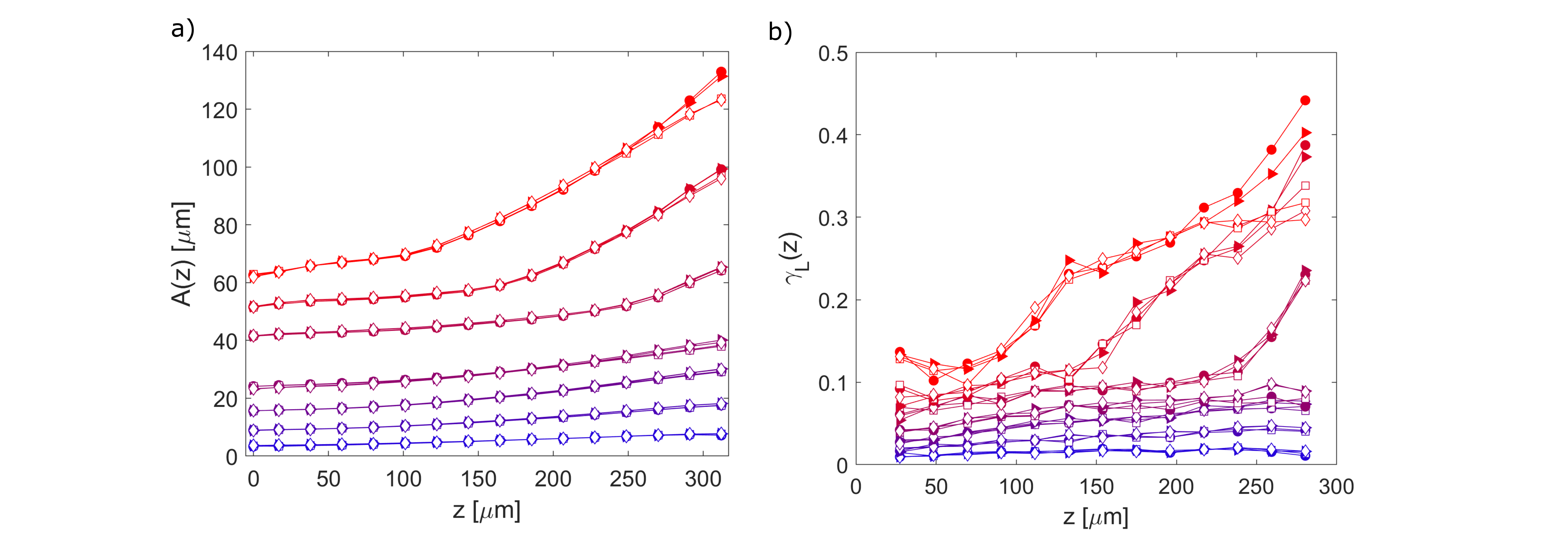}
\caption{\textbf{Repeatability of the deformation profile for the emulsion.} (a) Amplitude $A(z)$ of the deformation profile and (b) local strain amplitude $\gamma_L(z)$ across the gap, for a range of imposed stress amplitudes increasing from blue to red. At each amplitude, four scans are overlaid: $\medbullet$ ($\Box$) denote the descending (ascending) $z$-scan performed at each step of the 1\textsuperscript{st} amplitude sweep, whereas $\blacktriangleright$ ($\bigtriangledown$) denote the descending (ascending)  $z$-scan performed at each step of the 2\textsuperscript{nd} amplitude sweep. Pre-conditioning of the sample is applied before series 1\textsuperscript{st} and between series 1\textsuperscript{st} and 2\textsuperscript{nd}. %
}
\label{fig:S5}
\end{figure*}

\begin{figure*}[htbp]
\centering
\includegraphics[width=0.72\linewidth]{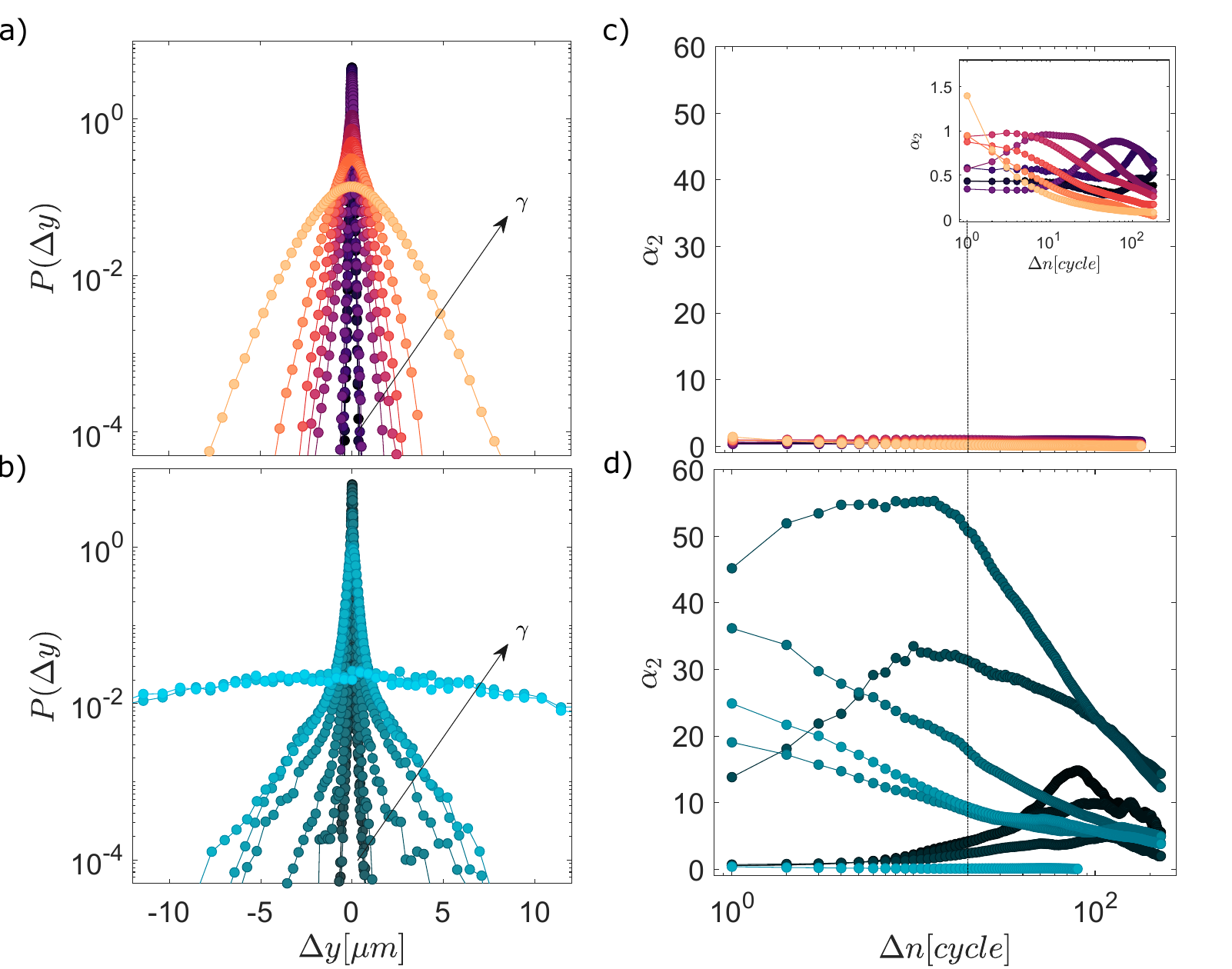}
\caption{\textbf{Gaussian \textit{vs} non-Gaussian dynamics} (a,b) Probability distribution functions (PDF) of the tracer particle displacements (radius $a_1=1.0$ \textmu m) along the vorticity direction at $\Delta n=20$  for different strain amplitudes $\gamma_L$, for Carbopol (a) and emulsion (b). The strain amplitudes correspond to those reported in the main text for Carbopol and in Fig.~\ref{fig:S4} for the emulsion, using the same colour code. (c-d) Non-Gaussian parameter $\alpha_2$ (Eq.~main text) as a function of the number of elapsed cycles $\Delta n$ for different strain amplitudes for Carbopol 5\% (c) and emulsion (d). The colour code is the same as in panels (a-b).The vertical dotted line corresponds to $\Delta n=20$, the number of elapsed cycles considered for the evaluation of the PDFs in panels (a) and(b).}
\label{fig:S6}
\end{figure*}

\begin{figure*}[htbp]
\centering
\includegraphics[width=0.9\linewidth]{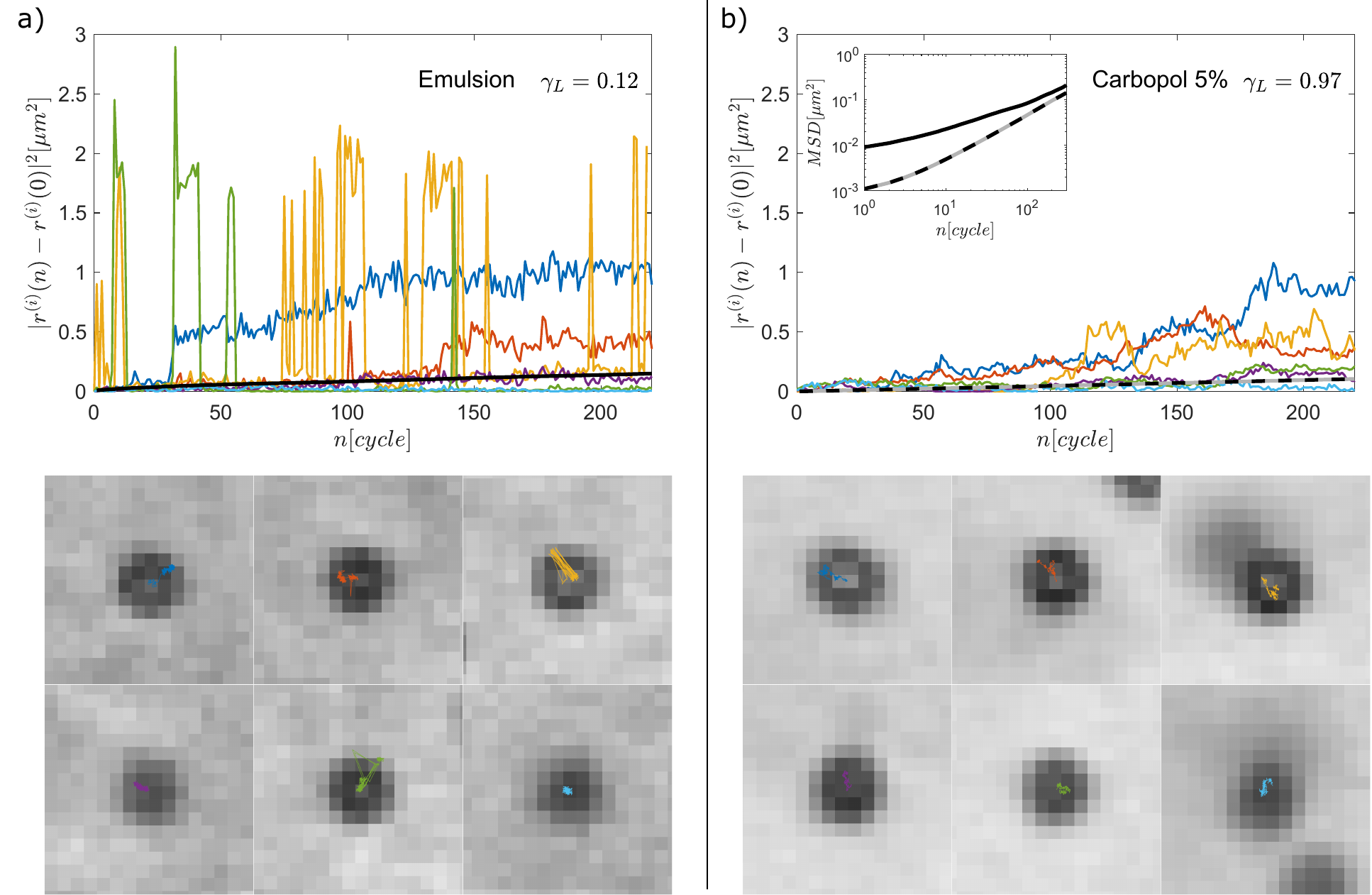}
\caption{\textbf{Short-time plateau in the mean square displacement.} a-b) Top panel: time-resolved square displacement of representative particle trajectories during oscillatory shear for a given local strain amplitude $\gamma_L = 0.12$ and $0.97$ for Emulsion and Carbopol 5$\%$, respectively. These two $\gamma_L$  values have been selected for their similar effective diffusion coefficient. The solid black line in (a) and the dashed line in (b) correspond to the mean square displacement calculated over all the particle trajectories. In the inset, the two $\mathrm{MSD}$s are compared. Bottom panel: images of the particles shown in the respective upper panel, the coloured lines mark the particle trajectories with the same colours as in top panel. Particle radius is $a_1=1.0$ \textmu m.}
\label{fig:S7}
\end{figure*}

\begin{figure*}[htbp]
\centering
\includegraphics[width=0.4\linewidth]{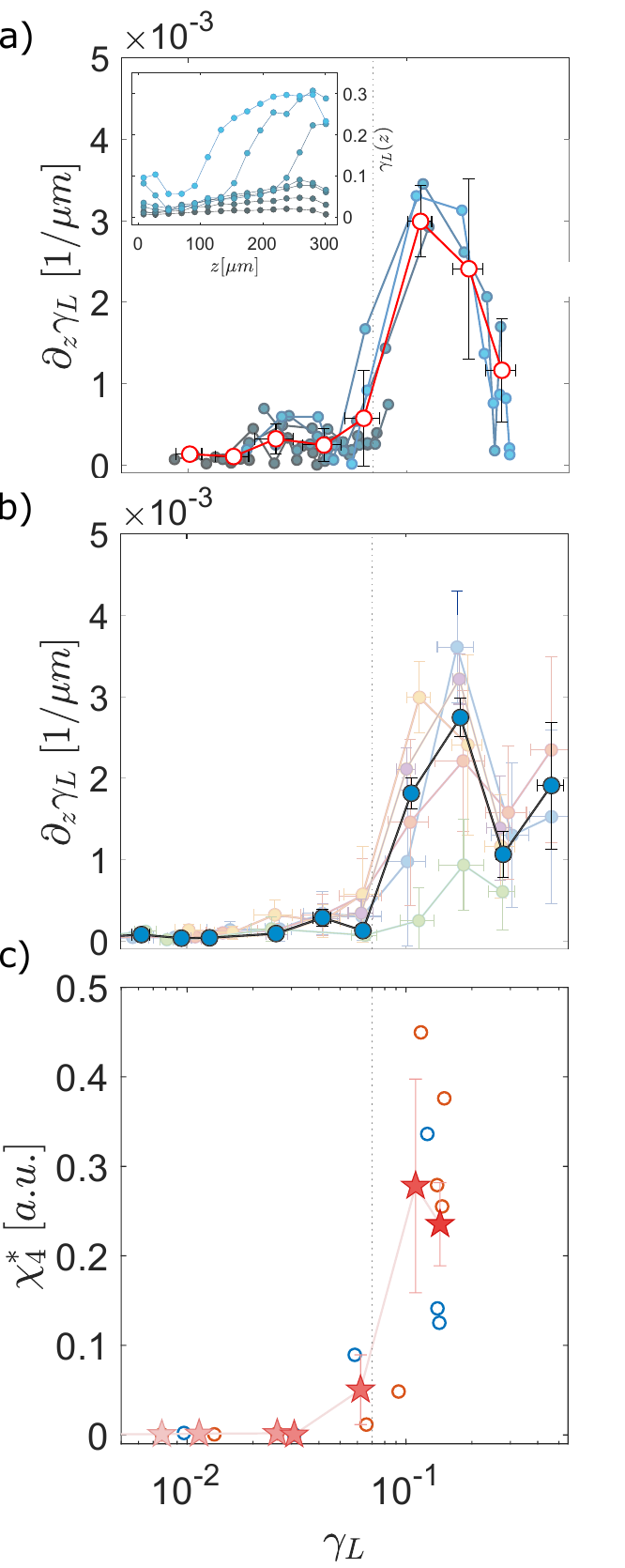}
\caption{\textbf{Correlation between shear-banding and dynamic heterogeneity} (a) Gradient of the local strain amplitude, $\partial_z\gamma_L(z)$, plotted against the local strain $\gamma_L$. Solid coloured symbols of the same colour represent the estimate obtained from a single deformation profile acquired at a given stress amplitude $\sigma_0$; each $\sigma_0$ therefore generates a set of points defining the local curve $\partial_z\gamma_L$. For an ideal, linear profile $\gamma_L$ is constant and $\partial_z\gamma_L$ vanishes. Colours ranging from dark blue to light blue indicate increasing stress amplitudes. Red open circles represent binned averages obtained from the solid dark symbols extracted from a single experimental run. Inset: local strain amplitude $\gamma_L$ obtained from the deformation profile, from which the gradient is calculated, as a function of the vertical position $z$. (b) Binned weighted average of $\partial_z\gamma_L$ as a function of $\gamma_L$, obtained by combining data from five independent experimental runs. Small coloured circles correspond to individual measurements, while blue solid circles denote the resulting weighted averages. (c) Peak values of the four-point dynamic susceptibility $\chi_4$. Red stars (identical to those shown in Fig.~4c of the main text) represent the average over two independent echo experiments performed at a fixed observation plane while varying the applied stress amplitude. Red and blue circles show the results from the two individual experimental runs used to compute the average.}
\label{fig:S8}
\end{figure*}

%\section{DDM: superposition vs coexistence}
%\label{sec:si_ddm}

%The centre-of-mass-subtracted particle-tracking dynamics imply that the two observed relaxation processes are superposed on the same particles, i.e. the intermediate scattering function should be a product of a ballistic and a diffusive decay, rather than a population-weighted sum of two decays associated with two coexisting populations of particles~\cite{aime2023unified}. DDM~\cite{cerbino2008differential,giavazzi2009scattering,aime2019probingII,edera2021deformation,richards2021characterising} provides a reciprocal-space, tracking-free test of this statement on the very same image sequences.

\begin{figure*}[htbp]
\centering
\includegraphics[width=1\linewidth]{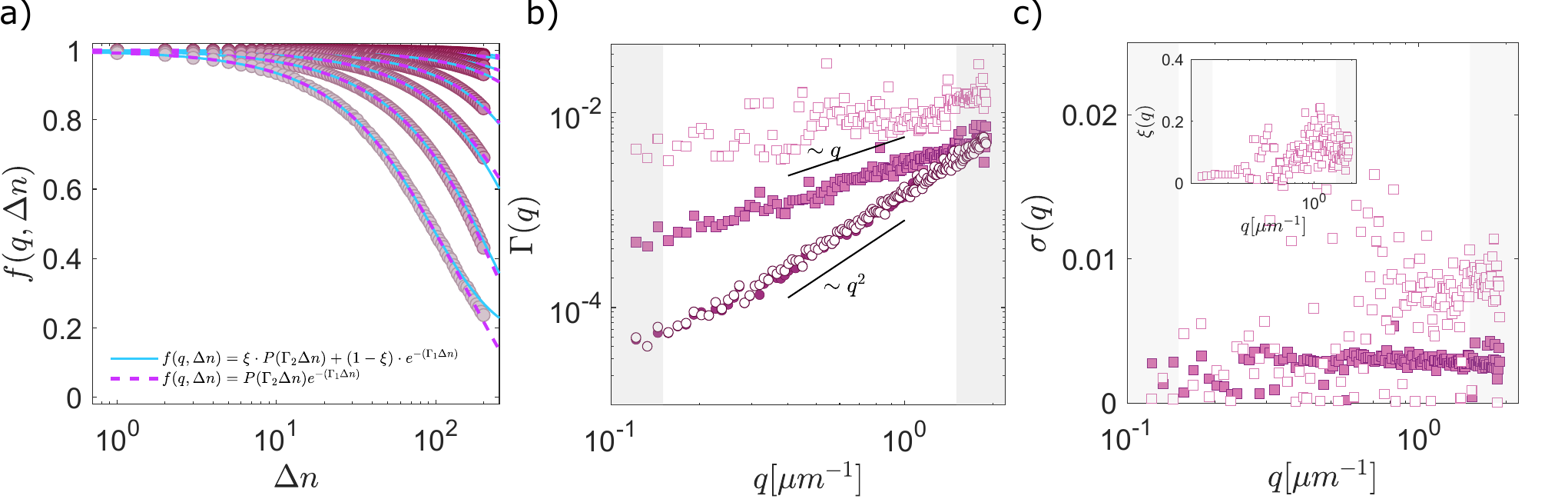}
\caption{\textbf{Product vs sum model of the DDM intermediate scattering function.} (a) Intermediate scattering function $f(q, \Delta n)$ of the Carbopol~5~\% (seeded with tracer particles of radius $a_1=1.0$ \textmu m) at $\gamma_L = 0.87$, for some representative wavevectors $q$ in the vorticity direction in the range $q\in[0.12-2.0]$\textmu m$^{-1}$. Data (symbols) are fitted by two competing models. Magenta dashed lines represent the best fitting curves assuming a product model $f(q,\Delta n)=P(\Gamma_2(q)\Delta n)\exp(-\Gamma_1(q)\Delta n)$, which describes a superposition of a Gaussian diffusive displacement and a ballistic contribution $P(\Gamma_2(q)\Delta n)$, which corresponds to a closed-form analytical expression obtained assuming a Flory--Schulz distribution of particle velocities. Continuous cyan lines correspond to the fitting curve assuming a sum (coexistence) model of the form: $f(q,\Delta n)=\xi \cdot P(\Gamma_2(q)\Delta n)+ (1-\xi)\cdot  \exp[-\Gamma_1(q)\Delta n]$, corresponding to two coexisting populations in the spirit of Aime \textit{et al.}~\cite{aime2023unified}, a sum decomposition being appropriate when distinct fractions of the sample relax at distinct rates without interconverting over the measurement. (b) Relaxation rates $\Gamma_1$ (circles) and $\Gamma_2$ (squares) estimated from the superposition model (filled symbols) and the coexistence model (open symbols). (c) Standard deviation $\sigma$ of the Flory-Schulz distribution of the particle velocities estimated from the superposition model (filled symbols) and the coexistence model (open symbols), respectively. Inset: estimated relative amplitude $\xi$ of the ballistic component in the coexistence model, in which alone it is defined.}
\label{fig:S9}
\end{figure*}

% The ESI body consists entirely of full-width (figure*/table*) floats and
% contains no ordinary body text. Starred floats can never be placed on the
% page currently being built, so without the \clearpage below LaTeX fills
% page 1 with the only non-float material available -- the bibliography --
% and defers every figure to the pages after it. \clearpage flushes all
% pending floats before the references are typeset.
\clearpage

\balance

%%%REFERENCES%%%
\clearpage
\balance

% Supplementary-information bibliography, with numbering restarted.
\providecommand{\noopsort}[1]{}\providecommand{\singleletter}[1]{#1}%
\providecommand*{\mcitethebibliography}{\thebibliography}
\csname @ifundefined\endcsname{endmcitethebibliography}
{\let\endmcitethebibliography\endthebibliography}{}


\begin{mcitethebibliography}{104}
\providecommand*{\natexlab}[1]{#1}
\providecommand*{\mciteSetBstSublistMode}[1]{}
\providecommand*{\mciteSetBstMaxWidthForm}[2]{}
\providecommand*{\mciteBstWouldAddEndPuncttrue}
  {\def\EndOfBibitem{\unskip.}}
\providecommand*{\mciteBstWouldAddEndPunctfalse}
  {\let\EndOfBibitem\relax}
\providecommand*{\mciteSetBstMidEndSepPunct}[3]{}
\providecommand*{\mciteSetBstSublistLabelBeginEnd}[3]{}
\providecommand*{\EndOfBibitem}{}
\mciteSetBstSublistMode{f}
\mciteSetBstMaxWidthForm{subitem}
{(\emph{\alph{mcitesubitemcount}})}
\mciteSetBstSublistLabelBeginEnd{\mcitemaxwidthsubitemform\space}
{\relax}{\relax}

\bibitem[Bonn \emph{et~al.}(2017)Bonn, Denn, Berthier, Divoux, and
  Manneville]{bonn2017yield}
D.~Bonn, M.~M. Denn, L.~Berthier, T.~Divoux and S.~Manneville, \emph{Rev. Mod.
  Phys.}, 2017, \textbf{89}, 035005\relax
\mciteBstWouldAddEndPuncttrue
\mciteSetBstMidEndSepPunct{\mcitedefaultmidpunct}
{\mcitedefaultendpunct}{\mcitedefaultseppunct}\relax
\EndOfBibitem
\bibitem[Nicolas \emph{et~al.}(2018)Nicolas, Ferrero, Martens, and
  Barrat]{nicolas2018deformation}
A.~Nicolas, E.~E. Ferrero, K.~Martens and J.-L. Barrat, \emph{Rev. Mod. Phys.},
  2018, \textbf{90}, 045006\relax
\mciteBstWouldAddEndPuncttrue
\mciteSetBstMidEndSepPunct{\mcitedefaultmidpunct}
{\mcitedefaultendpunct}{\mcitedefaultseppunct}\relax
\EndOfBibitem
\bibitem[Joshi and Petekidis(2018)]{Joshi2018YieldSF}
Y.~M. Joshi and G.~Petekidis, \emph{Rheologica Acta}, 2018, \textbf{57},
  521--549\relax
\mciteBstWouldAddEndPuncttrue
\mciteSetBstMidEndSepPunct{\mcitedefaultmidpunct}
{\mcitedefaultendpunct}{\mcitedefaultseppunct}\relax
\EndOfBibitem
\bibitem[Coussot \emph{et~al.}(2002)Coussot, Nguyen, Huynh, and
  Bonn]{coussot2002viscosity}
P.~Coussot, Q.~D. Nguyen, H.~T. Huynh and D.~Bonn, \emph{J. Rheol.}, 2002,
  \textbf{46}, 573--589\relax
\mciteBstWouldAddEndPuncttrue
\mciteSetBstMidEndSepPunct{\mcitedefaultmidpunct}
{\mcitedefaultendpunct}{\mcitedefaultseppunct}\relax
\EndOfBibitem
\bibitem[Dahbi \emph{et~al.}(2010)Dahbi, Alexander, Trappe, Dhont, and
  Schurtenberger]{dahbi2010rheology}
L.~Dahbi, M.~Alexander, V.~Trappe, J.~K.~G. Dhont and P.~Schurtenberger,
  \emph{J. Colloid Interface Sci.}, 2010, \textbf{342}, 564--570\relax
\mciteBstWouldAddEndPuncttrue
\mciteSetBstMidEndSepPunct{\mcitedefaultmidpunct}
{\mcitedefaultendpunct}{\mcitedefaultseppunct}\relax
\EndOfBibitem
\bibitem[Cerbino and Trappe(2023)]{cerbino2023introduction}
R.~Cerbino and V.~Trappe, \emph{Physica A}, 2023, \textbf{631}, 128653\relax
\mciteBstWouldAddEndPuncttrue
\mciteSetBstMidEndSepPunct{\mcitedefaultmidpunct}
{\mcitedefaultendpunct}{\mcitedefaultseppunct}\relax
\EndOfBibitem
\bibitem[Berthier \emph{et~al.}(2025)Berthier, Biroli, Manning, and
  Zamponi]{berthier2025yielding}
L.~Berthier, G.~Biroli, L.~Manning and F.~Zamponi, \emph{Nat. Rev. Phys.},
  2025, \textbf{7}, 313--330\relax
\mciteBstWouldAddEndPuncttrue
\mciteSetBstMidEndSepPunct{\mcitedefaultmidpunct}
{\mcitedefaultendpunct}{\mcitedefaultseppunct}\relax
\EndOfBibitem
\bibitem[Nelson and Ewoldt(2017)]{nelson2017design}
A.~Z. Nelson and R.~H. Ewoldt, \emph{Soft Matter}, 2017, \textbf{13},
  7578--7594\relax
\mciteBstWouldAddEndPuncttrue
\mciteSetBstMidEndSepPunct{\mcitedefaultmidpunct}
{\mcitedefaultendpunct}{\mcitedefaultseppunct}\relax
\EndOfBibitem
\bibitem[Divoux \emph{et~al.}(2024)Divoux, Agoritsas, Aime, Barentin, Barrat,
  Benzi, Berthier, Bi, Biroli, Bonn,\emph{et~al.}]{divoux2024ductile}
T.~Divoux, E.~Agoritsas, S.~Aime, C.~Barentin, J.-L. Barrat, R.~Benzi,
  L.~Berthier, D.~Bi, G.~Biroli, D.~Bonn \emph{et~al.}, \emph{Soft Matter},
  2024, \textbf{20}, 6868--6888\relax
\mciteBstWouldAddEndPuncttrue
\mciteSetBstMidEndSepPunct{\mcitedefaultmidpunct}
{\mcitedefaultendpunct}{\mcitedefaultseppunct}\relax
\EndOfBibitem
\bibitem[Ozawa \emph{et~al.}(2018)Ozawa, Berthier, Biroli, Rosso, and
  Tarjus]{ozawa2018random}
M.~Ozawa, L.~Berthier, G.~Biroli, A.~Rosso and G.~Tarjus, \emph{Proc. Natl.
  Acad. Sci. U.S.A.}, 2018, \textbf{115}, 6656--6661\relax
\mciteBstWouldAddEndPuncttrue
\mciteSetBstMidEndSepPunct{\mcitedefaultmidpunct}
{\mcitedefaultendpunct}{\mcitedefaultseppunct}\relax
\EndOfBibitem
\bibitem[Hyun \emph{et~al.}(2011)Hyun, Wilhelm, Klein, Cho, Nam, Ahn, Lee,
  Ewoldt, and McKinley]{hyun2011review}
K.~Hyun, M.~Wilhelm, C.~O. Klein, K.~S. Cho, J.~G. Nam, K.~H. Ahn, S.~J. Lee,
  R.~H. Ewoldt and G.~H. McKinley, \emph{Prog. Polym. Sci.}, 2011, \textbf{36},
  1697--1753\relax
\mciteBstWouldAddEndPuncttrue
\mciteSetBstMidEndSepPunct{\mcitedefaultmidpunct}
{\mcitedefaultendpunct}{\mcitedefaultseppunct}\relax
\EndOfBibitem
\bibitem[Miyazaki \emph{et~al.}(2006)Miyazaki, Wyss, Weitz, and
  Reichman]{miyazaki2006nonlinear}
K.~Miyazaki, H.~M. Wyss, D.~A. Weitz and D.~R. Reichman, \emph{Europhys.
  Lett.}, 2006, \textbf{75}, 915--921\relax
\mciteBstWouldAddEndPuncttrue
\mciteSetBstMidEndSepPunct{\mcitedefaultmidpunct}
{\mcitedefaultendpunct}{\mcitedefaultseppunct}\relax
\EndOfBibitem
\bibitem[Koumakis \emph{et~al.}(2012)Koumakis, Pamvouxoglou, Poulos, and
  Petekidis]{koumakis2012direct}
N.~Koumakis, A.~Pamvouxoglou, A.~S. Poulos and G.~Petekidis, \emph{Soft
  Matter}, 2012, \textbf{8}, 4271--4284\relax
\mciteBstWouldAddEndPuncttrue
\mciteSetBstMidEndSepPunct{\mcitedefaultmidpunct}
{\mcitedefaultendpunct}{\mcitedefaultseppunct}\relax
\EndOfBibitem
\bibitem[Donley \emph{et~al.}(2020)Donley, Singh, Shetty, and
  Rogers]{donley2020elucidating}
G.~J. Donley, P.~K. Singh, A.~Shetty and S.~A. Rogers, \emph{Proc. Natl. Acad.
  Sci. U.S.A.}, 2020, \textbf{117}, 21945--21952\relax
\mciteBstWouldAddEndPuncttrue
\mciteSetBstMidEndSepPunct{\mcitedefaultmidpunct}
{\mcitedefaultendpunct}{\mcitedefaultseppunct}\relax
\EndOfBibitem
\bibitem[Rogers \emph{et~al.}(2011)Rogers, Erwin, Vlassopoulos, and
  Cloitre]{rogers2011sequence}
S.~A. Rogers, B.~M. Erwin, D.~Vlassopoulos and M.~Cloitre, \emph{J. Rheol.},
  2011, \textbf{55}, 435--458\relax
\mciteBstWouldAddEndPuncttrue
\mciteSetBstMidEndSepPunct{\mcitedefaultmidpunct}
{\mcitedefaultendpunct}{\mcitedefaultseppunct}\relax
\EndOfBibitem
\bibitem[Rogers(2017)]{rogers2017search}
S.~A. Rogers, \emph{Rheol. Acta}, 2017, \textbf{56}, 501--525\relax
\mciteBstWouldAddEndPuncttrue
\mciteSetBstMidEndSepPunct{\mcitedefaultmidpunct}
{\mcitedefaultendpunct}{\mcitedefaultseppunct}\relax
\EndOfBibitem
\bibitem[Kamani and Rogers(2024)]{kamani2024brittle}
K.~M. Kamani and S.~A. Rogers, \emph{Proc. Natl. Acad. Sci. U.S.A.}, 2024,
  \textbf{121}, e2401409121\relax
\mciteBstWouldAddEndPuncttrue
\mciteSetBstMidEndSepPunct{\mcitedefaultmidpunct}
{\mcitedefaultendpunct}{\mcitedefaultseppunct}\relax
\EndOfBibitem
\bibitem[Barlow \emph{et~al.}(2020)Barlow, Cochran, and
  Fielding]{barlow2020ductile}
H.~J. Barlow, J.~O. Cochran and S.~M. Fielding, \emph{Phys. Rev. Lett.}, 2020,
  \textbf{125}, 168003\relax
\mciteBstWouldAddEndPuncttrue
\mciteSetBstMidEndSepPunct{\mcitedefaultmidpunct}
{\mcitedefaultendpunct}{\mcitedefaultseppunct}\relax
\EndOfBibitem
\bibitem[Vasisht \emph{et~al.}(2020)Vasisht, Roberts, and
  Del~Gado]{vasisht2020emergence}
V.~V. Vasisht, G.~Roberts and E.~Del~Gado, \emph{Phys. Rev. E}, 2020,
  \textbf{102}, 010604\relax
\mciteBstWouldAddEndPuncttrue
\mciteSetBstMidEndSepPunct{\mcitedefaultmidpunct}
{\mcitedefaultendpunct}{\mcitedefaultseppunct}\relax
\EndOfBibitem
\bibitem[Ozawa \emph{et~al.}(2022)Ozawa, Berthier, Biroli, and
  Tarjus]{ozawa2022rare}
M.~Ozawa, L.~Berthier, G.~Biroli and G.~Tarjus, \emph{Phys. Rev. Res.}, 2022,
  \textbf{4}, 023227\relax
\mciteBstWouldAddEndPuncttrue
\mciteSetBstMidEndSepPunct{\mcitedefaultmidpunct}
{\mcitedefaultendpunct}{\mcitedefaultseppunct}\relax
\EndOfBibitem
\bibitem[Benzi \emph{et~al.}(2023)Benzi, Divoux, Barentin, Manneville,
  Sbragaglia, and Toschi]{benzi2023continuum}
R.~Benzi, T.~Divoux, C.~Barentin, S.~Manneville, M.~Sbragaglia and F.~Toschi,
  \emph{Europhys. Lett.}, 2023, \textbf{141}, 56001\relax
\mciteBstWouldAddEndPuncttrue
\mciteSetBstMidEndSepPunct{\mcitedefaultmidpunct}
{\mcitedefaultendpunct}{\mcitedefaultseppunct}\relax
\EndOfBibitem
\bibitem[Leocmach \emph{et~al.}(2014)Leocmach, Perge, Divoux, and
  Manneville]{leocmach2014creep}
M.~Leocmach, C.~Perge, T.~Divoux and S.~Manneville, \emph{Phys. Rev. Lett.},
  2014, \textbf{113}, 038303\relax
\mciteBstWouldAddEndPuncttrue
\mciteSetBstMidEndSepPunct{\mcitedefaultmidpunct}
{\mcitedefaultendpunct}{\mcitedefaultseppunct}\relax
\EndOfBibitem
\bibitem[Cochran \emph{et~al.}(2024)Cochran, Callaghan, Caven, and
  Fielding]{cochran2024slow}
J.~O. Cochran, G.~L. Callaghan, M.~J.~G. Caven and S.~M. Fielding, \emph{Phys.
  Rev. Lett.}, 2024, \textbf{132}, 168202\relax
\mciteBstWouldAddEndPuncttrue
\mciteSetBstMidEndSepPunct{\mcitedefaultmidpunct}
{\mcitedefaultendpunct}{\mcitedefaultseppunct}\relax
\EndOfBibitem
\bibitem[Keane \emph{et~al.}(2025)Keane, Nikoumanesh, Kamani, Rogers, and
  Poling-Skutvik]{keane2025universal}
D.~P. Keane, E.~Nikoumanesh, K.~M. Kamani, S.~A. Rogers and R.~Poling-Skutvik,
  \emph{Phys. Rev. Lett.}, 2025, \textbf{134}, 208202\relax
\mciteBstWouldAddEndPuncttrue
\mciteSetBstMidEndSepPunct{\mcitedefaultmidpunct}
{\mcitedefaultendpunct}{\mcitedefaultseppunct}\relax
\EndOfBibitem
\bibitem[Denisov \emph{et~al.}(2015)Denisov, Dang, Struth, Zaccone, Wegdam, and
  Schall]{denisov2015sharp}
D.~V. Denisov, M.~T. Dang, B.~Struth, A.~Zaccone, G.~H. Wegdam and P.~Schall,
  \emph{Sci. Rep.}, 2015, \textbf{5}, 14359\relax
\mciteBstWouldAddEndPuncttrue
\mciteSetBstMidEndSepPunct{\mcitedefaultmidpunct}
{\mcitedefaultendpunct}{\mcitedefaultseppunct}\relax
\EndOfBibitem
\bibitem[Parmar \emph{et~al.}(2019)Parmar, Kumar, and Sastry]{parmar2019strain}
A.~D.~S. Parmar, S.~Kumar and S.~Sastry, \emph{Phys. Rev. X}, 2019, \textbf{9},
  021018\relax
\mciteBstWouldAddEndPuncttrue
\mciteSetBstMidEndSepPunct{\mcitedefaultmidpunct}
{\mcitedefaultendpunct}{\mcitedefaultseppunct}\relax
\EndOfBibitem
\bibitem[Kawasaki and Berthier(2016)]{kawasaki2016macroscopic}
T.~Kawasaki and L.~Berthier, \emph{Phys. Rev. E}, 2016, \textbf{94},
  022615\relax
\mciteBstWouldAddEndPuncttrue
\mciteSetBstMidEndSepPunct{\mcitedefaultmidpunct}
{\mcitedefaultendpunct}{\mcitedefaultseppunct}\relax
\EndOfBibitem
\bibitem[Leishangthem \emph{et~al.}(2017)Leishangthem, Parmar, and
  Sastry]{leishangthem2017yielding}
P.~Leishangthem, A.~D.~S. Parmar and S.~Sastry, \emph{Nat. Commun.}, 2017,
  \textbf{8}, 14653\relax
\mciteBstWouldAddEndPuncttrue
\mciteSetBstMidEndSepPunct{\mcitedefaultmidpunct}
{\mcitedefaultendpunct}{\mcitedefaultseppunct}\relax
\EndOfBibitem
\bibitem[Parley \emph{et~al.}(2022)Parley, Sastry, and Sollich]{parley2022mean}
J.~T. Parley, S.~Sastry and P.~Sollich, \emph{Phys. Rev. Lett.}, 2022,
  \textbf{128}, 198001\relax
\mciteBstWouldAddEndPuncttrue
\mciteSetBstMidEndSepPunct{\mcitedefaultmidpunct}
{\mcitedefaultendpunct}{\mcitedefaultseppunct}\relax
\EndOfBibitem
\bibitem[Bhaumik \emph{et~al.}(2021)Bhaumik, Foffi, and
  Sastry]{bhaumik2021role}
H.~Bhaumik, G.~Foffi and S.~Sastry, \emph{Proc. Natl. Acad. Sci. U.S.A.}, 2021,
  \textbf{118}, e2100227118\relax
\mciteBstWouldAddEndPuncttrue
\mciteSetBstMidEndSepPunct{\mcitedefaultmidpunct}
{\mcitedefaultendpunct}{\mcitedefaultseppunct}\relax
\EndOfBibitem
\bibitem[Priezjev(2018)]{priezjev2018yielding}
N.~V. Priezjev, \emph{Comput. Mater. Sci.}, 2018, \textbf{150}, 162--168\relax
\mciteBstWouldAddEndPuncttrue
\mciteSetBstMidEndSepPunct{\mcitedefaultmidpunct}
{\mcitedefaultendpunct}{\mcitedefaultseppunct}\relax
\EndOfBibitem
\bibitem[Yeh \emph{et~al.}(2020)Yeh, Ozawa, Miyazaki, Kawasaki, and
  Berthier]{yeh2020glass}
W.-T. Yeh, M.~Ozawa, K.~Miyazaki, T.~Kawasaki and L.~Berthier, \emph{Phys. Rev.
  Lett.}, 2020, \textbf{124}, 225502\relax
\mciteBstWouldAddEndPuncttrue
\mciteSetBstMidEndSepPunct{\mcitedefaultmidpunct}
{\mcitedefaultendpunct}{\mcitedefaultseppunct}\relax
\EndOfBibitem
\bibitem[Pollard and Fielding(2022)]{pollard2022yielding}
J.~Pollard and S.~M. Fielding, \emph{Phys. Rev. Res.}, 2022, \textbf{4},
  043037\relax
\mciteBstWouldAddEndPuncttrue
\mciteSetBstMidEndSepPunct{\mcitedefaultmidpunct}
{\mcitedefaultendpunct}{\mcitedefaultseppunct}\relax
\EndOfBibitem
\bibitem[Keim \emph{et~al.}(2019)Keim, Paulsen, Zeravcic, Sastry, and
  Nagel]{paulsen2019memory}
N.~C. Keim, J.~D. Paulsen, Z.~Zeravcic, S.~Sastry and S.~R. Nagel, \emph{Rev.
  Mod. Phys.}, 2019, \textbf{91}, 035002\relax
\mciteBstWouldAddEndPuncttrue
\mciteSetBstMidEndSepPunct{\mcitedefaultmidpunct}
{\mcitedefaultendpunct}{\mcitedefaultseppunct}\relax
\EndOfBibitem
\bibitem[Paulsen and Keim(2025)]{keim2024mechanical}
J.~D. Paulsen and N.~C. Keim, \emph{Annu. Rev. Condens. Matter Phys.}, 2025,
  \textbf{16}, 61--81\relax
\mciteBstWouldAddEndPuncttrue
\mciteSetBstMidEndSepPunct{\mcitedefaultmidpunct}
{\mcitedefaultendpunct}{\mcitedefaultseppunct}\relax
\EndOfBibitem
\bibitem[H{\'e}braud \emph{et~al.}(1997)H{\'e}braud, Lequeux, Munch, and
  Pine]{hebraud1997yielding}
P.~H{\'e}braud, F.~Lequeux, J.~P. Munch and D.~J. Pine, \emph{Phys. Rev.
  Lett.}, 1997, \textbf{78}, 4657--4660\relax
\mciteBstWouldAddEndPuncttrue
\mciteSetBstMidEndSepPunct{\mcitedefaultmidpunct}
{\mcitedefaultendpunct}{\mcitedefaultseppunct}\relax
\EndOfBibitem
\bibitem[H{\"o}hler \emph{et~al.}(1997)H{\"o}hler, Cohen-Addad, and
  Hoballah]{hohler1997periodic}
R.~H{\"o}hler, S.~Cohen-Addad and H.~Hoballah, \emph{Phys. Rev. Lett.}, 1997,
  \textbf{79}, 1154--1157\relax
\mciteBstWouldAddEndPuncttrue
\mciteSetBstMidEndSepPunct{\mcitedefaultmidpunct}
{\mcitedefaultendpunct}{\mcitedefaultseppunct}\relax
\EndOfBibitem
\bibitem[Petekidis \emph{et~al.}(2002)Petekidis, Moussa{\"i}d, and
  Pusey]{petekidis2002rearrangements}
G.~Petekidis, A.~Moussa{\"i}d and P.~N. Pusey, \emph{Phys. Rev. E}, 2002,
  \textbf{66}, 051402\relax
\mciteBstWouldAddEndPuncttrue
\mciteSetBstMidEndSepPunct{\mcitedefaultmidpunct}
{\mcitedefaultendpunct}{\mcitedefaultseppunct}\relax
\EndOfBibitem
\bibitem[Laurati \emph{et~al.}(2014)Laurati, Egelhaaf, and
  Petekidis]{laos_dws_colloidal_gels_2014}
M.~Laurati, S.~U. Egelhaaf and G.~Petekidis, \emph{Journal of Rheology}, 2014,
  \textbf{58}, 1395--1417\relax
\mciteBstWouldAddEndPuncttrue
\mciteSetBstMidEndSepPunct{\mcitedefaultmidpunct}
{\mcitedefaultendpunct}{\mcitedefaultseppunct}\relax
\EndOfBibitem
\bibitem[Smith \emph{et~al.}(2007)Smith, Petekidis, Egelhaaf, and
  Poon]{smith2007yielding}
P.~A. Smith, G.~Petekidis, S.~U. Egelhaaf and W.~C.~K. Poon, \emph{Physical
  Review E}, 2007, \textbf{76}, 041402\relax
\mciteBstWouldAddEndPuncttrue
\mciteSetBstMidEndSepPunct{\mcitedefaultmidpunct}
{\mcitedefaultendpunct}{\mcitedefaultseppunct}\relax
\EndOfBibitem
\bibitem[Derakhshandeh \emph{et~al.}(2013)Derakhshandeh, Petekidis,
  Shafiei~Sabet, Hamad, and Hatzikiriakos]{Derakhshandeh2013Ageing}
B.~Derakhshandeh, G.~Petekidis, S.~Shafiei~Sabet, W.~Y. Hamad and S.~G.
  Hatzikiriakos, \emph{Journal of Rheology}, 2013, \textbf{57}, 131--148\relax
\mciteBstWouldAddEndPuncttrue
\mciteSetBstMidEndSepPunct{\mcitedefaultmidpunct}
{\mcitedefaultendpunct}{\mcitedefaultseppunct}\relax
\EndOfBibitem
\bibitem[Knowlton \emph{et~al.}(2014)Knowlton, Pine, and
  Cipelletti]{knowlton2014microscopic}
E.~D. Knowlton, D.~J. Pine and L.~Cipelletti, \emph{Soft Matter}, 2014,
  \textbf{10}, 6931--6940\relax
\mciteBstWouldAddEndPuncttrue
\mciteSetBstMidEndSepPunct{\mcitedefaultmidpunct}
{\mcitedefaultendpunct}{\mcitedefaultseppunct}\relax
\EndOfBibitem
\bibitem[Rogers \emph{et~al.}(2014)Rogers, Chen, Andrzejewski, Narayanan,
  Ramakrishnan, Leheny, and Harden]{rogers2014echoes}
M.~C. Rogers, K.~Chen, L.~Andrzejewski, S.~Narayanan, S.~Ramakrishnan, R.~L.
  Leheny and J.~L. Harden, \emph{Phys. Rev. E}, 2014, \textbf{90}, 062310\relax
\mciteBstWouldAddEndPuncttrue
\mciteSetBstMidEndSepPunct{\mcitedefaultmidpunct}
{\mcitedefaultendpunct}{\mcitedefaultseppunct}\relax
\EndOfBibitem
\bibitem[Rogers \emph{et~al.}(2018)Rogers, Chen, Pagenkopp, Mason, Narayanan,
  Harden, and Leheny]{rogers2018microscopic}
M.~C. Rogers, K.~Chen, M.~J. Pagenkopp, T.~G. Mason, S.~Narayanan, J.~L. Harden
  and R.~L. Leheny, \emph{Phys. Rev. Mater.}, 2018, \textbf{2}, 095601\relax
\mciteBstWouldAddEndPuncttrue
\mciteSetBstMidEndSepPunct{\mcitedefaultmidpunct}
{\mcitedefaultendpunct}{\mcitedefaultseppunct}\relax
\EndOfBibitem
\bibitem[Aime \emph{et~al.}(2018)Aime, Ramos, and
  Cipelletti]{aime2018microscopic}
S.~Aime, L.~Ramos and L.~Cipelletti, \emph{Proc. Natl. Acad. Sci. U.S.A.},
  2018, \textbf{115}, 3587--3592\relax
\mciteBstWouldAddEndPuncttrue
\mciteSetBstMidEndSepPunct{\mcitedefaultmidpunct}
{\mcitedefaultendpunct}{\mcitedefaultseppunct}\relax
\EndOfBibitem
\bibitem[Aime and Cipelletti(2019)]{aime2019probingI}
S.~Aime and L.~Cipelletti, \emph{Soft Matter}, 2019, \textbf{15},
  200--212\relax
\mciteBstWouldAddEndPuncttrue
\mciteSetBstMidEndSepPunct{\mcitedefaultmidpunct}
{\mcitedefaultendpunct}{\mcitedefaultseppunct}\relax
\EndOfBibitem
\bibitem[Larobina \emph{et~al.}(2021)Larobina, Pommella, Philippe, Nagazi, and
  Cipelletti]{larobina2021enhanced}
D.~Larobina, A.~Pommella, A.-M. Philippe, M.~Y. Nagazi and L.~Cipelletti,
  \emph{Proc. Natl. Acad. Sci. U.S.A.}, 2021, \textbf{118}, e2103995118\relax
\mciteBstWouldAddEndPuncttrue
\mciteSetBstMidEndSepPunct{\mcitedefaultmidpunct}
{\mcitedefaultendpunct}{\mcitedefaultseppunct}\relax
\EndOfBibitem
\bibitem[Aime \emph{et~al.}(2023)Aime, Truzzolillo, Pine, Ramos, and
  Cipelletti]{aime2023unified}
S.~Aime, D.~Truzzolillo, D.~J. Pine, L.~Ramos and L.~Cipelletti, \emph{Nat.
  Phys.}, 2023, \textbf{19}, 1673--1679\relax
\mciteBstWouldAddEndPuncttrue
\mciteSetBstMidEndSepPunct{\mcitedefaultmidpunct}
{\mcitedefaultendpunct}{\mcitedefaultseppunct}\relax
\EndOfBibitem
\bibitem[Aime and Cipelletti(2019)]{aime2019probingII}
S.~Aime and L.~Cipelletti, \emph{Soft Matter}, 2019, \textbf{15},
  213--226\relax
\mciteBstWouldAddEndPuncttrue
\mciteSetBstMidEndSepPunct{\mcitedefaultmidpunct}
{\mcitedefaultendpunct}{\mcitedefaultseppunct}\relax
\EndOfBibitem
\bibitem[Edera \emph{et~al.}(2021)Edera, Brizioli, Zanchetta, Petekidis,
  Giavazzi, and Cerbino]{edera2021deformation}
P.~Edera, M.~Brizioli, G.~Zanchetta, G.~Petekidis, F.~Giavazzi and R.~Cerbino,
  \emph{Soft Matter}, 2021, \textbf{17}, 8553--8566\relax
\mciteBstWouldAddEndPuncttrue
\mciteSetBstMidEndSepPunct{\mcitedefaultmidpunct}
{\mcitedefaultendpunct}{\mcitedefaultseppunct}\relax
\EndOfBibitem
\bibitem[Richards \emph{et~al.}(2021)Richards, Martinez, and
  Arlt]{richards2021characterising}
J.~A. Richards, V.~A. Martinez and J.~Arlt, \emph{Soft Matter}, 2021,
  \textbf{17}, 8838--8849\relax
\mciteBstWouldAddEndPuncttrue
\mciteSetBstMidEndSepPunct{\mcitedefaultmidpunct}
{\mcitedefaultendpunct}{\mcitedefaultseppunct}\relax
\EndOfBibitem
\bibitem[Tamborini \emph{et~al.}(2014)Tamborini, Cipelletti, and
  Ramos]{tamborini2014plasticity}
E.~Tamborini, L.~Cipelletti and L.~Ramos, \emph{Phys. Rev. Lett.}, 2014,
  \textbf{113}, 078301\relax
\mciteBstWouldAddEndPuncttrue
\mciteSetBstMidEndSepPunct{\mcitedefaultmidpunct}
{\mcitedefaultendpunct}{\mcitedefaultseppunct}\relax
\EndOfBibitem
\bibitem[Chen \emph{et~al.}(2020)Chen, Rogers, Narayanan, Harden, and
  Leheny]{chen2020microscopic}
Y.~Chen, S.~A. Rogers, S.~Narayanan, J.~L. Harden and R.~L. Leheny, \emph{Phys.
  Rev. Mater.}, 2020, \textbf{4}, 035602\relax
\mciteBstWouldAddEndPuncttrue
\mciteSetBstMidEndSepPunct{\mcitedefaultmidpunct}
{\mcitedefaultendpunct}{\mcitedefaultseppunct}\relax
\EndOfBibitem
\bibitem[Donley \emph{et~al.}(2023)Donley, Narayanan, Wade, Park, Leheny,
  Harden, and Rogers]{donley2023investigation}
G.~J. Donley, S.~Narayanan, M.~A. Wade, J.~D. Park, R.~L. Leheny, J.~L. Harden
  and S.~A. Rogers, \emph{Proc. Natl. Acad. Sci. U.S.A.}, 2023, \textbf{120},
  e2215517120\relax
\mciteBstWouldAddEndPuncttrue
\mciteSetBstMidEndSepPunct{\mcitedefaultmidpunct}
{\mcitedefaultendpunct}{\mcitedefaultseppunct}\relax
\EndOfBibitem
\bibitem[Jaiswal \emph{et~al.}(2016)Jaiswal, Procaccia, Rainone, and
  Singh]{jaiswal2016mechanical}
P.~K. Jaiswal, I.~Procaccia, C.~Rainone and M.~Singh, \emph{Phys. Rev. Lett.},
  2016, \textbf{116}, 085501\relax
\mciteBstWouldAddEndPuncttrue
\mciteSetBstMidEndSepPunct{\mcitedefaultmidpunct}
{\mcitedefaultendpunct}{\mcitedefaultseppunct}\relax
\EndOfBibitem
\bibitem[Nagamanasa \emph{et~al.}(2014)Nagamanasa, Gokhale, Sood, and
  Ganapathy]{hima2014experimental}
K.~H. Nagamanasa, S.~Gokhale, A.~K. Sood and R.~Ganapathy, \emph{Phys. Rev. E},
  2014, \textbf{89}, 062308\relax
\mciteBstWouldAddEndPuncttrue
\mciteSetBstMidEndSepPunct{\mcitedefaultmidpunct}
{\mcitedefaultendpunct}{\mcitedefaultseppunct}\relax
\EndOfBibitem
\bibitem[Fielding(2014)]{fielding2014shear}
S.~M. Fielding, \emph{Rep. Prog. Phys.}, 2014, \textbf{77}, 102601\relax
\mciteBstWouldAddEndPuncttrue
\mciteSetBstMidEndSepPunct{\mcitedefaultmidpunct}
{\mcitedefaultendpunct}{\mcitedefaultseppunct}\relax
\EndOfBibitem
\bibitem[Divoux \emph{et~al.}(2016)Divoux, Fardin, Manneville, and
  Lerouge]{divoux2016shear}
T.~Divoux, M.~A. Fardin, S.~Manneville and S.~Lerouge, \emph{Annu. Rev. Fluid
  Mech.}, 2016, \textbf{48}, 81--103\relax
\mciteBstWouldAddEndPuncttrue
\mciteSetBstMidEndSepPunct{\mcitedefaultmidpunct}
{\mcitedefaultendpunct}{\mcitedefaultseppunct}\relax
\EndOfBibitem
\bibitem[Ovarlez \emph{et~al.}(2009)Ovarlez, Rodts, Chateau, and
  Coussot]{ovarlez2009phenomenology}
G.~Ovarlez, S.~Rodts, X.~Chateau and P.~Coussot, \emph{Rheol. Acta}, 2009,
  \textbf{48}, 831--844\relax
\mciteBstWouldAddEndPuncttrue
\mciteSetBstMidEndSepPunct{\mcitedefaultmidpunct}
{\mcitedefaultendpunct}{\mcitedefaultseppunct}\relax
\EndOfBibitem
\bibitem[Radhakrishnan and Fielding(2016)]{radhakrishnan2016shear}
R.~Radhakrishnan and S.~M. Fielding, \emph{Phys. Rev. Lett.}, 2016,
  \textbf{117}, 188001\relax
\mciteBstWouldAddEndPuncttrue
\mciteSetBstMidEndSepPunct{\mcitedefaultmidpunct}
{\mcitedefaultendpunct}{\mcitedefaultseppunct}\relax
\EndOfBibitem
\bibitem[Goyon \emph{et~al.}(2008)Goyon, Colin, Ovarlez, Ajdari, and
  Bocquet]{goyon2008spatial}
J.~Goyon, A.~Colin, G.~Ovarlez, A.~Ajdari and L.~Bocquet, \emph{Nature}, 2008,
  \textbf{454}, 84--87\relax
\mciteBstWouldAddEndPuncttrue
\mciteSetBstMidEndSepPunct{\mcitedefaultmidpunct}
{\mcitedefaultendpunct}{\mcitedefaultseppunct}\relax
\EndOfBibitem
\bibitem[Bocquet \emph{et~al.}(2009)Bocquet, Colin, and
  Ajdari]{bocquet2009kinetic}
L.~Bocquet, A.~Colin and A.~Ajdari, \emph{Phys. Rev. Lett.}, 2009,
  \textbf{103}, 036001\relax
\mciteBstWouldAddEndPuncttrue
\mciteSetBstMidEndSepPunct{\mcitedefaultmidpunct}
{\mcitedefaultendpunct}{\mcitedefaultseppunct}\relax
\EndOfBibitem
\bibitem[Parisi \emph{et~al.}(2017)Parisi, Procaccia, Rainone, and
  Singh]{parisi2017shear}
G.~Parisi, I.~Procaccia, C.~Rainone and M.~Singh, \emph{Proc. Natl. Acad. Sci.
  U.S.A.}, 2017, \textbf{114}, 5577--5582\relax
\mciteBstWouldAddEndPuncttrue
\mciteSetBstMidEndSepPunct{\mcitedefaultmidpunct}
{\mcitedefaultendpunct}{\mcitedefaultseppunct}\relax
\EndOfBibitem
\bibitem[Dimitriou \emph{et~al.}(2012)Dimitriou, Casanellas, Ober, and
  McKinley]{dimitriou2012rheopiv}
C.~J. Dimitriou, L.~Casanellas, T.~J. Ober and G.~H. McKinley, \emph{Rheol.
  Acta}, 2012, \textbf{51}, 395--411\relax
\mciteBstWouldAddEndPuncttrue
\mciteSetBstMidEndSepPunct{\mcitedefaultmidpunct}
{\mcitedefaultendpunct}{\mcitedefaultseppunct}\relax
\EndOfBibitem
\bibitem[Moorcroft and Fielding(2013)]{moorcroft2013criteria}
R.~L. Moorcroft and S.~M. Fielding, \emph{Phys. Rev. Lett.}, 2013,
  \textbf{110}, 086001\relax
\mciteBstWouldAddEndPuncttrue
\mciteSetBstMidEndSepPunct{\mcitedefaultmidpunct}
{\mcitedefaultendpunct}{\mcitedefaultseppunct}\relax
\EndOfBibitem
\bibitem[Agrawal \emph{et~al.}(2025)Agrawal, Spicer, and
  Garc{\'\i}a-Tu{\~n}{\'o}n]{agrawal2025connecting}
R.~Agrawal, P.~T. Spicer and E.~Garc{\'\i}a-Tu{\~n}{\'o}n, \emph{J. Colloid
  Interface Sci.}, 2025, \textbf{699}, 138226\relax
\mciteBstWouldAddEndPuncttrue
\mciteSetBstMidEndSepPunct{\mcitedefaultmidpunct}
{\mcitedefaultendpunct}{\mcitedefaultseppunct}\relax
\EndOfBibitem
\bibitem[Aime and Truzzolillo(2025)]{aime2025emergent}
S.~Aime and D.~Truzzolillo, \emph{Phys. Rev. Mater.}, 2025, \textbf{9},
  055602\relax
\mciteBstWouldAddEndPuncttrue
\mciteSetBstMidEndSepPunct{\mcitedefaultmidpunct}
{\mcitedefaultendpunct}{\mcitedefaultseppunct}\relax
\EndOfBibitem
\bibitem[Villa \emph{et~al.}(2022)Villa, Edera, Brizioli, Trappe, Giavazzi, and
  Cerbino]{villa2022quantitative}
S.~Villa, P.~Edera, M.~Brizioli, V.~Trappe, F.~Giavazzi and R.~Cerbino,
  \emph{Front. Phys.}, 2022, \textbf{10}, 1013805\relax
\mciteBstWouldAddEndPuncttrue
\mciteSetBstMidEndSepPunct{\mcitedefaultmidpunct}
{\mcitedefaultendpunct}{\mcitedefaultseppunct}\relax
\EndOfBibitem
\bibitem[Piau(2007)]{piau2007carbopol}
J.-M. Piau, \emph{J. Non-Newtonian Fluid Mech.}, 2007, \textbf{144},
  1--29\relax
\mciteBstWouldAddEndPuncttrue
\mciteSetBstMidEndSepPunct{\mcitedefaultmidpunct}
{\mcitedefaultendpunct}{\mcitedefaultseppunct}\relax
\EndOfBibitem
\bibitem[G{\'e}raud \emph{et~al.}(2013)G{\'e}raud, Bocquet, and
  Barentin]{geraud2013confined}
B.~G{\'e}raud, L.~Bocquet and C.~Barentin, \emph{Eur. Phys. J. E}, 2013,
  \textbf{36}, 30\relax
\mciteBstWouldAddEndPuncttrue
\mciteSetBstMidEndSepPunct{\mcitedefaultmidpunct}
{\mcitedefaultendpunct}{\mcitedefaultseppunct}\relax
\EndOfBibitem
\bibitem[Oelschlaeger \emph{et~al.}(2022)Oelschlaeger, Marten, P{\'e}ridont,
  and Willenbacher]{oelschlaeger2022imaging}
C.~Oelschlaeger, J.~Marten, F.~P{\'e}ridont and N.~Willenbacher, \emph{J.
  Rheol.}, 2022, \textbf{66}, 749--760\relax
\mciteBstWouldAddEndPuncttrue
\mciteSetBstMidEndSepPunct{\mcitedefaultmidpunct}
{\mcitedefaultendpunct}{\mcitedefaultseppunct}\relax
\EndOfBibitem
\bibitem[N'gouamba \emph{et~al.}(2019)N'gouamba, Goyon, and
  Coussot]{n2019elastoplastic}
E.~N'gouamba, J.~Goyon and P.~Coussot, \emph{Phys. Rev. Fluids}, 2019,
  \textbf{4}, 123301\relax
\mciteBstWouldAddEndPuncttrue
\mciteSetBstMidEndSepPunct{\mcitedefaultmidpunct}
{\mcitedefaultendpunct}{\mcitedefaultseppunct}\relax
\EndOfBibitem
\bibitem[N'gouamba(2020)]{ngoua2020}
E.~N'gouamba, \emph{PhD thesis}, 2020\relax
\mciteBstWouldAddEndPuncttrue
\mciteSetBstMidEndSepPunct{\mcitedefaultmidpunct}
{\mcitedefaultendpunct}{\mcitedefaultseppunct}\relax
\EndOfBibitem
\bibitem[Besseling \emph{et~al.}(2009)Besseling, Isa, Weeks, and
  Poon]{besseling2009quantitative}
R.~Besseling, L.~Isa, E.~R. Weeks and W.~C.~K. Poon, \emph{Adv. Colloid
  Interface Sci.}, 2009, \textbf{146}, 1--17\relax
\mciteBstWouldAddEndPuncttrue
\mciteSetBstMidEndSepPunct{\mcitedefaultmidpunct}
{\mcitedefaultendpunct}{\mcitedefaultseppunct}\relax
\EndOfBibitem
\bibitem[Thevenaz \emph{et~al.}(1998)Thevenaz, Ruttimann, and
  Unser]{thevenaz1998pyramid}
P.~Thevenaz, U.~E. Ruttimann and M.~Unser, \emph{IEEE transactions on image
  processing}, 1998, \textbf{7}, 27--41\relax
\mciteBstWouldAddEndPuncttrue
\mciteSetBstMidEndSepPunct{\mcitedefaultmidpunct}
{\mcitedefaultendpunct}{\mcitedefaultseppunct}\relax
\EndOfBibitem
\bibitem[Pelletier \emph{et~al.}(2009)Pelletier, Gal, Fournier, and
  Kilfoil]{pelletier2009microrheology}
V.~Pelletier, N.~Gal, P.~Fournier and M.~L. Kilfoil, \emph{Phys. Rev. Lett.},
  2009, \textbf{102}, 188303\relax
\mciteBstWouldAddEndPuncttrue
\mciteSetBstMidEndSepPunct{\mcitedefaultmidpunct}
{\mcitedefaultendpunct}{\mcitedefaultseppunct}\relax
\EndOfBibitem
\bibitem[Cerbino and Trappe(2008)]{cerbino2008differential}
R.~Cerbino and V.~Trappe, \emph{Phys. Rev. Lett.}, 2008, \textbf{100},
  188102\relax
\mciteBstWouldAddEndPuncttrue
\mciteSetBstMidEndSepPunct{\mcitedefaultmidpunct}
{\mcitedefaultendpunct}{\mcitedefaultseppunct}\relax
\EndOfBibitem
\bibitem[Giavazzi \emph{et~al.}(2009)Giavazzi, Brogioli, Trappe, Bellini, and
  Cerbino]{giavazzi2009scattering}
F.~Giavazzi, D.~Brogioli, V.~Trappe, T.~Bellini and R.~Cerbino, \emph{Phys.
  Rev. E}, 2009, \textbf{80}, 031403\relax
\mciteBstWouldAddEndPuncttrue
\mciteSetBstMidEndSepPunct{\mcitedefaultmidpunct}
{\mcitedefaultendpunct}{\mcitedefaultseppunct}\relax
\EndOfBibitem
\bibitem[Berthier(2011)]{berthier2011dynamic}
L.~Berthier, \emph{Physics}, 2011, \textbf{4}, 42\relax
\mciteBstWouldAddEndPuncttrue
\mciteSetBstMidEndSepPunct{\mcitedefaultmidpunct}
{\mcitedefaultendpunct}{\mcitedefaultseppunct}\relax
\EndOfBibitem
\bibitem[Keys \emph{et~al.}(2007)Keys, Abate, Glotzer, and
  Durian]{keys2007measurement}
A.~S. Keys, A.~R. Abate, S.~C. Glotzer and D.~J. Durian, \emph{Nat. Phys.},
  2007, \textbf{3}, 260--264\relax
\mciteBstWouldAddEndPuncttrue
\mciteSetBstMidEndSepPunct{\mcitedefaultmidpunct}
{\mcitedefaultendpunct}{\mcitedefaultseppunct}\relax
\EndOfBibitem
\bibitem[Brizioli \emph{et~al.}(2022)Brizioli, Sentjabrskaja, Egelhaaf,
  Laurati, Cerbino, and Giavazzi]{brizioli2022reciprocal}
M.~Brizioli, T.~Sentjabrskaja, S.~U. Egelhaaf, M.~Laurati, R.~Cerbino and
  F.~Giavazzi, \emph{Front. Phys.}, 2022, \textbf{10}, 893777\relax
\mciteBstWouldAddEndPuncttrue
\mciteSetBstMidEndSepPunct{\mcitedefaultmidpunct}
{\mcitedefaultendpunct}{\mcitedefaultseppunct}\relax
\EndOfBibitem
\bibitem[Wilson \emph{et~al.}(2011)Wilson, Martinez, Schwarz-Linek, Tailleur,
  Bryant, Pusey, and Poon]{wilson2011differential}
L.~G. Wilson, V.~A. Martinez, J.~Schwarz-Linek, J.~Tailleur, G.~Bryant,
  P.~Pusey and W.~C. Poon, \emph{Physical review letters}, 2011, \textbf{106},
  018101\relax
\mciteBstWouldAddEndPuncttrue
\mciteSetBstMidEndSepPunct{\mcitedefaultmidpunct}
{\mcitedefaultendpunct}{\mcitedefaultseppunct}\relax
\EndOfBibitem
\bibitem[Meeker \emph{et~al.}(2004)Meeker, Bonnecaze, and
  Cloitre]{meeker2004slip}
S.~P. Meeker, R.~T. Bonnecaze and M.~Cloitre, \emph{J. Rheol.}, 2004,
  \textbf{48}, 1295--1320\relax
\mciteBstWouldAddEndPuncttrue
\mciteSetBstMidEndSepPunct{\mcitedefaultmidpunct}
{\mcitedefaultendpunct}{\mcitedefaultseppunct}\relax
\EndOfBibitem
\bibitem[Seth \emph{et~al.}(2011)Seth, Mohan, Locatelli-Champagne, Cloitre, and
  Bonnecaze]{seth2011micromechanical}
J.~R. Seth, L.~Mohan, C.~Locatelli-Champagne, M.~Cloitre and R.~T. Bonnecaze,
  \emph{Nat. Mater.}, 2011, \textbf{10}, 838--843\relax
\mciteBstWouldAddEndPuncttrue
\mciteSetBstMidEndSepPunct{\mcitedefaultmidpunct}
{\mcitedefaultendpunct}{\mcitedefaultseppunct}\relax
\EndOfBibitem
\bibitem[Zhang \emph{et~al.}(2017)Zhang, Lorenceau, Basset, Bourouina, Rouyer,
  Goyon, and Coussot]{zhang2017wall}
X.~Zhang, E.~Lorenceau, P.~Basset, T.~Bourouina, F.~Rouyer, J.~Goyon and
  P.~Coussot, \emph{Phys. Rev. Lett.}, 2017, \textbf{119}, 208004\relax
\mciteBstWouldAddEndPuncttrue
\mciteSetBstMidEndSepPunct{\mcitedefaultmidpunct}
{\mcitedefaultendpunct}{\mcitedefaultseppunct}\relax
\EndOfBibitem
\bibitem[Vleminckx \emph{et~al.}(2020)Vleminckx, Jofore, Moldenaers, and
  Clasen]{vleminckx2020effect}
G.~Vleminckx, B.~D. Jofore, P.~Moldenaers and C.~Clasen, \emph{Rheol. Acta},
  2020, \textbf{59}, 435--453\relax
\mciteBstWouldAddEndPuncttrue
\mciteSetBstMidEndSepPunct{\mcitedefaultmidpunct}
{\mcitedefaultendpunct}{\mcitedefaultseppunct}\relax
\EndOfBibitem
\bibitem[Woodbridge \emph{et~al.}(2026)Woodbridge, Amini, Lundell, Tammisola,
  Juel, Poole, and Fonte]{woodbridge2026subyield}
A.~Woodbridge, K.~Amini, F.~Lundell, O.~Tammisola, A.~Juel, R.~J. Poole and
  C.~P. Fonte, \emph{Phys. Rev. Lett.}, 2026, \textbf{136}, 164001\relax
\mciteBstWouldAddEndPuncttrue
\mciteSetBstMidEndSepPunct{\mcitedefaultmidpunct}
{\mcitedefaultendpunct}{\mcitedefaultseppunct}\relax
\EndOfBibitem
\bibitem[Hemingway and Fielding(2020)]{hemingway2020interplay}
E.~J. Hemingway and S.~M. Fielding, \emph{J. Rheol.}, 2020, \textbf{64},
  1147--1159\relax
\mciteBstWouldAddEndPuncttrue
\mciteSetBstMidEndSepPunct{\mcitedefaultmidpunct}
{\mcitedefaultendpunct}{\mcitedefaultseppunct}\relax
\EndOfBibitem
\bibitem[Chan \emph{et~al.}(2023)Chan, Varchanis, Haward, and
  Shen]{chan2023perspective}
S.~T. Chan, S.~Varchanis, S.~J. Haward and A.~Q. Shen, \emph{J. Rheol.}, 2023,
  \textbf{67}, 949--963\relax
\mciteBstWouldAddEndPuncttrue
\mciteSetBstMidEndSepPunct{\mcitedefaultmidpunct}
{\mcitedefaultendpunct}{\mcitedefaultseppunct}\relax
\EndOfBibitem
\bibitem[Coussot \emph{et~al.}(2009)Coussot, Tocquer, Lanos, and
  Ovarlez]{coussot2009macroscopic}
P.~Coussot, L.~Tocquer, C.~Lanos and G.~Ovarlez, \emph{J. Non-Newtonian Fluid
  Mech.}, 2009, \textbf{158}, 85--90\relax
\mciteBstWouldAddEndPuncttrue
\mciteSetBstMidEndSepPunct{\mcitedefaultmidpunct}
{\mcitedefaultendpunct}{\mcitedefaultseppunct}\relax
\EndOfBibitem
\bibitem[Koumakis \emph{et~al.}(2016)Koumakis, Brady, and
  Petekidis]{koumakis2016amorphous}
N.~Koumakis, J.~F. Brady and G.~Petekidis, \emph{J. Non-Newtonian Fluid Mech.},
  2016, \textbf{233}, 119--132\relax
\mciteBstWouldAddEndPuncttrue
\mciteSetBstMidEndSepPunct{\mcitedefaultmidpunct}
{\mcitedefaultendpunct}{\mcitedefaultseppunct}\relax
\EndOfBibitem
\bibitem[Mungan \emph{et~al.}(2019)Mungan, Sastry, Dahmen, and
  Regev]{mungan2019networks}
M.~Mungan, S.~Sastry, K.~Dahmen and I.~Regev, \emph{Phys. Rev. Lett.}, 2019,
  \textbf{123}, 178002\relax
\mciteBstWouldAddEndPuncttrue
\mciteSetBstMidEndSepPunct{\mcitedefaultmidpunct}
{\mcitedefaultendpunct}{\mcitedefaultseppunct}\relax
\EndOfBibitem
\bibitem[Szulc \emph{et~al.}(2022)Szulc, Mungan, and
  Regev]{szulc2022cooperative}
A.~Szulc, M.~Mungan and I.~Regev, \emph{J. Chem. Phys.}, 2022, \textbf{156},
  164506\relax
\mciteBstWouldAddEndPuncttrue
\mciteSetBstMidEndSepPunct{\mcitedefaultmidpunct}
{\mcitedefaultendpunct}{\mcitedefaultseppunct}\relax
\EndOfBibitem
\bibitem[Regev \emph{et~al.}(2013)Regev, Lookman, and
  Reichhardt]{regev2013onset}
I.~Regev, T.~Lookman and C.~Reichhardt, \emph{Phys. Rev. E}, 2013, \textbf{88},
  062401\relax
\mciteBstWouldAddEndPuncttrue
\mciteSetBstMidEndSepPunct{\mcitedefaultmidpunct}
{\mcitedefaultendpunct}{\mcitedefaultseppunct}\relax
\EndOfBibitem
\bibitem[Regev and Lookman(2019)]{regev2019critical}
I.~Regev and T.~Lookman, \emph{J. Phys.: Condens. Matter}, 2019, \textbf{31},
  045101\relax
\mciteBstWouldAddEndPuncttrue
\mciteSetBstMidEndSepPunct{\mcitedefaultmidpunct}
{\mcitedefaultendpunct}{\mcitedefaultseppunct}\relax
\EndOfBibitem
\bibitem[Keim and Arratia(2014)]{keim2014mechanical}
N.~C. Keim and P.~E. Arratia, \emph{Phys. Rev. Lett.}, 2014, \textbf{112},
  028302\relax
\mciteBstWouldAddEndPuncttrue
\mciteSetBstMidEndSepPunct{\mcitedefaultmidpunct}
{\mcitedefaultendpunct}{\mcitedefaultseppunct}\relax
\EndOfBibitem
\bibitem[Reichhardt \emph{et~al.}(2023)Reichhardt, Regev, Dahmen, Okuma, and
  Reichhardt]{reichhardt2023reversible}
C.~Reichhardt, I.~Regev, K.~Dahmen, S.~Okuma and C.~J.~O. Reichhardt,
  \emph{Phys. Rev. Res.}, 2023, \textbf{5}, 021001\relax
\mciteBstWouldAddEndPuncttrue
\mciteSetBstMidEndSepPunct{\mcitedefaultmidpunct}
{\mcitedefaultendpunct}{\mcitedefaultseppunct}\relax
\EndOfBibitem
\bibitem[Kalafatakis and Cerbino(2026)]{kalafatakis2025}
N.~Kalafatakis and R.~Cerbino, \emph{Journal of Rheology}, 2026, \textbf{70},
  165--181\relax
\mciteBstWouldAddEndPuncttrue
\mciteSetBstMidEndSepPunct{\mcitedefaultmidpunct}
{\mcitedefaultendpunct}{\mcitedefaultseppunct}\relax
\EndOfBibitem
\bibitem[Kamani \emph{et~al.}(2023)Kamani, Donley, Rao, Grillet, Roberts,
  Shetty, and Rogers]{kamani2023understanding}
K.~M. Kamani, G.~J. Donley, R.~Rao, A.~M. Grillet, C.~Roberts, A.~Shetty and
  S.~A. Rogers, \emph{J. Rheol.}, 2023, \textbf{67}, 331--352\relax
\mciteBstWouldAddEndPuncttrue
\mciteSetBstMidEndSepPunct{\mcitedefaultmidpunct}
{\mcitedefaultendpunct}{\mcitedefaultseppunct}\relax
\EndOfBibitem
\bibitem[Lema{\^i}tre and Caroli(2009)]{lemaitre2009rate}
A.~Lema{\^i}tre and C.~Caroli, \emph{Phys. Rev. Lett.}, 2009, \textbf{103},
  065501\relax
\mciteBstWouldAddEndPuncttrue
\mciteSetBstMidEndSepPunct{\mcitedefaultmidpunct}
{\mcitedefaultendpunct}{\mcitedefaultseppunct}\relax
\EndOfBibitem
\bibitem[Jacob \emph{et~al.}(2019)Jacob, Moghimi, and
  Petekidis]{jacob2019rheological}
A.~R. Jacob, E.~Moghimi and G.~Petekidis, \emph{Phys. Fluids}, 2019,
  \textbf{31}, 087103\relax
\mciteBstWouldAddEndPuncttrue
\mciteSetBstMidEndSepPunct{\mcitedefaultmidpunct}
{\mcitedefaultendpunct}{\mcitedefaultseppunct}\relax
\EndOfBibitem
\bibitem[Graham(1995)]{graham1995wall}
M.~D. Graham, \emph{J. Rheol.}, 1995, \textbf{39}, 697--712\relax
\mciteBstWouldAddEndPuncttrue
\mciteSetBstMidEndSepPunct{\mcitedefaultmidpunct}
{\mcitedefaultendpunct}{\mcitedefaultseppunct}\relax
\EndOfBibitem
\bibitem[Atal{\i}k and Keunings(2004)]{atalik2004occurrence}
K.~Atal{\i}k and R.~Keunings, \emph{J. Non-Newtonian Fluid Mech.}, 2004,
  \textbf{122}, 107--116\relax
\mciteBstWouldAddEndPuncttrue
\mciteSetBstMidEndSepPunct{\mcitedefaultmidpunct}
{\mcitedefaultendpunct}{\mcitedefaultseppunct}\relax
\EndOfBibitem
\bibitem[Long \emph{et~al.}(2021)Long, Hui, Gong, and
  Bouchbinder]{long2021fracture}
R.~Long, C.-Y. Hui, J.~P. Gong and E.~Bouchbinder, \emph{Annu. Rev. Condens.
  Matter Phys.}, 2021, \textbf{12}, 71--94\relax
\mciteBstWouldAddEndPuncttrue
\mciteSetBstMidEndSepPunct{\mcitedefaultmidpunct}
{\mcitedefaultendpunct}{\mcitedefaultseppunct}\relax
\EndOfBibitem
\end{mcitethebibliography}

\begin{mcitethebibliography}{7}
\providecommand*{\natexlab}[1]{#1}
\providecommand*{\mciteSetBstSublistMode}[1]{}
\providecommand*{\mciteSetBstMaxWidthForm}[2]{}
\providecommand*{\mciteBstWouldAddEndPuncttrue}
  {\def\EndOfBibitem{\unskip.}}
\providecommand*{\mciteBstWouldAddEndPunctfalse}
  {\let\EndOfBibitem\relax}
\providecommand*{\mciteSetBstMidEndSepPunct}[3]{}
\providecommand*{\mciteSetBstSublistLabelBeginEnd}[3]{}
\providecommand*{\EndOfBibitem}{}
\mciteSetBstSublistMode{f}
\mciteSetBstMaxWidthForm{subitem}
{(\emph{\alph{mcitesubitemcount}})}
\mciteSetBstSublistLabelBeginEnd{\mcitemaxwidthsubitemform\space}
{\relax}{\relax}

\bibitem[Kamani and Rogers(2024)]{kamani2024brittle}
K.~M. Kamani and S.~A. Rogers, \emph{Proc. Natl. Acad. Sci. U.S.A.}, 2024,
  \textbf{121}, e2401409121\relax
\mciteBstWouldAddEndPuncttrue
\mciteSetBstMidEndSepPunct{\mcitedefaultmidpunct}
{\mcitedefaultendpunct}{\mcitedefaultseppunct}\relax
\EndOfBibitem
\bibitem[Miyazaki \emph{et~al.}(2006)Miyazaki, Wyss, Weitz, and
  Reichman]{miyazaki2006nonlinear}
K.~Miyazaki, H.~M. Wyss, D.~A. Weitz and D.~R. Reichman, \emph{Europhys.
  Lett.}, 2006, \textbf{75}, 915--921\relax
\mciteBstWouldAddEndPuncttrue
\mciteSetBstMidEndSepPunct{\mcitedefaultmidpunct}
{\mcitedefaultendpunct}{\mcitedefaultseppunct}\relax
\EndOfBibitem
\bibitem[Donley \emph{et~al.}(2020)Donley, Singh, Shetty, and
  Rogers]{donley2020elucidating}
G.~J. Donley, P.~K. Singh, A.~Shetty and S.~A. Rogers, \emph{Proc. Natl. Acad.
  Sci. U.S.A.}, 2020, \textbf{117}, 21945--21952\relax
\mciteBstWouldAddEndPuncttrue
\mciteSetBstMidEndSepPunct{\mcitedefaultmidpunct}
{\mcitedefaultendpunct}{\mcitedefaultseppunct}\relax
\EndOfBibitem
\bibitem[Keane \emph{et~al.}(2025)Keane, Nikoumanesh, Kamani, Rogers, and
  Poling-Skutvik]{keane2025universal}
D.~P. Keane, E.~Nikoumanesh, K.~M. Kamani, S.~A. Rogers and R.~Poling-Skutvik,
  \emph{Phys. Rev. Lett.}, 2025, \textbf{134}, 208202\relax
\mciteBstWouldAddEndPuncttrue
\mciteSetBstMidEndSepPunct{\mcitedefaultmidpunct}
{\mcitedefaultendpunct}{\mcitedefaultseppunct}\relax
\EndOfBibitem
\bibitem[Villa \emph{et~al.}(2022)Villa, Edera, Brizioli, Trappe, Giavazzi, and
  Cerbino]{villa2022quantitative}
S.~Villa, P.~Edera, M.~Brizioli, V.~Trappe, F.~Giavazzi and R.~Cerbino,
  \emph{Front. Phys.}, 2022, \textbf{10}, 1013805\relax
\mciteBstWouldAddEndPuncttrue
\mciteSetBstMidEndSepPunct{\mcitedefaultmidpunct}
{\mcitedefaultendpunct}{\mcitedefaultseppunct}\relax
\EndOfBibitem
\bibitem[Wilson \emph{et~al.}(2011)Wilson, Martinez, Schwarz-Linek, Tailleur,
  Bryant, Pusey, and Poon]{wilson2011differential}
L.~G. Wilson, V.~A. Martinez, J.~Schwarz-Linek, J.~Tailleur, G.~Bryant,
  P.~Pusey and W.~C. Poon, \emph{Physical review letters}, 2011, \textbf{106},
  018101\relax
\mciteBstWouldAddEndPuncttrue
\mciteSetBstMidEndSepPunct{\mcitedefaultmidpunct}
{\mcitedefaultendpunct}{\mcitedefaultseppunct}\relax
\EndOfBibitem
\bibitem[Aime \emph{et~al.}(2023)Aime, Truzzolillo, Pine, Ramos, and
  Cipelletti]{aime2023unified}
S.~Aime, D.~Truzzolillo, D.~J. Pine, L.~Ramos and L.~Cipelletti, \emph{Nat.
  Phys.}, 2023, \textbf{19}, 1673--1679\relax
\mciteBstWouldAddEndPuncttrue
\mciteSetBstMidEndSepPunct{\mcitedefaultmidpunct}
{\mcitedefaultendpunct}{\mcitedefaultseppunct}\relax
\EndOfBibitem
\end{mcitethebibliography}
\end{document}